\documentclass[twocolumn]{aastex63}
\turnoffeditone

\usepackage{amsmath}
\usepackage{enumerate}

\accepted{June 11, 2020}
\submitjournal{ApJ Supplement}

\shorttitle{Forecasting Chemical Abundance Precision}
\shortauthors{Sandford et al.}

\graphicspath{{./}{figures/}}

\begin{document}

\title{Forecasting Chemical Abundance Precision \\
                for Extragalactic Stellar Archaeology}

\correspondingauthor{Nathan Sandford}
\email{nathan\_sandford@berkeley.edu}

\author[0000-0002-7393-3595]{Nathan R. Sandford}
\affiliation{Department of Astronomy, University of California Berkeley, Berkeley, CA 94720, USA}

\author[0000-0002-6442-6030]{Daniel R. Weisz}
\affiliation{Department of Astronomy, University of California Berkeley, Berkeley, CA 94720, USA}

\author[0000-0001-5082-9536]{Yuan-Sen Ting}
\affiliation{Institute for Advanced Study, Princeton, NJ 08540, USA}
\affiliation{Department of Astrophysical Sciences, Princeton University, Princeton, NJ 08544, USA}
\affiliation{Observatories of the Carnegie Institution of Washington, 813 Santa Barbara Street, Pasadena, CA 91101, USA}
\affiliation{Research School of Astronomy and Astrophysics, Australia National University, Cotter Road, ACT 2611, Canberra, Australia}

\begin{abstract}  
Increasingly powerful and multiplexed spectroscopic facilities promise detailed chemical abundance patterns for millions of resolved stars in galaxies beyond the Milky Way (MW).  Here, we employ the Cram\'er-Rao Lower Bound (CRLB) to forecast the precision to which stellar abundances for metal-poor, low-mass stars outside the MW can be measured for 41 current (e.g., Keck, MMT, VLT, DESI) and planned (e.g., MSE, JWST, ELTs) spectrograph configurations.  We show that moderate resolution ($R\lesssim5000$) spectroscopy at blue-optical wavelengths ($\lambda\lesssim4500$ \AA) (i) enables the recovery of 2-4 times as many elements as red-optical spectroscopy ($5000\lesssim\lambda\lesssim10000$ \AA) at similar or higher resolutions ($R\sim 10000$) and (ii) can constrain the abundances of several neutron capture elements to $\lesssim$0.3 dex. We further show that high-resolution ($R\gtrsim 20000$), low S/N ($\sim$10 pixel$^{-1}$) spectra contain rich abundance information when modeled with full spectral fitting techniques.  We demonstrate that JWST/NIRSpec and ELTs can recover (i) $\sim$10 and 30 elements, respectively, for metal-poor red giants throughout the Local Group and (ii) [Fe/H] and [$\alpha$/Fe] for resolved stars in galaxies out to several Mpc with modest integration times.  We \edit1{\deleted{find}\added{show}} that select literature abundances are within a factor of $\sim$2 \edit1{\added{(or better)}} of our CRLBs. 
We suggest that, like ETCs, CRLBs should be used when planning stellar spectroscopic observations.  We include an open source python package, \texttt{Chem-I-Calc}, that allows users to compute CRLBs for spectrographs of their choosing.




\end{abstract}


\section{Introduction} \label{sec:intro}

Absorption features imprinted in the spectrum of a star encode its physical structure and chemical composition. In turn, the chemical composition of individual stars trace the chemistry of the interstellar medium (ISM) at their birth\footnote{Modulo mixing and gravitational settling.}, providing a detailed fossil record of a galaxy’s chemical evolution over cosmic time. Various enrichment processes (e.g., core-collapse and thermonuclear supernovae, stellar winds, neutron star mergers, and gas inflows) each leave a unique chemical signature on their environment, which are captured in the abundance patterns of stars observed today \citep{tinsley:1980}.
Accordingly, the spectra of resolved stars provide a wealth of information on everything from the formation histories of galaxies to detailed nuclear and quantum physics.

However, translating stellar spectra to stellar composition is a non-trivial undertaking that relies on $\sim$200 years of advancement in atomic and stellar physics, astronomical instrumentation, and computational methods.
The field of stellar spectroscopy and chemical abundance measurements has had a rich history since the first recorded Solar spectrum by \citet{fraunhofer:1817} and the subsequent identification of specific elemental absorption features nearly 50 years later \citep[e.g.,][]{kirchhoff_bunsen:1860, kirchhoff:1860, kirchhoff:1863, huggins:1864}. 
As chronicled in \citet{hearnshaw:2010}, it was another $\sim$70 years until the first quantitative abundance measurements were made. Such measurements were only possible after breakthroughs in theoretical physics (e.g., atomic/ionization theory and stellar atmospheres), development of new instrumentation (e.g., blazed gratings, coud\'e spectrographs, and Schmidt cameras), and substantial investment in laboratory experiments (e.g., transition wavelengths, oscillator strengths, and opacities). Together, these advances enabled the pioneering abundance work of \citet{payne:1925, russell:1929, unsold:1938, unsold:1942, stromgren:1940, aller:1942, aller:1946, greenstein:1948} and \citet{wright:1948} upon which modern stellar spectroscopy is founded.

Since the first half of the 20\textsuperscript{th} century,
high-resolution ($R > 10,000$) spectroscopy with broad optical wavelength coverage and high S/N ($>$30 pixel$^{-1}$) has been the gold standard for measuring precise stellar atmospheric parameters and detailed chemical abundance patterns \citep{nissen:2018}. These spectra provide clean, unblended absorption features that can typically be fit with equivalent widths (EW)\footnote{See \citet{minnaert:1934} for an early discussion of equivalent widths.}. 
At the same time, such high-resolution studies are often limited to small numbers of bright stars due to high-dispersion, low throughput, and poor multiplexing capabilities.

In comparison, low- and medium-resolution spectrographs provide the opportunity to observe more and fainter stars, but are burdened with the cost of having (sometimes heavily) blended features that prohibit the use of conventional EW techniques.

As a means around this challenge, a number of studies have employed spectral indices for low-resolution chemical abundance measurements. One especially common index is centered around the Ca II triplet at $\sim$9000 \AA\ \citep[e.g.,][and references therein]{cenarro:2001a, cenarro:2001b, cenarro:2002}. In this method, the strength of a blended spectral feature (e.g., the Ca II triplet) is calibrated to abundance measurements from high-resolution studies \citep[e.g.,][]{olszewski:1991, rutledge:1997, carrera:2013} or to theoretical (i.e., \textit{ab-initio}) spectra generated from stellar atmosphere and spectrum synthesis models \citep{baschek:1959, fischel:1964, bell:1970, bell:1976}\footnote{Similar to how EWs are calibrated for high-resolution studies.}.
However, spectral indices provide only bulk metal abundances (requiring assumptions of chemical abundance patterns) and are restricted to the parameter space of their calibrating stars or models \citep{battaglia:2008, koch:2008a, starkenburg:2010}. 

As computational resources and stellar models continued to improve, it became possible to directly compare theoretical (\textit{ab initio}) spectra to observed spectra on a pixel-by-pixel basis \citep[pioneering examples include][]{cayrel:1969, sneden:1973, sneden:1974, suntzeff:1981, carbon:1982, leep:1986, leep:1987, wallerstein:1987}. This technique leverages the full statistical power of the many absorption lines in a spectrum, yielding precise abundance measurements without use of EWs or spectral indices. These methods have proven powerful for the recovery of detailed abundance patterns from low- and medium-resolution spectra, which contain predominantly weak and blended absorption features.

In the last two decades, massively multiplexed stellar spectroscopic surveys (e.g., RAVE; \citealt{steinmetz:2006}, SEGUE; \citealt{yanny:2009}, LAMOST; \citealt{luo:2015}, GALAH; \citealt{desilva:2015}, APOGEE; \citealt{majewski:2017}, and DESI; \citealt{desi:2016_inst}) have collected millions of spectra of MW stars. Coupled with steady progress in theoretical and laboratory astrophysics, these surveys have revolutionized our ability to collect and interpret the spectra of stars (see reviews by \citealt{allende-prieto:2016, nissen:2018, jofre:2019}). Importantly, they have motivated the development of novel fitting techniques designed to efficiently fit the full spectrum of many stars. Some techniques are data-driven (e.g., The Cannon; \citealt{ness:2015}), some are trained on \textit{ab initio} spectra (e.g., The Payne; \citealt{ting:2019}), and others adopt hybrid methods (e.g., The DD-Payne; \citealt{xiang:2019}). All employ sophisticated statistical techniques (e.g., neural networks, Bayesian inference, and/or machine learning), enabling the precise recovery of dozens of elemental abundances from both low- and high-resolution spectra in modest compute times.



However, extragalactic stellar spectroscopy has yet to experience the same tremendous gains in quantity and quality of abundance measurements as seen for spectroscopy of stars in the MW. This is primarily the result of stars in external galaxies being much fainter and thus more challenging to observe. Generally, only the few brightest stars ($m_V \lesssim 19.5$) in extragalactic systems can be observed at high-resolution, even when using 10m-class telescopes \citep[e.g.,][]{shetrone:1998, shetrone:2001, shetrone:2003, tolstoy:2003, venn:2004, fulbright:2004, walker:2007, walker:2009b, walker:2009a, walker:2015_m2fs, walker:2015_hectochelle, koch:2008a, koch:2008b, cohen:2009, aoki:2009, frebel:2010, frebel:2014, frebel:2016, starkenburg:2013, koch:2014, ji:2016b, ji:2016c, ji:2016a, spencer:2017, venn:2017, spite:2018, hill:2019, theler:2019}\footnote{To date, $\sim$10$^{4}$ stars outside the MW have measured [Fe/H] from $R>10000$ spectroscopy, though most have only been observed over a small ($\sim$100 \AA) range in wavelength.}.

Instead, highly multiplexed low- and moderate-resolution ($R < 10000$) spectrographs on large aperture telescopes have become the work-horse instruments of extragalactic stellar spectroscopy (e.g., DEIMOS; \citealt{faber:2003}). Over the past 20 years, tens of thousands of low- and medium-resolution spectra have been acquired for extragalactic stars. Since detailed abundance measurements were typically viewed as the purview of high-resolution spectroscopy, most of the spectra were taken for the purpose of measuring radial velocities and bulk metallicities with spectral indices \citep[e.g.,][]{suntzeff:1993, tolstoy:2004, pont:2004, battaglia:2006, munoz:2006, simon:2007, koch:2007a, koch:2007b, koch:2009, battaglia:2008, battaglia:2011, norris:2008, leaman:2009, shetrone:2009, kalirai:2010, hendricks:2014, ho:2015, slater:2015, simon:2015, simon:2017, martin:2016a, martin:2016b, swan:2016, li:2017, longeard:2020}.

The ground-breaking work of \citet{kirby:2009} was the first to demonstrate that \edit1{\deleted{accurate}\added{precise}} abundances could be recovered from moderate resolution spectra in external galaxies. Since then, the method has been further refined and applied to thousands of stars in LG galaxies, measuring up to $\sim$10 abundances in MW satellites and $\sim$5 abundances at the distance of M31 \citep[e.g.,][]{kirby:2010, kirby:2015c, kirby:2015b, kirby:2015a, kirby:2017b, kirby:2017a, kirby:2018, kirby:2020, duggan:2018, vargas:2013, vargas:2014a, vargas:2014b, gilbert:2019, escala:2019b, escala:2019a}.

Presently, the field of extragalactic stellar spectroscopy (and with it, the field of extragalactic chemical evolution) is poised for enormous growth. Current and future spectroscopic facilities on large aperture telescopes promise to increase the number of stars outside the MW with observed spectra by at least an order of magnitude. Already, existing spectrographs on 6+ m telescopes have been used to measure abundances of over $\sim$10$^{4}$ stars in LG dwarf galaxies and the halo of M31 \citep[see][and references therein]{suda:2017} and are capable of measuring thousands more.

In the next decade, dedicated spectroscopic surveys on large telescopes (e.g., PFS; \citealt{takada:2014}, MSE; \citealt{MSE:2019}, and FOBOS; \citealt{bundy:2019}) will homogeneously collect 100s of thousands of resolved star spectra in external galaxies. The next decade will also bring JWST and Extremely Large Telescopes (ELTs; e.g., GMT, E-ELT, and TMT), which will make possible spectroscopy of stars in the most distant, faint, and crowded environments in the LG and beyond that are inaccessible to current ground-based facilities.

To fully realize the scientific potential of upcoming massive datasets and to plan for observational campaigns further in the future, it is imperative that we can quantify what we expect to be able to measure from these spectra, and to what precision. While there exist preferred spectral wavelength regions, absorption features, and minimum S/N for abundance measurements, best practices are frequently informally passed down in the community. Comprehensive and quantitative analyses of the chemical information content of spectra given their wavelength coverage, resolution, and S/N are important planning tools, but are sparse in the literature \citep[e.g.,][]{caffau:2013, bedell:2014, hansen:2015, ruchti:2016, ting:2017, feeney:2019}.

In this paper, we employ \textit{ab initio} stellar spectra and the Cram\'er-Rao Lower Bound (CRLB) to quantify the chemical information content of stellar spectra in terms of the precision (not accuracy\footnote{\edit1{\added{See \citet{blanco-cuaresma:2019} and \citet{jofre:2019} for investigations of the systematics present in spectroscopically-derived elemental abundances.}}}) to which elemental abundances can be measured. We apply this method to realistic observing conditions of metal-poor, low-mass stars outside the MW for $>$40 instrument configurations on current (e.g., Keck, LBT, Magellan, MMT, and VLT) and future (e.g., JWST, GMT, TMT, E-ELT, and MSE) spectroscopic facilities.
For this exercise, we assume the use of full-spectrum fitting techniques and adopt many of the assumptions commonly used at present in this field (e.g., 1D LTE models). We note, however, that the techniques we present can readily be adapted for other choices (e.g., when large grids of non-LTE and/or 3D atmospheres become available).

The paper is organized as follows. In \S \ref{sec:information} we provide a technical description of the information content of spectra and how it can be quantified using CRLBs. 
\edit1{\deleted{
We summarize the scope of stars, instruments, and observing scenarios evaluated in this work and our method of stellar spectra generation and CRLB calculation in \S \ref{sec:methods}.}
\added{
In \S \ref{sec:methods} we summarize the scope of stars, instruments, and observing scenarios evaluated in this work, our method of stellar spectra generation, and the assumptions that went into our CRLB calculations.
}}
We report the forecasted stellar abundance precision for current and planned spectrographs in \S \ref{sec:existing} and \ref{sec:future} respectively. We discuss the highlights and caveats of our forecasts in \S \ref{sec:discussion}. In \S \ref{sec:chemicalc} we present \texttt{Chem-I-Calc}, an open-source python package for calculating CRLBs of spectroscopic chemical abundance measurements. We summarize our findings in \S \ref{sec:conclusion} and present a number of technical details in the appendices.
%
%
%
%

\section{Information Content of Spectra} \label{sec:information}
In this section we introduce the notion of a spectrum's information content and its relation to the maximal precision to which stellar labels\footnote{In this work we use ``stellar labels" to broadly encompass both atmospheric parameters (e.g., effective temperature, surface gravity and microturbulent velocity) and elemental abundances. We do not, however, include radial velocities in our analysis.} can be measured. We begin in \S \ref{sec:qualitative} with a qualitative description of the factors that play a role in the degree of information  contained in a stellar spectrum. This is followed by a quantitative description of the information content as represented by the CRLB in \S \ref{sec:quantitative}.

\subsection{A Qualitative Description of Spectral Information} \label{sec:qualitative}
The information content of a star's spectrum determines the precision to which we can measure its stellar labels---or more technically, how broad the stellar labels' posteriors are.
The amount of information and how constraining that information is depends on the following \edit1{\deleted{aspects}\added{intrinsic and observed properties}} of the spectrum:
\begin{enumerate}[(i)]
    \item \textbf{Wavelength Coverage}: How many (and which) spectral features are included in the spectrum.
    \item \textbf{Wavelength Sampling}: How many wavelength pixels are measured per resolution element.
    \item \textbf{Spectral Resolution}: How distinct the spectral features of one label are from those of another label.
    \item \textbf{Flux Covariance}: How uncertain/covariant is the flux in each spectral pixel.
    \item \textbf{Gradient Spectra}: How strongly spectral features respond to changes in the stellar labels.
\end{enumerate}

Aspects (i)-(iv) are determined by the instrument configuration and observing conditions. \edit1{Generally speaking, they set the size and quality of the spectral dataset in question, modulating the availability and accessibility of the spectrum's information}. Larger wavelength coverage and higher wavelength sampling both increase the amount of information-carrying pixels contained in a spectrum. Increased spectral resolution, or resolving power ($R=\lambda/\delta\lambda$), reduces the blending of spectral features and the covariance between stellar labels. Lower flux covariance (i.e., higher S/N) increases the constraining power of informative spectral features. These various characteristics can depend on one another as well \edit1{\deleted{(e.g., spectral resolution and affects the S/N)}\added{(e.g., spectral resolution and wavelength sampling affect the S/N and pixel-to-pixel flux covariance)}} and there are often trade-offs between them for a fixed instrument configuration or observational strategy.

\edit1{The gradient spectra, aspect (v), is \deleted{arguably }the most import factor in determining a star's spectral information content.} 
Generally speaking, it is the stellar labels which result in the largest spectral gradients that have the highest information content and therefore can be recovered to the highest precision. 
In a $\chi^2$ sense, the more strongly a spectral feature responds to a change in stellar labels, the less the labels need to be offset from the true value to result in a large $\chi^2$ value. \edit1{\added{More technically phrased: the expectation of the negative second derivative of the spectrum with respect to the stellar labels gives the Fisher information matrix, which provides a lower bound on the covariance matrix of the stellar labels as discussed in \S \ref{sec:quantitative}.}}

\begin{figure}[ht!] 
	\includegraphics[width=0.5\textwidth]{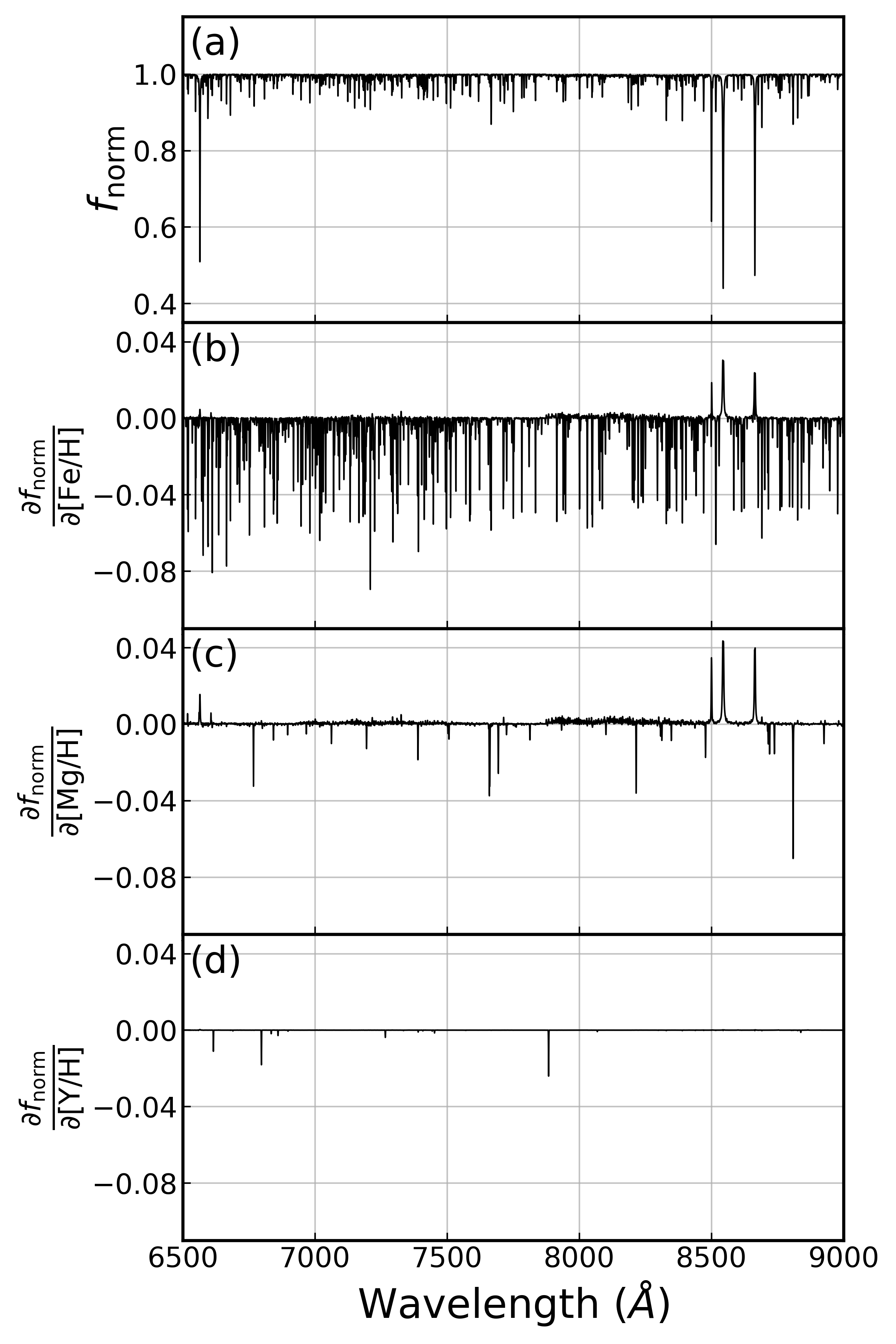}
    \caption{(a) Normalized flux of a \edit1{\added{synthetic}} $\log(Z)=-1.5$ RGB star at $R=6500$ \edit1{\added{generated using \texttt{atlas12} and \texttt{synthe} (see \S\ref{sec:gradients} for model details)}}. (b-d) Gradients of the normalized flux with respect to Fe, Mg, and Y respectively. Many features in the stellar spectrum respond strongly to changes in Fe, meaning that there is considerable information about the iron abundance contained in this spectrum. Changes in Y, on the other hand, cause very weak changes in only a few lines; as a result, the Y abundance would be difficult to recover precisely from this spectrum. Strong positive gradients for Fe and Mg can be seen at the location of the Ca II triplet, which is sensitive to the number of free electrons provided by Fe, Mg, and other electron donors.} 
    \label{fig:gradients}
\end{figure}

Figure \ref{fig:gradients} helps to build an intuition for the importance of spectral gradients. Here, we consider a moderate resolution ($R=6500$) \edit1{\added{\textit{ab initio}}} normalized spectrum of a metal poor ($\log(Z)=-1.5$) red giant branch (RGB) star\footnote{It is important to remember that the gradient spectrum of a star depends on the star's labels. Cool stars, giant stars, and metal-rich stars all have stronger gradients than hot stars, dwarf stars, and metal-poor stars, meaning that it is easier to precisely recover their stellar labels.}
and the partial derivative of that spectrum with respect to
Fe, Mg, and Y\footnote{Unless otherwise stated, elemental abundances are assumed to be in the form of standard Solar-scaled abundance ratios with respect to H i.e., $[\text{X}/\text{H}]=\log_{10}(\text{X}/\text{H}) -\log_{10}(\text{X}/\text{H})_\odot$, where $(\text{X}/\text{H})_\odot$ is the Solar abundance ratio.}.
\edit1{\added{This spectrum and its derivatives were generated using the \texttt{ATLAS12} and \texttt{synthe} models \citep{kurucz:1970, kurucz:1993, kurucz:2005, kurucz:2013, kurucz:2017, kurucz:1981}, which we describe in more detail in \S\ref{sec:gradients}. The locations and strengths of certain features in the spectral gradient may depend on the adopted stellar atmosphere and radiative transfer models, an issue we discuss in \S\ref{sec:caveats}.}}

\edit1{\added{As depicted in panel (b) of Figure \ref{fig:gradients},}}
Fe contributes strongly to a large number of absorption features between 6500 and 9000 $\text{\AA}$, including over 200 lines with changes of $>$1\%/dex and nearly 50 lines with changes of $>$5\%/dex. The large number of information rich lines is the reason why Fe is one of the most readily recovered elements for cool, low-mass stars.

Compared to Fe, Mg contributes to only 20 features at the $>$1\%/dex level and only one that is $>$5\%/dex (at $\lambda$8809 \AA). As a result, it is not as well-constrained as Fe. 
Finally, Y exhibits only three features with gradients larger than 1\%/dex, illustrating the challenge of recovering its abundance, even with favorable telescope (high spectral resolution) and observational (high S/N) configurations. 

Visually exploring the gradients is a particularly informative exercise. For example, there are clear peaks (i.e., positive deviations in the gradient) in the gradient spectra of Fe and Mg at $\sim$8500 \AA. These peaks are not due to Fe or Mg transitions, but rather the Ca II triplet which is sensitive to the number density of free electrons that Fe and Mg contribute. Y, unlike Fe and Mg, is not a key electron donor and thus does not yield a strong gradient at the location of the Ca II triplet.
In this manner, elements that change a star's atmospheric structure or otherwise indirectly affect the line formation of other elements may be measured---even in the absence of strong absorption features of the element in question (e.g., O can be recovered from spectra that contain few, or no, O lines due to its important role in the CNO molecular network; see \citealt{ting:2018}). \edit1{\added{Such measurements, however, require a high degree of trust in the stellar atmosphere and radiative transfer models being used.}}

\subsection{Quantifying Information Content with CRLBs} \label{sec:quantitative}
A main goal of this paper is to quantify the information content encapsulated in the gradient spectrum, modulated by commonly used instrumental setups and realistic observational considerations. To do this, we employ the CRLB \citep{frechet:1943, rao:1945, darmois:1945, cramer:1946}, a formal metric for quantifying information content, which we now describe mathematically.

Suppose that we wish to quantify the information content of a stellar spectra observed using a spectrograph with a wavelength coverage of $\lambda_0\leq\lambda\leq\lambda_N$, a resolving power $R$, and a wavelength sampling of $\Delta\lambda=\lambda/nR$, where $n$ is the number of pixels per resolution element. Let $f_{\text{obs}}(\lambda)$ be the star's continuum normalized flux and $\Sigma$ be the covariance matrix of the normalized flux. 
\edit1{\deleted{We assume that the precision uncertainty due to imperfect continuum normalization is negligible such that $\Sigma$ is due entirely to photon noise and thus is a function of solely exposure time, instrument throughput, observing conditions, and the star's brightness\footnote{Reliably determining the (pseudo-)continuum in practice is challenging and is a potential source of systematic errors (see \S \ref{sec:caveats}). However, self-consistently normalizing both the observed and model spectra can mitigate these systematics. Evaluating these effects is beyond the scope of this paper, which focuses on the possible precision (and not accuracy) of abundance measurements.}. If there are no correlations between adjacent wavelength pixels, then $\Sigma$ is just a diagonal matrix with the variance, $\sigma^2$, at each wavelength point on the diagonal.}}

To make any assessment about the information contained within this spectrum requires a model that relates the star's physical characteristics (e.g., $T_\text{eff}$, $\log(g)$, [Fe/H], [X/Fe]) to its observed spectrum. Suppose we have such a model, $f(\lambda, \theta)$, which predicts the normalized flux of a star at each wavelength, $\lambda$, given a set of stellar labels, $\theta$. The nature of this model, whether it be data-driven \citep[e.g.,][]{ness:2015}, \textit{ab initio} \citep[e.g.,][]{ting:2019}, or a combination of the two \citep[e.g.,][]{xiang:2019}, is unimportant provided that it is generative (i.e., it predicts a normalized flux that mimics the observed spectrum from a set of stellar labels) and differentiable in $\theta$ (i.e., the spectrum varies smoothly as the star's labels change). 

We can then quantify the precision of our measurements by evaluating the log-likelihood of the data given our model\edit1{\deleted{\footnote{We assume a multivariate Gaussian likelihood, which is standard in the fitting of stellar spectra.}}},
\begin{align}
    \ln &L(D|\boldsymbol{\theta}) = \nonumber\\ &-\frac{1}{2}\sum_{i=0}^{N}\Big[\left(f_{\text{obs}}(\lambda_i)-f(\lambda_i,\boldsymbol{\theta})\right)^{T} 
    \Sigma^{-1} 
    \left(f_{\text{obs}}(\lambda_i)-f(\lambda_i,\boldsymbol{\theta})\right)\nonumber\\
    & + \ln\left(2\pi|\Sigma|\right)\Big].
    \label{eq:loglike}
\end{align}
for all $\theta$ (i.e., over all stellar labels).

The precision to which these labels can be recovered is given by the width of this likelihood function.
In practice, however, evaluating the likelihood over a sufficiently large region of parameter space is computationally expensive (and sometimes infeasible) given the high-dimensional nature of spectral fitting\footnote{Note that the number of grid points needed to fully sample the likelihood scales exponentially with the number of dimensions.}. \edit1{\deleted{If one assumes priors on the stellar labels (uniform or otherwise) a Markov chain Monte Carlo (MCMC) method can be used to more efficiently sample the full posterior.}\added{If one assumes priors on the stellar labels (uniform or otherwise) a Markov chain Monte Carlo (MCMC) method can be employed, which enables more efficient sampling of the full posterior than evaluating the likelihood at a grid of labels.}} However, it ultimately still succumbs to the curse of dimensionality when the simultaneous fitting of $>$20 elemental abundances is required. Since we require our model to be differentiable, this can be made more tractable with alternative sampling techniques like the Hamiltonian Monte Carlo algorithm \citep{duane:1987}. Even so, this is still a very computationally expensive exercise to do for every instrument and observational combination.

A more efficient way to obtain the width of the distribution (and in turn the precision on each label) is with the CRLB. Within astrophysics, the CRLB has been used extensively in cosmological contexts \citep[e.g.,][]{albrecht:2006, adshead:2008, wang:2010, becker:2012, betoule:2014, font-ribera:2014, king:2014, eriksen:2015}, but has only recently been applied to \edit1{\added{abundance measurements from full-spectrum}} stellar spectroscopy \citep{ting:2016, ting:2017}\footnote{\citet{ireland:2005} first applied the CRLB formalism to stellar spectroscopy in their analysis of the limiting precision of Solar emission lines. \citet{hansen:2015} later used CRLBs to quantify the precision of EW measurements of blended stellar absorption lines. \edit1{\deleted{CRLBs have also been used to quantify astrometric precision \citep[e.g.,][]{mendez:2013, echeverria:2016}.}}}. 

Formally, the CRLB is the highest possible precision achievable for a set of observations and can be derived from the Fisher information matrix (FIM),
\begin{equation}
    F_{\alpha\beta} = E\left[\frac{\partial^2\left[-\ln L(D|\theta)\right]}{\partial\theta_\alpha\partial\theta_\beta}\right]_{\hat{\theta}},
    \label{eq:FIM_like}
\end{equation}
where $E[.]$ denotes the expectation value, $\hat{\theta}$ is the maximum likelihood estimate, and $\alpha$ and $\beta$ are each a specific \edit1{\deleted{stellar}} label.
In simpler terms, the FIM describes how fast the likelihood function declines for each \edit1{\deleted{stellar}} label around the maximum likelihood point. The steeper the decline, the narrower the distribution, and the more precisely a label can be measured.

Using the Cram\'er-Rao inequality, this curvature can be related directly to the width of the Gaussian likelihood. Specifically, the inverse of the FIM gives the lower bound on the covariance matrix of the \edit1{\deleted{stellar}} labels
\begin{equation}
K_{\alpha\beta} \geq (F^{-1})_{\alpha\beta}
\label{eq:K_FIM}
\end{equation}
or in terms of measurement uncertainty, 
\begin{equation}
    \sigma_{\alpha} \geq \sqrt{(F^{-1})_{\alpha\alpha}}.
    \label{eq:CRLB_FIM}
\end{equation}
This lower bound on the measurement uncertainty, $\sigma_\alpha$, is the CRLB for the \edit1{\deleted{stellar}} label $\alpha$.

\edit1{\added{
In order to apply CRLBs to the fitting of stellar spectra, we must make two fundamental assumptions:
\begin{enumerate}[(i)]
\item The observed spectra have Gaussian noise, and the likelihood of the spectra given our model is well described by a multivariate Gaussian.
\item The spectral models accurately reproduce the observed spectra (i.e., the fitting is  free of systematic errors and $\hat{\theta}$ is an unbiased estimator of a star's true labels)\footnote{The CRLB can be generalized to relax the assumption that $\hat{\theta}$ is an unbiased estimator (see Appendix \ref{app:biased}), but this requires knowing the bias of $\hat{\theta}$ as a function of the stellar labels, which is beyond the scope of this paper.}.
\end{enumerate}
}}
\edit1{\added{
Assuming Gaussianity (i) is standard practice in the fitting of stellar spectra with $\text{S/N} >10$ pixel$^{-1}$ and enables substituting Equation \ref{eq:loglike} for the log-likelihood in Equation \ref{eq:FIM_like}.}}

\edit1{\added{
Though rarely strictly true, the assumption of accurate models (ii) is commonplace across all of astronomy and astrophysics. Model fidelity is a necessary assumption in all matters of parameter estimation, and so we too assume the stellar models to be correct though we know them to have flaws and over-simplifications (e.g., 1D LTE atmospheres, mixing length theory, incomplete linelists, miscalibrated oscillator strengths).
It is important to remember that the CRLBs we calculate are predictions of precision, not accuracy. And while they may be challenging to achieve in practice due to various systematics (see \S\ref{sec:caveats} for further discussion), they nevertheless provide useful guidance for stellar abundance work (see \S\ref{sec:comparison} and Appendix \ref{app:lamost_compare} for a comparison of CRLBs with the abundance precision measured in practice).
}}

\edit1{
\deleted{
In the absence of systematic uncertainties
(e.g., the models are perfect; see \S \ref{sec:caveats} for further discussion),
we can replace $f_{\text{obs}}(\lambda_i)$ in Equation \ref{eq:loglike} with $f(\lambda_i,\hat{\theta})$ and note that the maximum likelihood estimate, $\hat{\theta}$, corresponds to the true stellar labels. Furthermore, having assumed a multivariate normal likelihood in Equation \ref{eq:loglike}, we can re-write Equation \ref{eq:FIM_like} in terms of the gradient spectra as
}\added{
Under the assumption of perfect models we can replace $f_{\text{obs}}(\lambda_i)$ in Equation \ref{eq:loglike} with $f(\lambda_i,\hat{\theta})$, noting that $\hat{\theta}$, as an unbiased estimator, corresponds to the true stellar labels. Combined with the assumption of a multivariate Gaussian log-likelihood, we can re-write Equation \ref{eq:FIM_like} in terms of the gradient spectra as
}}
\begin{align}
    F_{\alpha\beta} = \left[\frac{\partial f(\lambda, \boldsymbol{\theta})}{\partial\theta_\alpha}\right]_{\hat{\theta}}^T
    &\Sigma^{-1}
    \left[\frac{\partial f(\lambda, \boldsymbol{\theta})}{\partial \theta_\beta}\right]_{\hat{\theta}}\nonumber\\
     &+ \frac{1}{2}\text{tr}\left(\Sigma^{-1}\frac{\partial\Sigma}{\partial \theta_\alpha}\Sigma^{-1}\frac{\partial\Sigma}{\partial \theta_\beta}\right)
     \label{eq:FIM_grad_long}
\end{align}
as worked out in \citet{kay:1993}.
Since in the context of stellar spectra the covariance matrix of the normalized flux, $\Sigma$, is independent of the stellar labels, the second term in Equation \ref{eq:FIM_grad_long} vanishes, leaving the FIM as the quadrature sum of the gradient spectra across all wavelength pixels, weighted by the uncertainty of the normalized flux:
\begin{equation}
    F_{\alpha\beta} = \left[\frac{\partial f(\lambda, \theta)}{\partial \theta_\alpha}\right]_{\hat{\theta}}^T
    \Sigma^{-1}
    \left[\frac{\partial f(\lambda, \theta)}{\partial \theta_\beta}\right]_{\hat{\theta}}.
    \label{eq:FIM_grad}
\end{equation}

Using this form of the FIM, we can now write the CRLB in terms of the spectral gradients as
\begin{equation}
   \sigma_{\alpha}
   = \left(\left[\frac{\partial f(\lambda, \theta)}{\partial \theta_\alpha}\right]_{\hat{\theta}}^T
    \Sigma^{-1}
    \left[\frac{\partial f(\lambda, \theta)}{\partial     \theta_\alpha}\right]_{\hat{\theta}} \right)^{-1/2}.
    \label{eq:CRLB}
\end{equation}
Equation \ref{eq:CRLB} shows that the CRLB is sensitive to the factors that affect the information content of spectra as discussed in \S \ref{sec:qualitative}. 
More specifically, if the gradient of the spectrum with respect to a given label is high $\left(\frac{\partial f(\lambda, \theta)}{\partial \theta_\alpha} \text{ is large}\right)$, then $\sigma_{\alpha}$ is small and more precise measurements are possible.

Similarly, having high S/N $\left(\Sigma^{-1} \text{ is large}\right)$ in informative regions of the spectrum will also result in small $\sigma_{\alpha}$ and high possible precision. \edit1{\deleted{Higher}\added{Larger wavelength coverage and higher}} wavelength sampling means summing over more \edit1{\deleted{informative}}pixels and thus higher precision\edit1{\added{, provided that the pixels are informative and not highly correlated}}. The importance of instrumental resolution is embedded in the matrix multiplication, where higher resolution gradients lead to deeper spectral features and less blended features, resulting in smaller covariances between stellar labels. 

An analytic description of the resolution-dependence of the CRLBs is presented in \citet{ting:2017}, which we summarize here:
\begin{enumerate}[(i)]
    \item The \textit{rms} depth per pixel (and information) of an absorption feature in the gradient spectrum scales as $R$.
    \item For fixed exposure time and stellar flux, the S/N scales as $R^{-1/2}$ due to Poisson statistics.
    \item For fixed number of detector pixels, the wavelength range scales as $1/R$. Assuming that absorption features are evenly distributed in wavelength space, the information content scales as $R^{-1/2}$ since information adds in quadrature.
    \item Together, the simple arguments in (i)-(iii) show that to first order the stellar label precision is independent of spectral resolving power.
\end{enumerate}
\edit1{\added{We add to this analytic description that, similar to (iii), the information content scales as $n^{-1/2}$, where $n$ is the number of independent pixels per resolution element. In the extreme case that all $n$ pixels in a resolution element are 100\% correlated, the CRLB will be $\sqrt{n}$ larger than if the pixels were entirely uncorrelated. We present a more detailed exploration of the effects of sampling and pixel-to-pixel correlation on the CRLBs in Appendix \ref{app:sampling}.}}

For a given spectral model (i.e., 1D LTE, as we employ in this work, or 3D non-LTE when they become widely available), forecasting abundance precision is reduced to a matter of calculating derivatives and multiplying matrices. Furthermore, because most spectra have \edit1{\deleted{large $N_{\text{pix}}$}\added{thousands, if not tens of thousands, of pixels}}, the central limit theorem can be used to show that the CRLB becomes theoretically attainable (i.e., Equation \ref{eq:K_FIM} becomes an equality \edit1{\added{if all assumptions hold}}). 
CRLBs are thus an incredibly valuable tool for efficiently exploring the possible precision of a large number of instrumental and observational scenarios when the high-dimensionality of the problem makes more rigorous sampling techniques costly or unfeasible.

\subsubsection{Incorporating Prior Information} \label{sec:including_priors}
In many cases, there may be additional knowledge of the star's properties beyond the spectra in hand. For example, in an extragalactic context, we may know the distance to the star's host galaxy quite well and/or we may have photometry of the star. Such information can give external constraints on the luminosity, surface gravity, temperature, and even metallicity of a star, and can be used to improve the spectral fitting process. We now demonstrate how this information can be included in the CRLB calculation.

While the CRLB was initially derived in a frequentist context, a Bayesian equivalent of the CRLB can be formulated for application to scenarios in which prior information on the stellar labels is available.
This is done by replacing the log-Likelihood in Equation \ref{eq:loglike} with the full Bayesian probability
\edit1{\replaced{
\begin{equation*}
    \ln P(\theta) = \ln\Pi(\theta) + \ln L(\theta|D),
\end{equation*}
}{
\begin{equation}
    \ln P(\theta|D) = \ln\Pi(\theta) + \ln L(D|\theta),
\end{equation}
}}
where $\Pi(\theta)$ is the prior on the stellar labels. This results in the following equation for the Bayesian FIM:
\begin{equation}
    F_{\text{Bayes}} = F_{\text{spec}} + F_{\text{prior}}
\label{eq:FIM_bayes}
\end{equation}
Appendix A of \citet{echeverria:2016} presents a detailed derivations of Equation \ref{eq:FIM_bayes}.

The first term on the right hand side of the equation is the standard spectral gradient FIM found previously (Equation \ref{eq:FIM_grad}). The second term on the right hand side of the equation is the FIM of the prior and encapsulates the additional information included in the prior. It can be shown that for Gaussian priors with standard deviation  
$\sigma_{\text{prior}, \alpha}$ for each stellar label, the prior FIM is the diagonal matrix
\begin{equation}
F_{\text{prior}, \alpha\alpha} = \left(\frac{1}{\sigma_{\text{prior}, \alpha}}\right)^2.
\end{equation}
As a result, we can write the Bayesian CRLB of a stellar label, $\alpha$, with Gaussian priors as 
\begin{eqnarray}
    \sigma_{\text{Bayes},\alpha}& = &\sqrt{(F_{\text{spec}} + F_{\text{prior}})^{-1}_{\alpha\alpha}} \\
    & = &\left(\left[\frac{\partial f(\lambda, \theta)}{\partial \theta_\alpha}\right]_{\hat{\theta}}^T
    \Sigma^{-1}
    \left[\frac{\partial f(\lambda, \theta)}{\partial     \theta_\alpha}\right]_{\hat{\theta}} + \frac{1}{\sigma_{\text{prior},\alpha}^2}\right)^{-1/2}.
\label{eq:CRLB_bayes}
\end{eqnarray}
As a check, we note that in the case of weak priors or strongly informative data, the CRLBs approach the value predicted by Equation \ref{eq:CRLB}, while in the case of strong priors or uninformative data, the CRLBs approach the standard deviation of the priors.

\subsubsection{Combining Information From Multiple Spectra} \label{sec:composite}
The CRLB can also be applied to the context in which multiple disjoint spectra of the same star exist across different wavelength ranges and resolutions, but are to be fit together. Such cases commonly arise for multi-armed spectrographs (e.g.,  Keck/LRIS, LBT/MODS, and DESI) and for echelle spectrographs, which observe multiple discrete orders of the stellar spectrum (e.g., VLT/FLAMES-GIRAFFE).

Replacing the log-likelihood in Equation \ref{eq:FIM_like} with the sum of the \edit1{\added{log-}}likelihoods for each spectra and following through the previous derivation (Equations \ref{eq:FIM_grad_long}-\ref{eq:CRLB}) reveals that the relevant FIM for the joint fitting is simply the sum of the individual spectra's FIM. This is equivalent to concatenating the gradient spectra and covariance matrices of each observation together and using these combined quantities in Equation \ref{eq:CRLB}.
This can be done for arbitrary combinations of stellar spectra provided that the covariance of overlapping wavelength ranges is properly accounted for (as done in \citealt{czekala:2015}) otherwise the number of independent information-carrying pixels is artificially inflated.

\section{Methods} \label{sec:methods}
In this section, we outline our process of generating synthetic stellar spectral gradients and using them to compute CRLBs for a variety of stars, observing scenarios, and spectrographs.
We begin by describing the non-exhaustive scope of instruments (\S \ref{sec:instruments}) and stellar targets (\S \ref{sec:stars}) considered in this work. 
In \S \ref{sec:snr}, we describe the determination of realistic S/N estimates for each spectrograph and stellar target. Lastly, we walk through our methodology for generating gradient spectra in \S \ref{sec:gradients}.
The technical details of the matrix multiplication and inversion used to calculate the CRLBs can be found in Appendix \ref{app:crlb_calc}.

\subsection{Observational Scope} \label{sec:observations}
While the CRLB is broadly applicable to the entire field of resolved star spectroscopy, we choose to focus this work on forecasting the precision possible for spectroscopy of stars outside of the Milky Way (MW). In general, this limits the scope of this work to large aperture ground- and space-based telescopes observing faint, metal-poor red giant branch (RGB) stars at low- and moderate-resolution ($R<10000$). In the rest of this section, we describe in detail our choice of targets, instruments, and observing conditions.

\subsubsection{Properties of Reference Stars} \label{sec:stars}

\begin{figure}[ht!] 
	\includegraphics[width=0.5\textwidth]{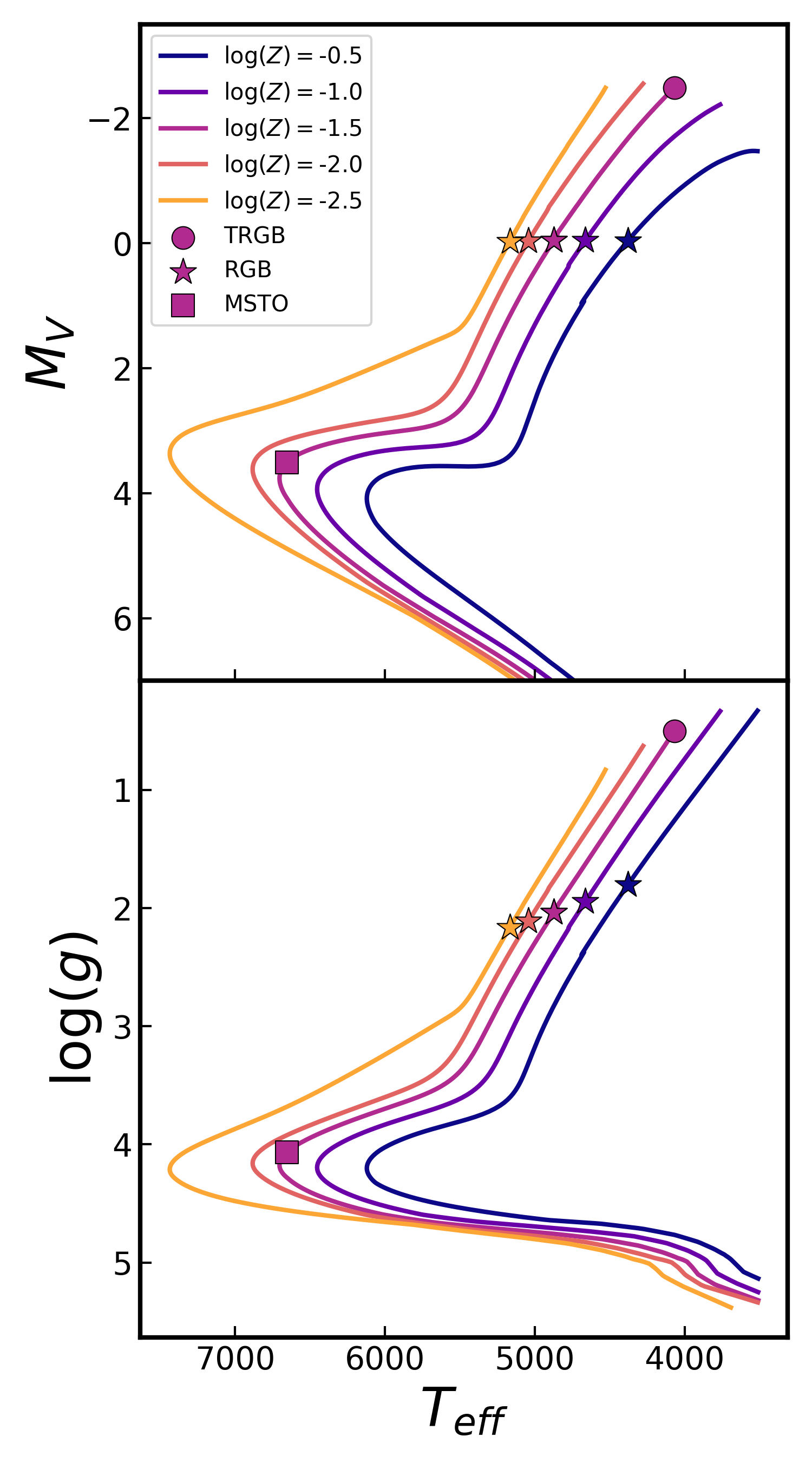}
    \caption{Hertzsprung-Russell (top) and Kiel (bottom) diagrams of the seven reference stars considered in this work (see Table \ref{tab:ref_stars}). Shapes denote stellar evolutionary phase and colors denote metallicity. The five RGB stars of differing metallicity were chosen to have the same V-band absolute magnitude and thus lie on slightly different portions of the RGB. Solid lines are MIST isochrones of a 10 Gyr-old main sequence and red giant branch.} 
    \label{fig:cmd}
\end{figure}

In this work, we limit our analysis to the stars predominantly accessible to spectroscopic campaigns of extragalactic stellar populations: metal-poor RGB stars. We also consider how the CRLBs vary from this fiducial star along several axes, including apparent magnitude, metallicity, and evolutionary phase as described below. The stellar labels used for these reference stars can be found in Table \ref{tab:ref_stars}. Their position in the Kiel and Hertzprung-Russell diagrams can be seen in Figure \ref{fig:cmd}.

For each of the stellar targets considered in this work, we determine the effective temperature and surface gravity of the star using a MIST isochrone corresponding to the star's age, metallicity, and absolute magnitude \citep{paxton:2011, paxton:2013, paxton:2015, dotter:2016, choi:2016}. As was done in \citet{ting:2017}, we assume a microturbulent velocity for each star using the \edit1{\deleted{following}} relationship between microturbulent velocity and surface gravity \edit1{\added{found by}} \citet{holtzman:2015}:
\begin{equation} \label{eq:logg_vmicro}
    v_{\text{turb}} = 2.478 - 0.325\log(g)~\text{km/s}
\end{equation}

\begin{deluxetable}{cDCCCC}[ht!]
    \decimals
	\caption{Stellar labels of the stars considered in this work. \label{tab:ref_stars}}
	\tablehead{\colhead{Phase} & \multicolumn2c{$M_{V}$} & \colhead{$T_\text{eff}$ (K)} & \colhead{$\log(g)$} & \colhead{$v_\text{turb}$ (km/s)} & \colhead{$\log(Z)$}}
	\startdata
		 RGB & -0.5 & 4200 & 1.5 & 2.0 & -0.5 \\
		 RGB & -0.5 & 4530 & 1.7 & 1.9 & -1.0 \\
		\textbf{RGB} & \mathbf{-0}.\mathbf{5} & \textbf{4750} & \textbf{1.8} & \textbf{1.9} & \mathbf{-1}.\mathbf{5}  \\
		 RGB & -0.5 & 4920 & 1.9 & 1.9 & -2.0 \\ 
		 RGB & -0.5 & 5050 & 1.9 & 1.9 & -2.5 \\
		 \tableline
		 MSTO & 3.5 & 6650 & 4.1 & 1.2 & -1.5  \\
		 TRGB & -2.5 & 4070 & 0.5 & 2.3 & -1.5  \\
	\enddata
	\tablecomments{The bold line designates the fiducial stellar reference used throughout this study. All stars have Solar abundance patterns. $T_\text{eff}$ and $\log(g)$ are determined from MIST isochrones given the star's age (10 Gyr), metallicity, and absolute magnitude. $v_{\text{turb}}$ is found using the scaling relationship presented in \citet{holtzman:2015}. For $\log(Z)=-1.5$, $M_V=-0.5$ corresponds to a star roughly halfway up the RGB; for more metal poor stars, the same magnitude corresponds to stars lower on the RGB closer to the main sequence turn-off (see Figure \ref{fig:cmd}).}
\end{deluxetable}

\paragraph{Fiducial Star}
We adopt as our fiducial stellar reference a star that is roughly halfway up the RGB with a V-band absolute magnitude of $M_{V,\text{Vega}}=-0.5$ ($M_{g,\text{AB}}\sim-0.2$). This choice splits the difference between the brighter but rarer stars at the tip of the RGB (TRGB) and the more numerous but fainter main sequence turn-off (MSTO) stars. Furthermore, we assume that this fiducial star is 10 Gyr old, has a metallicity of $\log(Z/Z_\odot)=-1.5$, and has Solar abundance patterns.

\paragraph{Apparent Magnitude}
As can be seen from Equation \ref{eq:CRLB}, the CRLB scales inversely proportional to the S/N of the spectrum. We consider our fiducial star with apparent magnitudes $m_V=18$, 19.5, and 21, but at fixed stellar evolutionary phase, to avoid conflating the effects of S/N and the star's atmospheric parameters. This amounts to observing an identical star at distances of $\sim$50, 100, and 200 kpc, which are typical distances to nearby MW satellites. When not evaluating the effects of S/N on the chemical abundance precision, we assume the star is located at a distance of 100 kpc ($m_V=19.5$).

\paragraph{Metallicity}
We also investigate how the the information content of a RGB star's spectrum changes as its metallicity decreases from $\log(Z)=-0.5$ to $-2.5$. Because the shape of the red giant branch changes as a function of metallicity, we make this comparison at fixed $M_{V}$ instead of fixed evolutionary phase. As a result, the lower metallicity stars considered in this work are located further down the red giant branch (i.e., have higher effective temperature and surface gravity; see Figure \ref{fig:cmd}).

\paragraph{Evolutionary Phase}
To isolate the effect of stellar evolutionary phase on the chemical abundance precision, we compare the CRLBs of our fiducial RGB star to that of a MSTO or RGB star of the same metallicity and apparent brightness.

\subsubsection{Instruments} \label{sec:instruments}
Because the stars we consider in this work are so faint ($m_V=19.5$), we limit our forecasts to instruments, both existing and planned, that can efficiently acquire spectra with modest S/N ($>$15 pixel$^{-1}$) in reasonable amounts of time ($<$1 night).

In practice, this includes instruments on ground-based telescopes with $>$5 m apertures and large-aperture space telescopes. This excludes most of the spectrographs responsible for large MW surveys (e.g., RAVE; \citealt{steinmetz:2006}, SEGUE; \citealt{yanny:2009}, LAMOST; \citealt{luo:2015}, GALAH; \citealt{desilva:2015}, and APOGEE; \citealt{majewski:2017}) and most spectrographs with very high resolving powers ($R>50000$). We do not include any instruments with very low resolving powers ($R<1000$), though there is reason to believe that the information content accessible to very low-resolution grism spectroscopy is still considerable \citep{bailer-jones:2000}. 

Lastly, the linelists\footnote{\url{http://kurucz.harvard.edu/}} we use to generate synthetic spectra are limited in extent to wavelengths between 3000 \AA\ and 1.8 $\mu$m. As such, we exclude instruments observing in the UV and IR despite the significant chemical information that these wavelength regimes contain \citep[e.g.,][]{garcia-perez:2016, roederer:2019, ting:2019}. 

Even with the aforementioned restrictions, the list of spectrographs already on sky suitable for extragalactic stellar spectroscopy is extensive. As shown in Table \ref{tab:instruments}, we consider 12 existing spectrographs at five world-class observing facilities as well as 9 spectrographs that will be coming online within the next decade. Each of these instruments features numerous choices of observing modes, dispersive elements, and other specifications. This flexibility enables a broad range of science, but makes an exhaustive evaluation of each observing configuration infeasible. Instead, we consider only the setups that we believe most relevant to acquiring precise chemical abundances in extragalactic stellar populations for a total of 41 configurations\footnote{This list is extensive but far from complete. We encourage readers interested in spectrographs not listed in Table \ref{tab:instruments} to calculate their own chemical abundance precision using the \texttt{Chem-I-Calc} python package detailed in \S \ref{sec:chemicalc}.}. For each observational setup, we attempt to use realistic wavelength coverage, wavelength sampling, and resolving power as reported either in literature or in design documents.

Despite an extensive literature search, not all pertinent spectrograph details were readily available, and we had to make some assumptions. For example, for several instruments the number of pixels per resolution element could not be found; in these cases we adopt a fiducial wavelength sampling of 3 pixels/FWHM as assumed in \cite{ting:2017}. For multi-object spectrographs, we assume the nominal wavelength coverage for a star observed in the center of the instrument's field of view and ignore the variations in wavelength coverage incurred for off-center stars. Additionally, most instruments have wavelength dependent resolving powers, usually decreasing towards the blue. The manner in which the resolving power changes across the spectrum, known as the line-spread function (LSF), depends on the star's position in the slit and can vary slit to slit. For simplicity, we assume all instruments have a fixed LSF with a resolution approximately equal to the average across the entire spectrum.

Lastly, while we do compare and contrast the forecasted precision of these instruments, we emphasize that the ``best" instrument is largely of a science-dependent nature. There are numerous trade-offs between field of view and multiplexing (see Table \ref{tab:instruments2}), radial velocity precision, and detailed chemical abundance measurements. Balancing them is a matter of their relative importance to the science at hand.

\startlongtable
\begin{deluxetable*}{lccccccc}
    \centerwidetable
	\tablecaption{Spectroscopic configurations used in this work. \label{tab:instruments}}
	\tablehead{ \colhead{Telescope/Instrument}  & \colhead{Spectroscopic}   & \colhead{Wavelength}  & \colhead{$R$}
	                                & \colhead{Sampling}        & \colhead{Aperture}    & \colhead{Section} & \colhead{Reference$^\ddagger$} \vspace{-2mm}\\
	            \colhead{}                      & \colhead{Configuration}   & \colhead{Range (\AA)} & \colhead{$(\lambda/\Delta\lambda)$}
	                                & \colhead{(Pixels/FWHM)}   & \colhead{(m)}         & \colhead{}        & \colhead{}}
	
	\startdata
	    \multicolumn{7}{c}{Existing Instruments}\\
	    \tableline
		 Keck II/DEIMOS\tablenotemark{a}    & 1200G     & \phn\phn6500-9000 & \phn6500  & 4 & 10.0  & \ref{sec:d1200g}      & [1]   \\
		                                    & 1200B     & \phn\phn4000-6400 & \phn4000  & 4 & 10.0  & \ref{sec:blue_keck}   & [1]   \\
		                                    & 600ZD     & \phn\phn4100-9000 & \phn2500  & 5 & 10.0  & \ref{sec:blue_keck}   & [1]   \\
		                                    & 900ZD     & \phn\phn4000-7200 & \phn2500  & 5 & 10.0  & \ref{sec:blue_keck}   & [1]   \\
		Keck I/LRIS\tablenotemark{a}        & 600/4000  & \phn\phn3900-5500 & \phn1800  & 4 & 10.0  & \ref{sec:blue_keck}   & [2]   \\
		                                    & 1200/7500 & \phn\phn7700-9000 & \phn4000  & 5 & 10.0  & \ref{sec:blue_keck}   & [2]   \\
		Keck I/HIRESr\tablenotemark{b}      & B5 Decker & \phn\phn3900-8350 & 49000     & 3 & 10.0  & \ref{sec:single_slit} & [3]   \\
		                                    & C5 Decker & \phn\phn3900-8350 & 35000     & 3 & 10.0  & \ref{sec:single_slit} & [3]   \\
		\tableline
		LBT/MODS\tablenotemark{a}   & Blue Arm  & \phn\phn3200-5500 & \phn1850 & 4 & 11.8   & \ref{sec:other_mos} & [4] \\
		                            & Red Arm   & \phn5500-10500    & \phn2300 & 4 & 11.8   & \ref{sec:other_mos} & [4] \\
		\tableline
		Magellan/MIKEr\tablenotemark{b} & Blue (1".0 slit)  & \phn\phn3500-5000 & 28000 & 4             & \phn6.5   & \ref{sec:single_slit}     & [5]   \\
		                                & Red (1".0 slit)   & \phn5000-10000    & 22000 & 3             & \phn6.5   & \ref{sec:single_slit}     & [5]   \\
		Magellan/M2FS\tablenotemark{c}  & HiRes             & \phn\phn5130-5185 & 18000 & 3$^\dagger$   & \phn6.5   & \ref{sec:single_order}    & [6]   \\
		                                & MedRes            & \phn\phn5100-5315 & 10000 & 3$^\dagger$   & \phn6.5   & \ref{sec:single_order}    & [6]   \\
		\tableline
		MMT/Hectochelle\tablenotemark{c}    & RV31              & \phn\phn5160-5280 & 20000     & 6 & \phn6.5   & \ref{sec:single_order}    & [7]   \\
		MMT/Hectospec\tablenotemark{a}      & 270 mm$^{-1}$     & \phn\phn3900-9200 & \phn1500  & 5 & \phn6.5   & \ref{sec:other_mos}       & [8]   \\
		                                    & 600 mm$^{-1}$     & \phn\phn5300-7800 & \phn5000  & 5 & \phn6.5   & \ref{sec:other_mos}       & [8]   \\
		MMT/Binospec\tablenotemark{a}       & 270 mm$^{-1}$     & \phn\phn3900-9200 & \phn1300  & 4 & \phn6.5   & \ref{sec:other_mos}       & [9]   \\
			                                & 600 mm$^{-1}$     & \phn\phn4500-7000 & \phn2700  & 3 & \phn6.5   & \ref{sec:other_mos}       & [9]   \\
			                                & 1000 mm$^{-1}$    & \phn\phn3900-5400 & \phn3900  & 3 & \phn6.5   & \ref{sec:other_mos}       & [9]   \\
		\tableline
		VLT/MUSE\tablenotemark{d}           &  Nominal          & \phn\phn4800-9300 & \phn2500  & 3$^\dagger$   & \phn8.2   &  \ref{sec:other_mos}      & [10]  \\
		VLT/X-SHOOTER\tablenotemark{b}      & UVB (0".8 slits)  & \phn\phn3000-5500 & \phn6700  & 5             & \phn8.2   & \ref{sec:single_slit}     & [11]  \\
		                                    & VIS (0".7 slits)  & \phn5500-10200    & 11400     & 4             & \phn8.2   & \ref{sec:single_slit}     & [11]  \\
		                                    & NIR (0".9 slits)  & 10200-18000       & \phn5600  & 4             & \phn8.2   & \ref{sec:single_slit}     & [11]  \\
		VLT/FLAMES-UVES\tablenotemark{e}    &  r580              & \phn\phn4800-6800 & 40000     & 5             & \phn8.2   &                                              \ref{sec:single_slit}     &  [12]   \\
		VLT/FLAMES-                         & LR8               & \phn4200-11000    & \phn6500  & 3$^\dagger$   & \phn8.2   & \ref{sec:single_order}    & [13] \\
		\phm{VLT/}GIRAFFE\tablenotemark{c}  & HR10              & \phn\phn5340-5620 & 19800     & 3$^\dagger$   & \phn8.2   & \ref{sec:single_order}    & [13] \\
		                                    & HR13              & \phn\phn6120-6400 & 22500     & 3$^\dagger$   & \phn8.2   & \ref{sec:single_order}    & [13] \\
		                                    & HR14A             & \phn\phn6400-6620 & 28800     & 3$^\dagger$   & \phn8.2   & \ref{sec:single_order}    & [13] \\
		                                    & HR15              & \phn\phn6620-6960 & 19300     & 3$^\dagger$   & \phn8.2   & \ref{sec:single_order}    & [13] \\
		\tableline
	    \multicolumn{7}{c}{Future Instruments}\\
		\tableline
		JWST/NIRSpec\tablenotemark{a}   & G140M/F070LP  & \phn7000-12700    & \phn1000  & 3$^\dagger$   & \phn6.5   & \ref{sec:jwst}    & [14]  \\
		                                & G140M/F100LP  & \phn9700-18400    & \phn1000  & 3$^\dagger$   & \phn6.5   & \ref{sec:jwst}    & [14]  \\
		                                & G140H/F070LP  & \phn8100-12700    & \phn2700  & 3$^\dagger$   & \phn6.5   & \ref{sec:jwst}    & [14]  \\
		                                & G140H/F100LP  & \phn9700-18200    & \phn2700  & 3$^\dagger$   & \phn6.5   & \ref{sec:jwst}    & [14]  \\
		\tableline
		GMT/GMACS\tablenotemark{a}      & Blue Arm (LR) & \phn\phn3200-5500     & \phn1000  & 3             & 24.5  & \ref{sec:elt} & [15]  \\
		                                & Blue Arm (MR) & \phn\phn3700-5500     & \phn2500  & 3             & 24.5  & \ref{sec:elt} & [15]  \\
		                                & Blue Arm (HR) & \phn\phn4200-5000     & \phn5000  & 3             & 24.5  & \ref{sec:elt} & [15]  \\
		                                & Red Arm (LR)  & \phn5500-10000        & \phn1000  & 3             & 24.5  & \ref{sec:elt} & [15]  \\
		                                & Red Arm (MR)  & \phn\phn6100-8900     & \phn2500  & 3             & 24.5  & \ref{sec:elt} & [15]  \\
		                                & Red Arm (HR)  & \phn\phn6700-8300     & \phn5000  & 3             & 24.5  & \ref{sec:elt} & [15]  \\
		GMT/G-CLEF\tablenotemark{a}     & Med Res       & \phn\phn3000-9000     & 35000     & 3             & 24.5  & \ref{sec:elt} & [16]  \\
		TMT/WFOS\tablenotemark{a}       & B1210         & \phn\phn3100-5500     & \phn1500  & 3$^\dagger$   & 30.0  & \ref{sec:elt} & [17]  \\
		                                & B2479         & \phn\phn3300-4750     & \phn3200  & 3$^\dagger$   & 30.0  & \ref{sec:elt} & [17]  \\
		                                & B3600         & \phn\phn3250-4100     & \phn5000  & 3$^\dagger$   & 30.0  & \ref{sec:elt} & [17]  \\
		                                & R680          & \phn5500-10000        & \phn1500  & 3$^\dagger$   & 30.0  & \ref{sec:elt} & [17]  \\
		                                & R1392         & \phn\phn5850-8400     & \phn3200  & 3$^\dagger$   & 30.0  & \ref{sec:elt} & [17]  \\
		                                & R2052         & \phn\phn5750-7250     & \phn5000  & 3$^\dagger$   & 30.0  & \ref{sec:elt} & [17]  \\
		E-ELT/MOSAIC\tablenotemark{a}   &  HMM-Vis      & \phn \phn4500-8000    & \phn5000  & 4             & 39.0  & \ref{sec:elt} & [18]  \\
		                                &  HMM-NIR      & \phn 8000-18000       & \phn5000  & 3             & 39.0  & \ref{sec:elt} & [18]  \\
		\tableline
		Subaru/PFS\tablenotemark{a}     & Blue Arm          & \phn\phn3800-6300 & \phn2300  & 4             & \phn8.2   & \ref{sec:surveys}               & [19]  \\
		                                & Red Arm (LR)      & \phn\phn6300-9400 & \phn3000  & 4             & \phn8.2   & \ref{sec:surveys}               & [19]  \\
		                                & Red Arm (MR)      & \phn\phn7100-8850 & \phn5000  & 4             & \phn8.2   & \ref{sec:surveys}               & [19]  \\
		                                & NIR Arm           & \phn9400-12600    & \phn4300  & 4             & \phn8.2   & \ref{sec:surveys}               & [19]  \\
		MSE\tablenotemark{a}            & Blue Arm (MR)     & \phn\phn3900-5000 & \phn5000  & 3             & 11.3      & \ref{sec:surveys}               & [20]  \\
		                                & Green Arm (MR)    & \phn\phn5750-6900 & \phn5000  & 3             & 11.3      & \ref{sec:surveys}               & [20]  \\
		                                & Red Arm (MR)      & \phn\phn7370-9000 & \phn5000  & 3             & 11.3      & \ref{sec:surveys}               & [20]  \\
		                                & All Arms (LR)     & \phn3600-13000    & \phn3000  & 3             & 11.3      & \ref{sec:surveys}               & [20]  \\
		Keck/FOBOS\tablenotemark{a}     & Proposed          & \phn3100-10000    & \phn3500  & 6             & 10.0      & \ref{sec:surveys}               & [21]  \\
		LAMOST\tablenotemark{a}         &                   & \phn\phn3700-9000 & \phn1800  & 3$^\dagger$   & 4.0       & App.\ \ref{app:lamost_compare}  & [22]  \\
		Mayall/DESI\tablenotemark{a}    & Blue Arm          & \phn\phn3600-5550 & \phn2500  & 3             & \phn4.0   & App.\ \ref{app:desi}            & [23]  \\
		                                & Red Arm           & \phn\phn5550-6560 & \phn3500  & 3             & \phn4.0   & App.\ \ref{app:desi}            & [23]  \\
		                                & Infrared Arm      & \phn\phn6560-9800 & \phn4500  & 3             & \phn4.0   & App.\ \ref{app:desi}            & [23]  \\
	\enddata    
	\tablecomments{This table lists the spectroscopic configurations we adopt for computing the chemical abundance precision as well as the section in which those precisions are presented. For each instrument, we adopt a constant resolution and number of pixels per resolution element across the wavelength range indicated. The instruments listed here span a large range in wavelength coverage (3200 \AA\ - 1.8 $\mu$m), resolving powers ($1000<R<49000$), and instrument designs. \vspace{2mm} \\ 
	$^\dagger$Sampling information was not found so a nominal value of 3 pixels/FWHM is assumed. \\
	$^\ddagger$[1] \citealt{faber:2003}, [2] \citealt{oke:1995}, [3] \citealt{vogt:1994}, [4] \citealt{pogge:2010}, [5] \citealt{bernstein:2003}, [6] \citealt{mateo:2012},
	           [7] \citealt{szentgyorgyi:2011}, [8] \citealt{fabricant:2005}, [9] \citealt{fabricant:2019}, [10] \citealt{bacon:2010}, [11] \citealt{vernet:2011},
	           [12] \citet{dekker:2000}, [13] \citealt{pasquini:2002}, [14] \citealt{bagnasco:2007}, [15] \citealt{depoy:2012}, [16] \citealt{szentgyorgyi:2016},
	           [17] \citealt{pazder:2006}, [18] \citealt{jagourel:2018}, [19] \citealt{tamura:2018}, [20] \citealt{MSE:2019}, [21] \citealt{bundy:2019}, 
	           [22] \citealt{cui:2012}, [23] \citealt{desi:2016_inst}\\
	$^a$Low-/Medium-Resolution Multi-Object Spectrograph\\
	$^b$Single-Slit Multi-Order Echelle Spectrograph\\
	$^c$Multi-Object Single-Order Echelle Spectrograph\\
	$^d$Integral Field Unit Spectrograph\\
	$^e$Multi-Object Multi-Order Echelle Spectrograph
	}
\end{deluxetable*}

\begin{deluxetable}{lcc}
	\tablecaption{Field of view and multiplexing of instruments. \label{tab:instruments2}}
	\tablehead{\colhead{Telescope/Instrument} & \colhead{Field of View} & \colhead{$N_\text{slits}$ or $N_\text{fibers}$}}
	
	\startdata
		Keck II/DEIMOS          & $16'\times4'.0$       & 100   \\
		Keck I/LRIS             & $6'.0\times7'.8$      & 40    \\
		Keck I/HIRESr           & ---                   & 1     \\
		Magellan/MIKEr          & ---                   & 1     \\
		Magellan/M2FS           & $30'.0$               & 250   \\
		MMT/Hectochelle         & $1^{\circ}.0$         & 240   \\
		MMT/Hectospec           & $1^{\circ}.0$         & 300   \\
		MMT/Binospec            & $16'.0\times15'.0$    & 150   \\
		VLT/MUSE                & $1'0\times1'.0$       & ---   \\
		VLT/X-SHOOTER           & ---                   & 1     \\
		VLT/FLAMES-UVES         & $25'.0$               & 8     \\
		VLT/FLAMES-GIRAFFE      & $25'.0$               & 130   \\
		LBT/MODS                & $6'.0\times6'.0$      & 50    \\
		JWST/NIRSpec            & $3'.0\times3'.0$      & 100   \\
		Mayall/DESI             & $2^{\circ}.8$         & 5000  \\
		Subaru/PFS              & $1^{\circ}.3$         & 2400  \\
		MSE                     & $9'.5$                & 3250  \\
		Keck/FOBOS              & $20'.0$               & 1800  \\
		GMT/GMACS               & $7'.4$                & 100   \\
		GMT/GMACS+MANIFEST      & $20'$                 & 100s  \\
		GMT/G-CLEF+MANIFEST     & $20'$                 & 40    \\
		TMT/WFOS                & $4'.2\times9'.6$      & 600   \\
		E-ELT/MOSAIC (HMM-Vis)  & $6'.0$                & 200   \\ 
		E-ELT/MOSAIC (HMM-NIR)  & $6'.0$                & 100   \\ 
	\enddata
	\tablecomments{$N_\text{slits}$ ($N_\text{fibers}$) is the approximate number of slits (fibers) that an instrument can handle in a single pointing. This can be used as a rough estimate for the number of stars a spectrograph can observe simultaneously. In practice, of course, not all slits/fibers can be placed on stars because some may be required for guiding, alignment, or sky-subtraction, while others may go unused simply due to the distribution of stars in the field. Single numbers for the FoV indicate the FoV's diameter, while pairs of number indicate the approximate rectangular dimensions of the FoV. For single-slit spectrographs, the field of view is irrelevant for resolved star spectroscopy. As an IFU, MUSE does not have a fixed number of fibers or slits to assign to stars.}
\end{deluxetable}

\subsubsection{Observing Conditions and Integration Time} \label{sec:snr}
\edit1{\added{
We assume the the flux covariance, $\Sigma$, is due entirely to photon noise and thus is a function of solely exposure time, instrument throughput, observing conditions, and the star's brightness, ignoring any uncertainty introduced by imperfect data reduction or continuum normalization}}\footnote{\edit1{\added{Reliably determining the (pseudo-)continuum in practice is challenging and is a potential source of systematic errors (see \S \ref{sec:caveats}). However, self-consistently normalizing both the observed and model spectra can mitigate these systematics. Evaluating these effects is beyond the scope of this paper.}}}.
Whenever possible, we use the exposure time calculator (ETC) specific to each instrument listed in Table \ref{tab:ETC}. This allows us to \edit1{\deleted{generate mock spectra}\added{adopt a flux covariance}} as specific as possible to each facility and accordingly compute realistic CRLBs. For instruments that do not have public ETCs, we scale the S/N from a similar instrument according to
\edit1{
\replaced{
\begin{equation}\label{eq:snr_scaling}
    \text{S/N}\propto DR^{-1/2},
\end{equation}
}{
\begin{equation}\label{eq:snr_scaling}
    \text{S/N}\propto D(nR)^{-1/2},
\end{equation}
}}
\edit1{\deleted{
where $D$ is the effective aperture of the telescope and $R$ is the instrument's resolving power.
}\added{
where $D$ is the effective aperture of the telescope, $R$ is the instrument's resolving power, and $n$ is the instrument's wavelength sampling.
}}

For our S/N calculations we assume an airmass of 1.1 and a seeing of 0".75 (or as close to these values as possible with each ETC). We assume read-noise is negligible such that the S/N of a single one-hour exposure is the same as that of four 15-minute exposures stacked together. 

Because not all ETCs provide the same stellar spectral energy distribution (SED), we use a K0I, K2V, or K0V spectral template (in preferential order when provided) to best match the SED of our fiducial RGB star. Additionally, we use a K0V spectral template for the RGB reference stars with $\log(Z)\leq-1.5$ and a K5V spectral template for the RGB stars with $\log(Z)>-1.5$.
For the $\log(Z)=-1.5$ MSTO and TRGB reference stars we use G5V and K5III/K5V stellar templates respectively.

Once calculated by the ETC, the S/N is interpolated onto the same wavelength grid as the stellar spectra corresponding to that instrument's resolving power, spectral sampling, and wavelength range. 

\edit1{\deleted{
Assuming no correlations between adjacent wavelength pixels we can write the covariance matrix of the normalized flux, $\Sigma$, as the diagonal matrix 
}\added{
Since most spectrographs are designed to slightly over-sample the spectrum ($\geq$3 pixels/FWHM), adjacent pixels are not completely uncorrelated, though most stellar abundance studies treat them as such (see however \citealt{czekala:2015}). For simplicity, we also assume no correlations between adjacent wavelength pixels so that we can write the covariance matrix of the normalized flux, $\Sigma$, as the diagonal matrix 
}}

\edit1{\replaced{
\begin{equation}\label{eq:snr_scaling}
    \Sigma = \left[
    \begin{array}{ccc}
    \sigma(\lambda_{1}) & & \\
     & \ddots & \\
     & & \sigma(\lambda_{N})\\
    \end{array}\right],
\end{equation}
}{
\begin{equation}\label{eq:covar_diag}
    \Sigma = \left[
    \begin{array}{ccc}
    \sigma^2(\lambda_{1}) & & \\
     & \ddots & \\
     & & \sigma^2(\lambda_{N})\\
    \end{array}\right],
\end{equation}
}}

\edit1{\deleted{
where $\sigma(\lambda_{i})=(\text{S/N})^{-1}$ is the variance in each pixel.
}\added{
where $\sigma^2(\lambda_{i})=(\text{S/N})^{-2}$ is the variance in each pixel. A more accurate treatment of the pixel-to-pixel covariance would effectively reduce the number of independent information-carrying pixels in the spectrum, increasing the CRLB slightly---recall that the CRLB is proportional to $n^{-1/2}$, where $n$ is the number of independent pixels per resolution element. A more in-depth analysis of pixel correlation and wavelength sampling is presented in Appendix \ref{app:sampling}.
}}

The large variety of resolving powers included in this work means that a universal ``observing strategy" can not be applied to all instruments. Instead, we consider separate observing setups for a fiducial spectrograph, low- and medium-resolution spectrographs ($R<10000$), high-resolution spectrographs ($R>10000$), and JWST/NIRSpec, which we describe below. 
A summary of all of the relevant assumptions used in the S/N calculation of each instrument is contained in Table \ref{tab:ETC}.

\paragraph{Fiducial Spectrograph}
To investigate the effects of exposure time, object brightness, and stellar evolutionary phase and metallicity, we adopt the 1200G grating on Keck/DEIMOS as our fiducial spectroscopic setup. We consider 1, 3 and 6 hour integration times and stars with $m_V=18$, 19.5, and 21. For comparisons of metallicity and stellar evolutionary phase we hold the integration time and apparent magnitude fixed at 1 hour and $m_V=19.5$ respectively.

\paragraph{Low- and Medium-Resolution Spectrographs}
For spectrographs with $R<10000$, we consider the baseline observing strategy to be 1 hour of integration of our fiducial $m_V=19.5$ RGB star. This is generally sufficient for spectrographs on 6+ meter telescopes to achieve $\text{S/N}>15$ pixel$^{-1}$ across the optical spectrum. In this category, we include Keck/DEIMOS, Keck/LRIS, MMT/Hectospec, MMT/Binospec, VLT/MUSE, LBT/MODS, Subaru/PFS, MSE, Keck/FOBOS, GMT/GMACS, TMT/WFOS, and E-ELT/MOSAIC. 

The GMACS ETC provides two sample settings, each of which assume a constant $\delta\lambda$ across both the blue and red channels, resulting in wavelength dependent resolutions. We choose the higher resolution setting ($\Delta\lambda=1.4$) and scale the S/N at each pixel according to $\text{S/N}\propto R^{-1/2}$ to match the constant resolving power we are attempting to emulate. Because ETCs do not yet exist for \edit1{\deleted{WFOS and}} MOSAIC, we scale the S/N from GMACS for \edit1{\deleted{WFOS and}} MOSAIC (HMM-Vis) and from JWST/NIRSpec for MOSAIC (HMM-NIR) according to Equation \ref{eq:snr_scaling}\footnote{By using the ETC of space-based NIRSpec for MOSAIC (HMM-NIR), we ignore a number of telluric features that affect observations in the NIR.}.


\paragraph{High-Resolution Spectrographs}
Due to the higher dispersion and generally lower throughput of high-resolution spectrographs, a single hour of integration is insufficient to achieve adequate S/N ($>$15 pixel$^{-1}$) for a $m_V=19.5$ RGB star. Instead we consider an integration of 6 hours ($\sim$1 night of observing). Instruments in this category include Keck/HIRES, Magellan/MIKE, Magellan/M2FS, MMT/Hectochelle, VLT/X-SHOOTER\footnote{Despite the more moderate resolution of the X-SHOOTER UVB and NIR arms, we include X-SHOOTER with the other high-resolution spectrographs due to its higher resolution VIS arm and single-slit echelle design.},
VLT/FLAMES-GIRAFFE, and GMT/G-CLEF. M2FS and Hectochelle do not have public ETCs so we scale the average S/N from the GIRAFFE HR10 ETC according to Equation \ref{eq:snr_scaling} and assume the S/N is roughly constant over the short wavelength range observed by these instruments.

\paragraph{JWST/NIRSpec}
The strength of JWST/NIRSpec is its high sensitivity and high angular resolution. The most likely use case will be to acquire spectra in distant and/or crowded environments, which may require longer integration times than our fiducial 1 hour setup for ground-based low-resolution instruments. Thus, for JWST only, we adopt a 6 hours of integration on a $m_V=21$ TRGB star\footnote{Specifically we assume three exposures each of which includes one integration of 170 groups (sub-integrations) for a total exposure time of 6 hours 5 minutes and 35 seconds.}. This scenario is chosen to mimic the observation of bright stars in the disk of M31 or in a galaxy at the edge of the Local Group.

\paragraph{Beyond 1 Mpc}
To investigate the distance to which JWST/NIRSpec and GMT/GMACS (as a representative ELT) can provide useful chemical measurements, we additionally hold the exposure time constant at 6 hours and systematically decrease the apparent magnitude of our target TRGB star from $m_V=21$ to 26. This corresponds to observing a TRGB star at distances between 0.5 and 5 Mpc.

\begin{deluxetable*}{ccccccccc}[ht!] \label{tab:ETC}
    \decimals
	\caption{ETC configurations used in this work.}
	\tablehead{\colhead{Instrument} & \colhead{$m_V$} & \colhead{$t_{exp}$} &\colhead{Airmass} &\colhead{Seeing} &\colhead{Slitwidth/}& \colhead{Spatial$\times$Spectral} & \colhead{Stellar} & \colhead{ETC} \vspace{-2mm}\\  
	\colhead{} & \colhead{} & \colhead{(hours)} & \colhead{} & \colhead{} & \colhead{Fiber Diameter} & \colhead{Binning} & \colhead{Template} & \colhead{}} 
	\startdata
		DEIMOS          & 18.0, 19.5, 21.0 & 1, 3, 6   & 1.1 & 0".75 & 0".75        & $1\times1$            & G5V, K0V, K5V     & 1\\
		LRIS$^\ddagger$ & 19.5             & 1         & 1.1 & 0".75 & 0".70        & $1\times1$            & K0V               & 2\\
		HIRESr (B5/C5)  & 19.5             & 6         & 1.1 & 0".75 & 0".86/1".10  & $2\times2$            & K0V               & 3\\
		MIKE            & 19.5             & 6         & 1.1 & 0".75 & 1".00        & $3\times1$            & K0V               & 4\\
		M2FS            & 19.5             & 6         & 1.1 & 0".75 & 1".20        & $2\times2$            & K2V               & 5$^\dagger$\\
		Hectochelle     & 19.5             & 6         & 1.1 & 0".75 & 1".00        & $3\times2$            & K2V               & 5$^\dagger$\\
		Hectospec       & 19.5             & 1         & 1.1 & 0".75 & 1".5         & $1\times1$            & K0V               & 6\\
		Binospec        & 19.5             & 1         & 1.1 & 0".75 & 1".0         & $1\times1$            & K0V               & 6\\
		MUSE            & 19.5             & 1         & 1.1 & 0".80 & ---$^*$      & $(3\times3)\times1$   & K2V               & 7\\
		X-SHOOTER       & 19.5             & 6         & 1.1 & 0".75 & 0".80/0".70/ & $1\times1$            & K2V               & 8\\
		(UVB/VIS/NIR)   &                  &           &     &       & 0".90        &                       &                   & \\
		UVES            & 19.5             & 6         & 1.1 & 0".80 & 1".00        & $1\times1$            & K2V               & 9\\
		GIRAFFE         & 19.5             & 6         & 1.1 & 0".75 & 1".20        & $1\times1$            & K2V               & 5\\
		MODS            & 19.5             & 1         & 1.1 & 0".75 & 0".70        & $1\times1$            & K2V               & 10\\
		NIRSpec         & 21.0-26.0        & 6         & --- & ---   & 0".2         & $1\times1$            & K5III             & 11\\
		PFS             & 19.5             & 1         & 1.1 & 0".75 & 1".05        & $1\times1$            & K2V               & 12\\
		MSE             & 19.5             & 1         & 1.0 & 0".75 & 0".80        & $1\times1$            & K2V               & 13\\
		FOBOS           & 19.5             & 1         & 1.1 & 0".75 & 0".80        & $1\times1$            & K2V               & 14\\
		GMACS           & 19.5, 21.0-26.0  & 1, 6      & 1.1 & 0".75 & 0".70        & $4\times4$            & K0V, K5V          & 15\\
		WFOS            & 19.5             & 1         & 1.1 & 0".75 & 0".75        & $1\times1$            & K0V               & 14\\
		MOSAIC (NIR/Vis)& 19.5             & 1         & 1.1 & 0".75 & 0".80/0".60  & $1\times1$            & K0I/V             & 10$^\dagger$/14$^\dagger$ \\
		G-CLEF          & 19.5             & 6         & 1.0 & 0".79 & 0".70        & $6\times9$            & K2V               & 16\\
	\enddata
	\tablecomments{Exposure times are chosen to mimic realistic observing strategies for each instrument. Multiple apparent magnitudes, exposure times, and stellar templates are used with the fiducial 1200G grating on the Keck/DEIMOS spectrograph to investigate their effects on chemical abundance precision. Stellar templates are chosen to best match the stellar energy distribution of the relevant reference star. \vspace{2mm} \\ 
	$^\dagger$S/N adapted from ETC of similar instrument according to Equation \ref{eq:snr_scaling}.\\
	$^\ddagger$The LRIS ETC does not include the 1200/7500 grating throughput so the 1200/9000 grating throughput is used in its place.\\
	$^*$As an IFU, MUSE does not have a definite fiber or slit size on the sky.\\
	$^1$DEIMOS ETC: \url{http://etc.ucolick.org/web_s2n/deimos}\\
	$^2$LRIS ETC: \url{http://etc.ucolick.org/web_s2n/lris}\\
	$^3$HIRES ETC: \url{http://etc.ucolick.org/web_s2n/hires}\\
	$^4$LCO ETC: \url{http://alyth.lco.cl/gblanc_www/lcoetc/lcoetc_sspec.html}\\
	$^5$GIRAFFE ETC: \url{https://www.eso.org/observing/etc/bin/gen/form?INS.NAME=GIRAFFE+INS.MODE=spectro}\\
	$^6$SAO ETC v0.5: \url{http://hopper.si.edu/etc-cgi/TEST/sao-etc}\\
	$^7$MUSE ETC: \url{eso.org/observing/etc/bin/gen/form?INS.NAME=MUSE+INS.MODE=swspectr}\\
	$^8$X-SHOOTER ETC: \url{https://www.eso.org/observing/etc/bin/gen/form?INS.NAME=X-SHOOTER+INS.MODE=spectro}\\
	$^9$UVES ETC: \url{https://www.eso.org/observing/etc/bin/gen/form?INS.NAME=UVES+INS.MODE=FLAMES}\\
	$^{10}$MODS Instrumental Sensitivity: \url{http://www.astronomy.ohio-state.edu/MODS/ObsTools/Docs/MODS1_InstSens.pdf}\\
	$^{11}$JWST ETC: \url{https://jwst.etc.stsci.edu/} (Workbooks available upon request.)\\
	$^{12}$PFS ETC and Spectrum Simulator: \url{https://github.com/Subaru-PFS/spt_ExposureTimeCalculator}\\
	$^{13}$MSE ETC: \url{http://etc-dev.cfht.hawaii.edu/mse/}\\
	$^{14}$FOBOS/WFOS ETC: \url{https://github.com/Keck-FOBOS/enyo}\\
	$^{15}$GMACS ETC v2.0: \url{http://instrumentation.tamu.edu/etc_gmacs/}\\
	$^{16}$G-CLEF ETC: \url{http://gclef.cfa.harvard.edu/etc/}\\
	}
\end{deluxetable*}

\subsection{Gradient Spectra} \label{sec:gradients}
\textit{Ab initio} spectra are generated using the same method as described in \cite{ting:2017}. Briefly, we first compute 1D LTE model atmospheres using the \texttt{atlas12} code maintained by R.\ Kurucz \citep{kurucz:1970, kurucz:1993, kurucz:2005, kurucz:2013, kurucz:2017, kurucz:1981}. 
We adopt Solar abundances from \cite{asplund:2009} and assume the standard mixing length theory with a mixing length of 1.25 and no overshooting for convection\footnote{We note that these are not identical assumptions to those made in the MIST isochrones used in \S \ref{sec:stars}. This may have a small impact on the consistency of the bolometric magnitudes of the reference stars, but shouldn't otherwise affect the results presented in this paper.}.
We then evaluate spectra for these atmospheres at a nominal resolution of $R=300,000$ using the \texttt{synthe} radiative transfer code (also maintained by R.\ Kurucz). The spectrum is then continuum normalized using the theoretical continuum from \texttt{synthe}\footnote{Again, the use of imperfectly continuum normalized spectra here should not dramatically change the results of this work as long as all spectra are self-consistently normalized.}. These high-resolution, normalized spectra are then subsequently convolved down to the average resolution of the relevant instrument (assuming a uniform Gaussian LSF) and finally sub-sampled onto a wavelength grid with $\Delta\lambda/nR$, where $n$ is the number of pixels per resolution element.

To calculate stellar spectral gradients for each label, we generate a grid of 200 mock spectra, each with one of 100 stellar labels offset from the star's reference labels (see Table \ref{tab:ref_stars}) by
\begin{eqnarray*}
    \Delta T_\text{eff}& = &\pm 50~\text{K}, \\
    \Delta \log g& = &\pm 0.1, \\
    \Delta v_{\text{turb}}& = &\pm 0.1~\text{km/s, or} \\
    \Delta [X/H]& = &\pm 0.05,
\end{eqnarray*}
where $X$ refers to elements with atomic numbers between 3 and 99. These step sizes are chosen to be small enough such that the spectral response to each label change is approximately linear, but large enough that the spectral responses remain dominant over numerical noise ($>0.1\%$). For each spectrum in which the abundance of an element is changed, the hydrogen mass fraction is re-normalized to compensate, while the helium mass fraction remains constant\footnote{We opt not to calculate gradients with respect to the helium fraction, but recognize that this may be of relevance to abundance measurements of hot ($T_\text{eff}>8500$ K) stars in globular clusters or other environments where light element variations are common (see review by \citealt{bastian:2018} and references \edit1{\replaced{therin}{therein}}).}.

As in \cite{ting:2017}, we re-evaluate the atmospheric structure whenever a stellar label is varied. While more computationally expensive, this is not only essential to capture the response of the spectrum with respect \edit1{\added{to the}} atmospheric parameters (i.e., $T_\text{eff}$, $\log(g)$, and $v_{micro}$), but is also important for certain elemental abundances that have substantial impact on the star's atmospheric structure (see \citealt{ting:2016} for details). For example, Mg and Fe are both major electron donors in the atmospheres of cool stars and effect the absorption features of many other elements (Figure \ref{fig:gradients}). While not necessary for all elemental abundances (e.g., Y, which contributes negligibly to the atmosphere's structure), we nevertheless recompute the stellar atmosphere in all cases for consistency.

The final step is to calculate the gradients via the finite difference method. In past work, \citet{ting:2017}
calculated an asymmetric approximation of the gradient of the spectrum with respect to each stellar label by considering the difference of the reference spectrum and the spectra with offsets in that label. In this work, we use a symmetric approximation of the gradient, using the two spectra offset positively and negatively from the reference spectra as we find it yields a more accurate instantaneous derivative at the location of the reference labels. Thus the gradient of the spectrum with respect to each stellar label, $\alpha$, evaluated at the reference point $\theta$ is
\begin{equation}
   \frac{\partial f(\lambda, \theta)}{\partial \theta_\alpha} = \frac{f(\lambda, \theta+\Delta\theta_\alpha) - f(\lambda, \theta-\Delta\theta_\alpha)}{2\Delta\theta_\alpha}.   
\end{equation}

\subsection{Summary of Assumptions} \label{sec:methods_assumptions}
\edit1{\added{
For reference, we provide a list of the simplifying assumptions employed throughout our methods. This does not include any assumptions inherent to the derivation of the CRLBs in \S\ref{sec:quantitative}.}}

\edit1{\added{
\textbf{Stellar Model Assumptions:}
\begin{itemize}
    \item \texttt{atlas12} stellar atmosphere model (1D LTE; mixing length of 1.25; no overshoot for convection)
    \item \texttt{synthe} radiative transfer code
    \item Perfectly normalized spectra
    \item MIST stellar isochrones
    \item Solar abundance patterns
    \item \citet{holtzman:2015} empirical relationship between surface gravity and microturbulent velocity
    \item SED approximated by a K0I, K2V, or K0V spectral template
\end{itemize}
}}

\edit1{\added{
\textbf{Instrument Assumptions:}
\begin{itemize}
    \item Gaussian LSF constant with wavelength
    \item Nominal wavelength sampling of 3 pixels/FWHM adopted when unknown
    \item No correlations between adjacent pixels
    \item Negligible read noise
    \item Same instrument throughput when scaling the S/N using Equation \ref{eq:snr_scaling}
\end{itemize}
}}

\section{Forecasted Precision of Existing Instruments} \label{sec:existing}
Having established how to calculate CRLBs, we are adequately positioned to forecast the chemical abundance precision of existing instruments. With an emphasis on extragalactic stellar spectroscopy, we begin with a thorough analysis of our fiducial instrument setup: the 1200G grating on Keck/DEIMOS
. We then proceed to forecast the precision of other low- and moderate-resolution multi-object spectrographs (MOS) on large ground-based telescopes, emphasizing those with wavelength coverage bluer than 5000 \AA. Finally, we investigate the capability of low S/N high-resolution spectroscopy for precise abundance measurements. With the exception of the analysis in \S \ref{sec:d1200g_priors}, we assume uniform priors on all stellar labels throughout this section.

\subsection{D1200G: A Fiducial Example} \label{sec:d1200g}
Though designed with galaxy spectra in mind, the DEIMOS spectrograph on the 10-m Keck telescope has been critical to our understanding of the resolved stellar populations and chemical evolution of dwarf galaxies. Over the past two decades, observational campaigns with DEIMOS have measured spectra of nearly 10,000 stars in roughly 60 Local Group dwarf galaxies and the halo of M31 \citep[e.g.,][]{chapman:2005, martin:2007, simon:2007, kirby:2010, collins:2013, vargas:2014a, vargas:2014b, martin:2016a, martin:2016b, kirby:2018}. The majority of these observations have been made with the 1200G grating centered at 7000 \AA\ (see Table \ref{tab:instruments} for details). We will refer to this observational setup as D1200G throughout this work.

In the years immediately following the commissioning of DEIMOS, its primary scientific application was the measurement of radial velocities \citep[e.g.,][]{chapman:2005, martin:2007, simon:2007}. Stellar chemistry was often a secondary goal, particularly since high-resolution spectroscopy was often assumed to be necessary for any reliable abundance determinations \citep[see ][and references therin]{tolstoy:2009}. \citet{kirby:2009} demonstrated that the D1200G setup on Keck (and medium-resolution spectroscopy more generally) could be used to recover accurate abundances. Since then, D1200G has become a predominant observing mode for resolved star abundance measurements in dwarf galaxies, making it an excellent fiducial setup for our CRLB calculations.

For this exercise we consider 1, 3, and 6 hours of integration on our fiducial [Fe/H$]=-1.5$ RGB with apparent magnitudes of $m_V=18$, 19.5, and 21.0 (or equivalently at 50, 100, and 200 kpc). The S/N in each case is calculated using the public exposure time calculator (ETC) according to the configurations in Table \ref{tab:ETC}.

\begin{figure*}[ht!]
	\includegraphics[width=\textwidth]{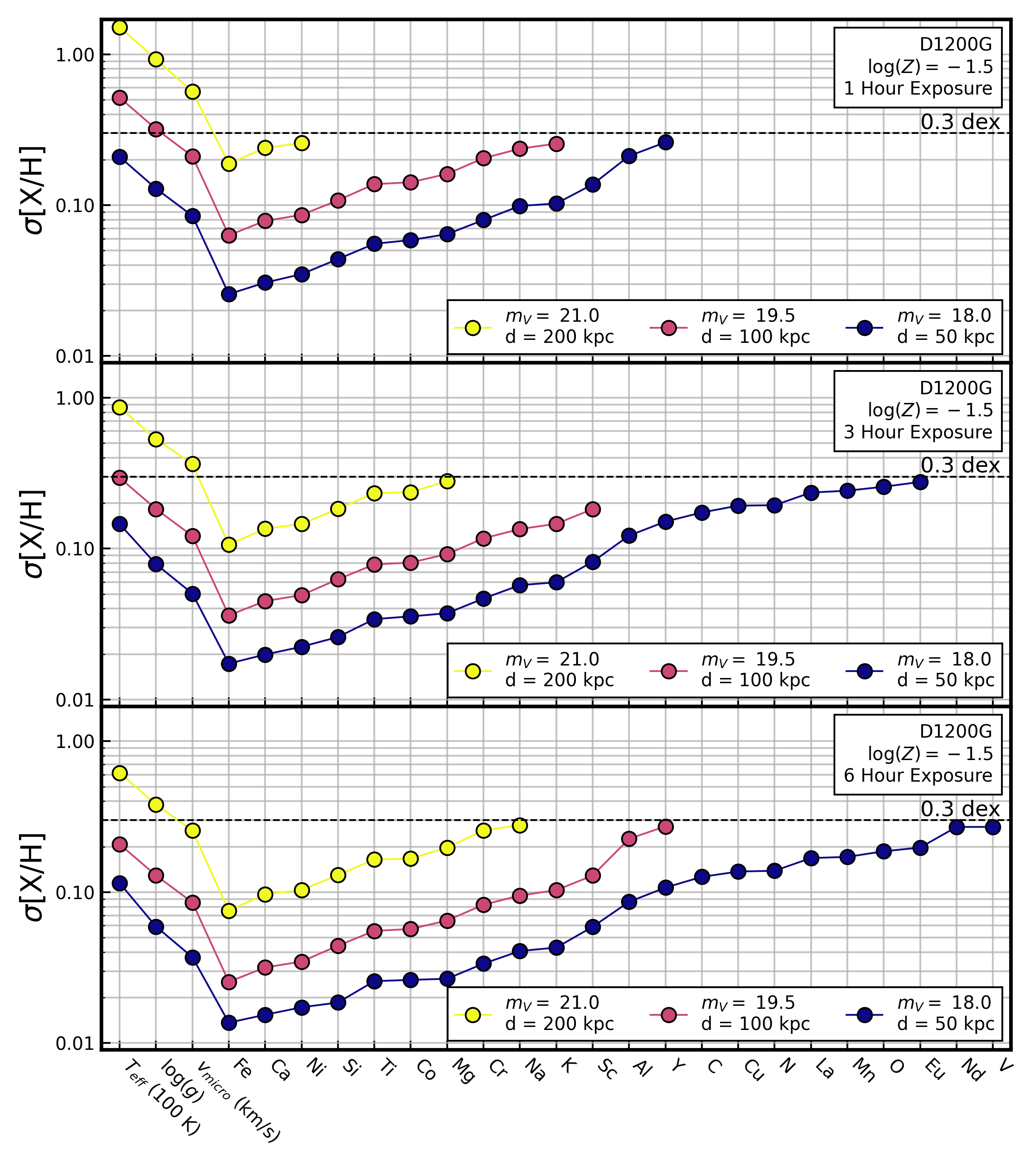}
    	\caption{CRLBs for 1, 3, and 6 hour exposures (top, middle, and bottom respectively) of a $\log(Z)=-1.5$, $M_V=-0.5$ RGB star (see Table \ref{tab:ref_stars}) using the 1200G grating on Keck/DEIMOS (see Table \ref{tab:instruments}). Each panel includes the CRLBs for the RGB star located at a distance of 50, 100, and 200 kpc. The elements are ordered by decreasing precision up to 0.3 dex.}
   	 \label{fig:crlb_d1200g}
\end{figure*}

The CRLBs for  D1200G are displayed in Figure \ref{fig:crlb_d1200g}. Throughout this work, we report precisions for Solar-scaled relative abundances with respect to hydrogen (i.e., $\sigma$[X/H])\footnote{The precision of abundances with respect to Fe (i.e., $\sigma$[X/Fe]) can be found by adding $\sigma\text{[X/H]}$ and $\sigma\text{[Fe/H]}$ in quadrature.}. We consider $\sigma_{CRLB}=0.3$ dex to be the worst precision that still enables useful science and thus restrict our analysis to those that can be recovered to this precision or better.
We forecast that one hour on D1200G is sufficient to measure 13 elements to better than 0.3 dex in RGB stars out to 50 kpc, 10 elements out to 100 kpc, and 3 elements out to 200 kpc.

As expected from the many features seen in the gradient spectrum (Figure \ref{fig:gradients}b), the Fe abundance is recovered to the highest precision. The many strong (and weak) Fe lines included in the D1200G spectrum lead to a precision of 0.02 dex at 50 kpc and to better than 0.2 dex at 200 kpc in only one hour of integration. Ni and Si are also precisely recovered due to their numerous features ($\sim$40 lines with gradients $>$1\%/dex) in the red-optical.
The high precision possible for Ca, however, is predominantly a result of the very strong Ca II triplet\footnote{We note that the Ca II triplet is produced in the chromosphere of stars and is subject to substantial non-LTE effects, especially at low metallicities and so must be treated with caution in practice \citep{jorgensen:1992, mashonkina:2007, starkenburg:2010}.} at $\lambda\lambda$8498, 8542, and 8662 \AA. Meanwhile, elements like Y have only a few weak lines within the D1200G wavelength range (Figure \ref{fig:gradients}d) and are thus only recoverable in nearby stars. 

Longer exposures provide better S/N, allowing for more precise measurements of more abundances. For a 3 hour observation, the number of elements measured to $<$0.3 dex increases to 20, 11, and 7 for RGB stars at 50, 100, and 200 kpc respectively. For a nearby 18$^\text{th}$-magnitude star the S/N is sufficient ($\sim$150 pixel$^{-1}$) to measure elements with only weak signatures in the spectrum. For example C, and N can be recovered from broad, weak CN molecular features between 7000 and 9000 \AA. Cu can be measured from two weak ($\sim$1\%/dex) absorption lines at $\lambda\lambda7935,8095$ \AA. Similarly, elements like La, Mn, O, and Eu have no more than 10 absorption lines with gradients $>$0.5\%/dex and only one or two lines with gradients $>$1\%/dex. However, given the high S/N of these observations, they can nevertheless be recovered to a precision of $<$0.3 dex.

At six hours of integration the S/N is approximately 200, 75, and 30 pixel$^{-1}$ for RGB stars at 50, 100, and 200 kpc respectively. This enables the recovery of 22, 13, and 9 elements to better than 0.3 dex for these stars. Only after 6 hours of exposures are the weak Y lines enough to measure its abundance out to 100 kpc. These extra three hours of integration are necessary to measure Nd and V in the 18$^\text{th}$-magnitude RGB from roughly a dozen very weak lines with gradients $<$0.5\%/dex.

In Figure \ref{fig:crlb_d1200g}, we also include the spectroscopic precision on the atmospheric parameters $T_\text{eff}$, $\log(g)$, and $v_\text{micro}$. With the continuum shape removed from our spectrum, the effective temperature can only be constrained by its impact on atomic and molecular transitions as seen in absorption features. Compared to changes in abundance, the effect of $T_\text{eff}$ on absorption lines is quite weak ($\sim$2\% per 100 K for H$\alpha$ and $<$1\% per 100 K for most other lines), but because it manifests in thousands of lines across the D1200G wavelength coverage it nonetheless allows for $T_\text{eff}$ to be recovered to better than 100 K in most of the scenarios considered here.
In contrast to $T_\text{eff}$, changes in $\log(g)$ effect fewer lines, but much more strongly. H$\alpha$ and the Ca II triplet are notable lines sensitive to the surface gravity in the red-optical. 
The microturbulent velocity lies somewhere between $T_\text{eff}$ and $\log(g)$, moderately impacting (1-4\% per km/s) $\sim$50 absorption features across the spectrum.

\subsubsection{Comparison to Literature Precision} \label{sec:comparison}
Our CRLBs formally represents the best achievable abundance precision via full spectral fitting, not necessarily what is obtained in practice (due to imperfect models, variable LSFs, masked or obscured features, etc.). It is therefore useful to compare our CRLB estimates to published abundance precisions from full spectral fitting to get a sense of how close to the CRLB current abundance measurements get.

For an illustrative comparison, we select abundances measured by \citet{kirby:2018}, who use a full spectral fitting technique (as opposed to EWs) for RGB stars in Local Group galaxies \citep{kirby:2009}. Because of the large variety in stellar targets and spectral quality, we make several cuts to the \citet{kirby:2018} sample in order to fairly compare the reported precision and our CRLBs. First, we consider only stars with $T_\text{eff}$ between 4500 and 5000 K, $\log(g)$ between 1.7 and 1.9, and [Fe/H] between -2.0 and -1.0. Second, we consider only stars that were observed to \edit1{\deleted{$45~\text{\AA}^{-1}< \text{S/N} < 75~\text{\AA}^{-1}$}\added{$35~\text{\AA}^{-1}< \text{S/N} < 65~\text{\AA}^{-1}$}}, which corresponds to roughly the mean S/N of a 1 hour exposure of a 19.5 magnitude star. These cuts leave the reported abundance precision of \edit1{\deleted{30}\added{33}} stars.

\edit1{\added{
Before we make a direct comparison, we modify our CRLB calculation to closely adhere to the choices made by \citet{kirby:2018}. For example, $\log(g)$ and $v_\text{micro}$ are not fit via spectroscopy, but held fixed at values determined by the star's photometry. This can lead to more precise recovery of abundances by removing their covariances with these labels. Similarly, only Fe, Ca, Ni, Si, Ti, Co, Mg, and Cr are fit, while all other abundances are fixed at Solar abundance value. These are not unreasonable assumptions since the information content of the spectra is dominated by these elements, and $\log(g)$ is typically better constrained with photometry than spectroscopy in extragalactic contexts where the distance is well constrained. We mimic this analysis by adopting a delta-function prior on all stellar labels that are not fit for by \citet{kirby:2018}.
}}

\edit1{\added{
In addition, \citet{kirby:2018} masks a handful of specific spectral regions that are contaminated by poorly modelled lines or strong telluric absorption features. Following \citet{kirby:2008} we mask 13 spectral regions including 
notable spectral features such as the Ca II triplet ($\lambda\lambda$6498, 8542, 8662) and the Mg I $\lambda$8807 line.
}}

\edit1{\added{
It is worth noting that there are several aspects of the method used by \citet{kirby:2018} that we cannot account for. First, they adopt a different set of stellar models and linelists than we do, albeit with similar 1D, LTE assumptions (e.g., ATLAS9 vs ATLAS12; see \citealt{kirby:2010}). Second, they fit stellar labels iteratively by looping through the labels and fitting each individually while holding the rest constant until convergence is achieved. It is possible that this approach may neglect some covariances between labels that are expected when all labels are fit simultaneously as assumed by the CRLB. Third, the specific wavelength coverage of each spectrum varies from the nominal depending on the star's location on DEIMOS's detector.
}}

\edit1{\deleted{Note}\added{Lastly, we note}} that the chemical abundance uncertainties reported by \citet{kirby:2018} include both a statistical and systematic uncertainty component added in quadrature. Because CRLBs are purely a measure of statistical precision and not accuracy, we subtract out in quadrature the systematic component (of order 0.2 dex for Co and 0.1 dex for all other elements) to make a better one-to-one comparison with the literature uncertainties. 

\begin{figure*}[ht!]
	\includegraphics[width=\textwidth]{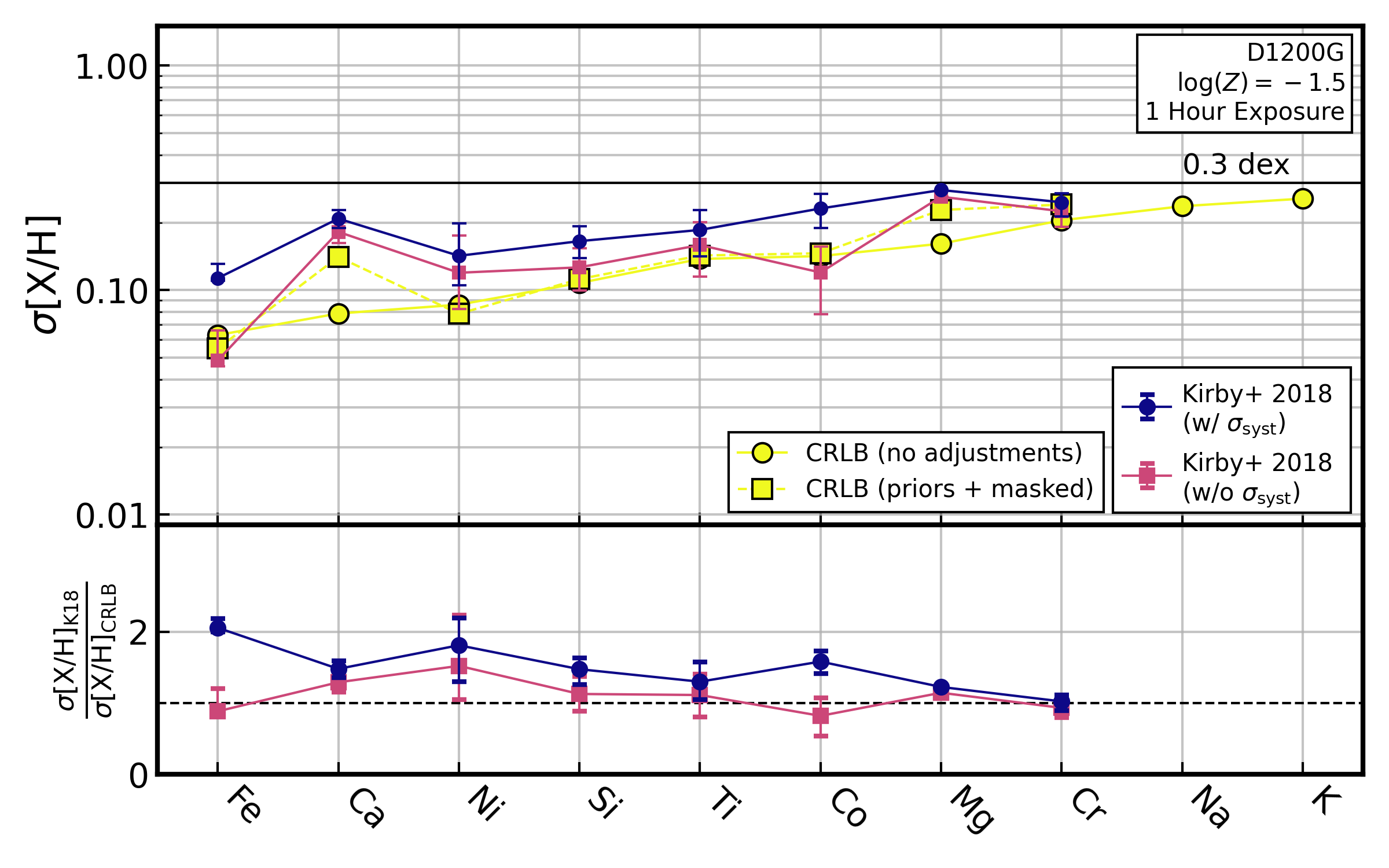}
    \caption{(Top) D1200G CRLBs for a 1 hour exposure of a 19.5 magnitude $\log(Z) = -1.5$ RGB star over-plotted with the uncertainties of abundances for 35 comparable RGB stars reported by \citet{kirby:2018}. The CRLBs represented by squares and dashed lines are calculated by fixing the same stellar labels and masking the same spectral features as \citet{kirby:2018}, while the  CRLBs represented by circles and solid lines are the same as those presented in Figure \ref{fig:crlb_d1200g}. Literature uncertainties include a systematic uncertainty and are only provided for stars with uncertainties less than 0.3 dex. Uncertainties for atmospheric parameters $T_\text{eff}$, $\log(g)$, and $v_\text{turb}$ are not provided. \citet{kirby:2018} did not measure [Na/Fe] or [K/Fe] abundances and therefore have no uncertainties to report for those elements. (Bottom) The ratio of the reported precision to the CRLBs that mimic the analysis techniques of \citet{kirby:2018}. Measurement precisions for most elements are within a factor of 2 larger than the CRLBs.} 
    \label{fig:DEIMOS_lit}
\end{figure*}

Figure \ref{fig:DEIMOS_lit} shows the reported precision of the \edit1{\deleted{30}\added{33}} stars from \citet{kirby:2018} plotted with our D1200G CRLBs\edit1{\added{---both with and without adjustments to match their specific analysis}}
\edit1{\deleted{
We find that most abundances reported by \citet{kirby:2018} are within a factor of $\sim$2 of our CRLBs.
For example, we predict a precision for Fe to be $\sim$0.06 dex, while \citet{kirby:2018} reports a precision of 0.07 dex.
Similarly, \citet{kirby:2018} finds a typical precision for Ni, Si, and Ti to be 0.15, 0.17, and 0.19 dex respectively,
while we predict CRLBs of 0.09, 0.11, and 0.14 dex.
On the other hand, the average precision for Ca (0.18 dex) is over twice that of the CRLB we predict (0.08 dex). The deviations for Co, Mg, and Cr are also small but are slightly misleading since only the abundances of a star that are recovered to better than 0.3 dex are reported, leaving only 8 stars with Co abundances, 1 stars with Mg abundances, and 6 stars with Cr abundances. This skews the average uncertainty to higher precisions.
}\added{
We find that the abundances reported by \citet{kirby:2018} are within a factor of $\sim$2 of our corresponding CRLBs. 
The precisions reported for Fe (0.05 dex), Co (0.12 dex), and Cr (0.22 dex) are slightly less than our predicted precisions (0.06, 0.14, and 0.20 dex respectively). This may be due to a slight overestimation of the systematic uncertainty on these labels or the underestimation of label degeneracies as a result of the iterative fitting. The reported precision for Co, Mg, and Cr, are
likely skewed to higher precision since only abundances recovered to better than 0.3 dex are reported, leaving only 8 stars with Co abundances, 1 star with Mg abundances, and 6 stars with Cr abundances.}}

\edit1{\added{
The biggest difference between the CRLBs calculated previously and those calculated to mimic the analysis of \citet{kirby:2018} is in the forecasted uncertainty of Ca and Mg, which increased from 0.07 and 0.16 dex to 0.14 and 0.22 dex respectively. This is the result of masking strong lines for these elements, which are both highly informative but challenging to model correctly. Fixing $\log(g)$ would have considerably improved the precision for Ca had the Ca I triplet not been masked due the feature's strong dependence on surface gravity. Instead, it only very slightly increases the precision of Fe and Ni from 0.06 and 0.09 dex to 0.05 and 0.08 dex respectively, but otherwise does not change the CRLB substantially. From this comparison, we can see the importance of folding in these effects to our ability to estimate the expected precision.
}}

\edit1{\deleted{
It is encouraging to see that full spectral fitting techniques can recover abundances generally within a factor of $\sim$2 of the CRLBs. There are several reasons why poorer precision in practice could be expected. Examples include poor model fidelity, imperfect calibrations, and masked or lost spectral regions (see \S \ref{sec:caveats} for further discussion). 
The larger difference seen for Ca is also not wholly unexpected. The Ca II triplet, which are some of the strongest gradient features for Ca, were excluded from the spectral analysis of \citet{kirby:2018} due to their strong non-LTE dependence. While this increases accuracy, it also results in a loss of information and reduced precision.
}\added{
While the reported uncertainty for most elements is slightly higher than the CRLB, it is encouraging to see them within a factor of $\sim$2. There are several reasons why poorer precision in practice could be expected. Examples include poor model fidelity, imperfect calibrations, and masked or lost spectral regions (see \S \ref{sec:caveats} for further discussion). While future comparisons with abundance precisions from full-spectrum fitting are necessary to more completely understand the prospects of achieving the CRLB in practice, this comparison with D1200G illustrate that the CRLBs at least provide a realistic benchmark for spectroscopic abundance precision. In Appendix \ref{app:lamost_compare}, we perform an analogous comparison with LAMOST and find similar agreement between our CRLBs and the literature abundance precision.
}}

\edit1{\deleted{
Overall, this comparison illustrates that the CRLBs provide a realistic benchmark and goal for the achievable precision of spectroscopic abundance measurements.}}

\subsubsection{CRLBs vs. [Fe/H]}
We now consider how the CRLB changes as a function of metallicity. To do this we compare the CRLBs for RGB stars with $\log(Z)=-0.5$, $-1.0$, $-1.5$, $-2.0$, and $-2.5$. In order to achieve similar observing conditions for each star, we make comparisons at fixed $m_V$ instead of at fixed stellar phase (or fixed location on the RGB; see Figure \ref{fig:cmd}). As a result of the RGB isochrone's metallicity dependent morphology, $T_\text{eff}$ and $\log(g)$ for these stars are all slightly different with more metal-poor stars having higher $T_\text{eff}$ and $\log(g)$ (Table \ref{tab:ref_stars}). The S/N for these stars are calculated for our fiducial observation of a 1 hour exposure of a star at 100 kpc ($m_V=19.5$) and the configurations summarized in Table \ref{tab:ETC}.

\begin{figure*}[ht!]
	\includegraphics[width=\textwidth]{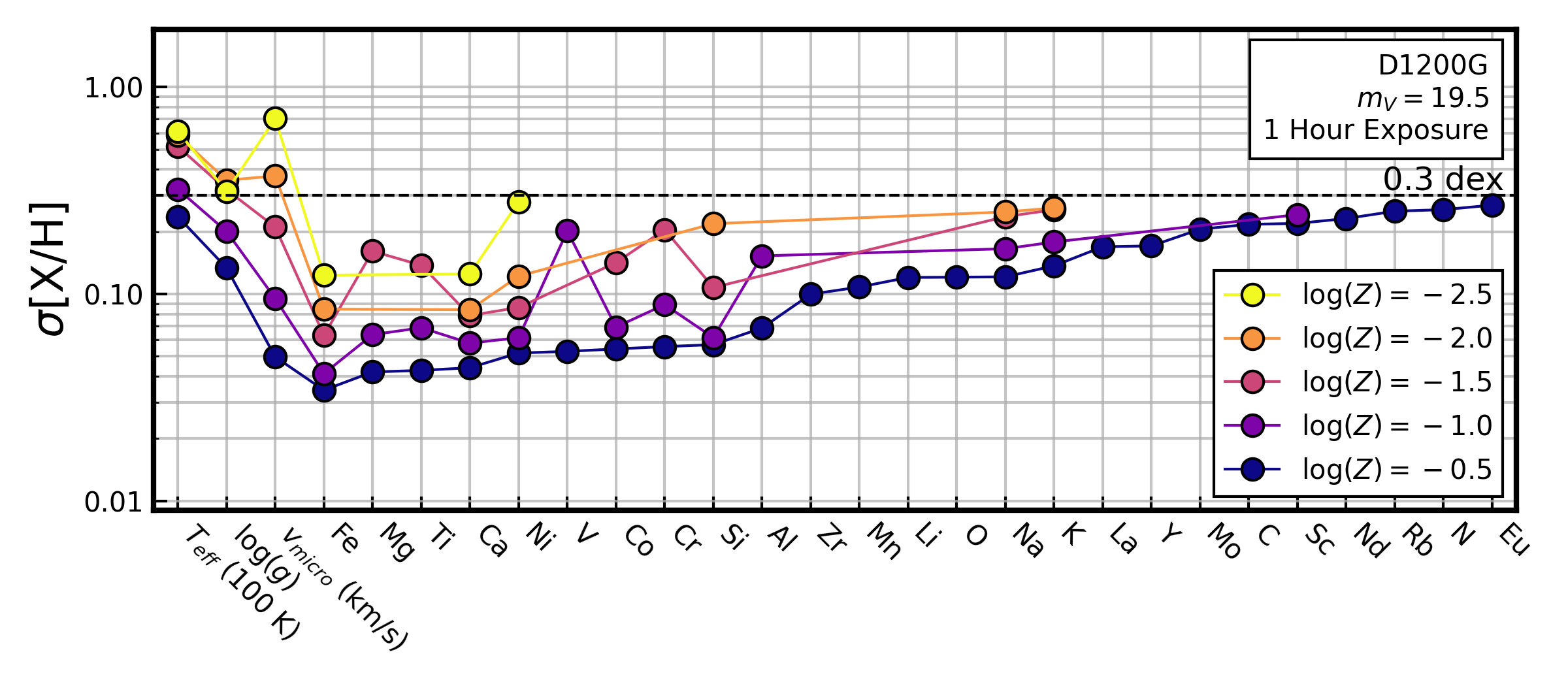}
    	\caption{D1200G CRLBs for a 1 hour exposure of RGB stars with metallicities of $\log(Z)=-0.5$, $-1.0$, $-1.5$, $-2.0$, and $-2.5$ at a distance of 100 kpc ($m_V=19.5$). Table \ref{tab:ref_stars} lists the atmosphere parameters for each star. In general, abundance recovery is less precise for lower metallicity stars due to weaker absorption features.}
   	 \label{fig:crlb_feh}
\end{figure*}

The CRLBs for the various metallicity stars are plotted in Figure \ref{fig:crlb_feh}. As expected, the achievable abundance precision decreases towards lower metallicity as there are fewer and weaker absorption features. However, the dependence of precision with metallicity is not uniform across all elements. For example, the precision of Fe steadily decreases from $\sim$0.03 dex to $\sim$0.1 dex as the metallicity decreases from $\log(Z)=-0.5$ to $-2.5$. The precision of V, however, decreases dramatically from $\sim0.05$ dex to $\sim0.2$ dex between $\log(Z)=-0.5$ and $-1.0$ as a result of its absorption features being strongly temperature dependent. At even lower metallicities (and slightly higher $T_\text{eff}$), V features are nearly entirely absent. 

Below $\log(Z)=-1.5$, the CRLBs for $T_\text{eff}$ and $\log(g)$ remain constant, or even improve. This seemingly counterintuitive result is due to increasingly prominent Paschen lines red-ward of 8200 \AA\ with increasing temperature. These lines are very sensitive to the star's $T_\text{eff}$ and $\log(g)$, allowing for precise measurements of these atmospheric parameters despite the lower metallicities.

\subsubsection{CRLBs vs. Stellar Phase} \label{sec:d1200g_phase}
Just as a star's spectral gradients vary as a function of metallicity, it also varies as a function of atmospheric structure (i.e., $\log(g)$, $T_\text{eff}$, and $v_{micro}$). As a result, we expect the achievable abundance precision at varying stellar phases to be different even at fixed metallicity and apparent magnitude. While we focus our analysis on a typical RGB star, stars from the main sequence turn-off (MSTO) to the tip of the red giant branch (TRGB) are also targets of extragalactic studies. 

Here, we consider the CRLBs for the $\log(Z)=-1.5$ RGB star considered previously with that of a MSTO and TRGB star at the same metallicity (see Table \ref{tab:ref_stars}). We once more consider a 1 hour integration of a $m_V=19.5$ star with the relevant ETC configuration in Table \ref{tab:ETC}.

\begin{figure*}[ht!]
	\includegraphics[width=\textwidth]{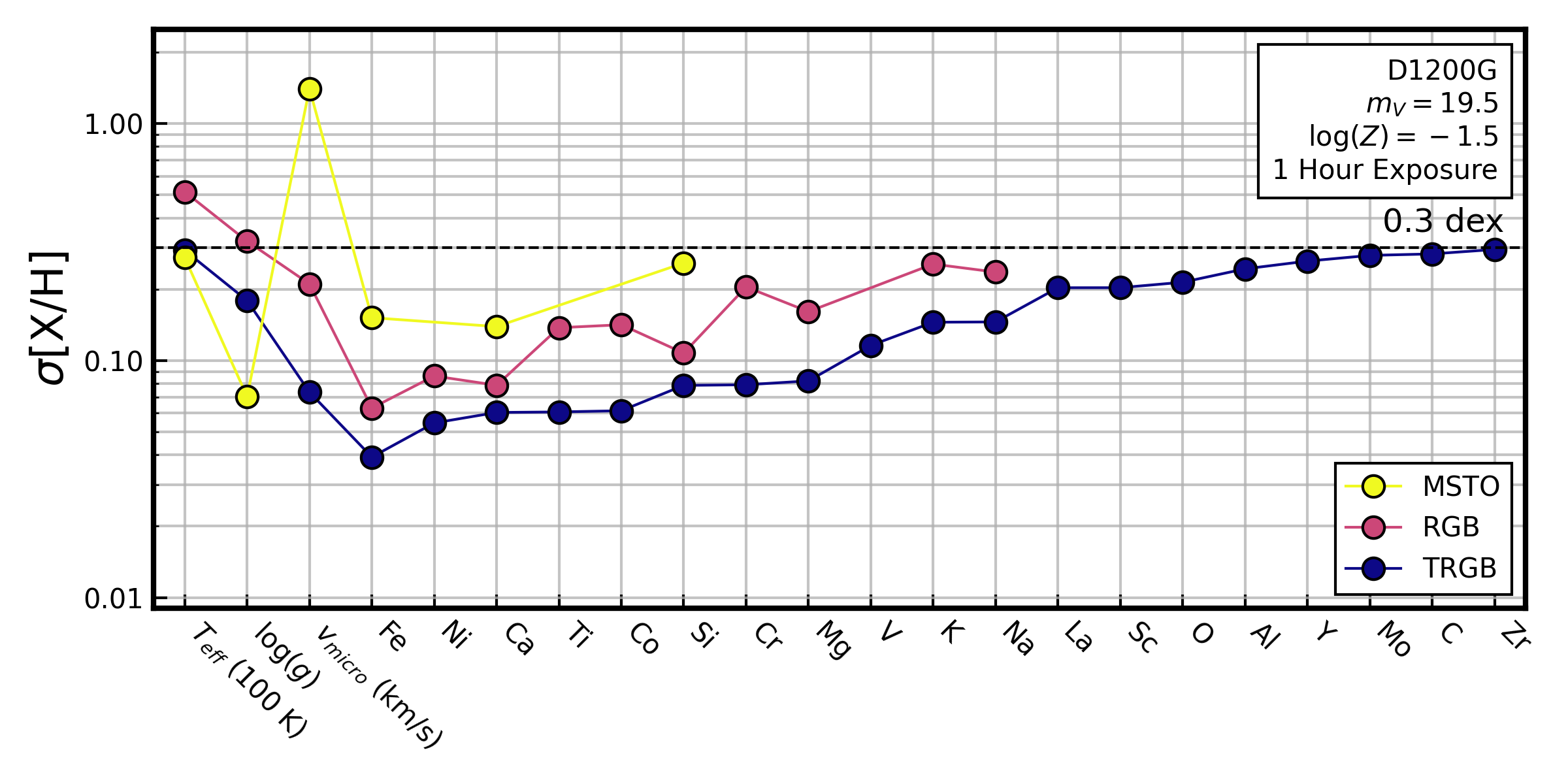}
    	\caption{D1200G CRLBs for a 1 hour exposure of $\log(Z)=-1.5$, $m_V=19.5$ MSTO, RGB, and TRGB stars. The atmosphere parameters for each star can be found in Table \ref{tab:ref_stars}. At low metallicities (such as $\log(Z)=-1.5$), abundance recovery is more precise for cool giants due to stronger absorption features and less precise for hot sub-giants, which have weaker absorption features.}
   	 \label{fig:crlb_phase}
\end{figure*}

The CRLBs of each of these stellar phases are plotted in Figure \ref{fig:crlb_phase}, illustrating that the chemical abundance precision is best for TRGB stars and worst for MSTO stars (all other things being equal). While only 3 elements can be measured to better than 0.3 dex from the spectrum of the MSTO star, 10 elements can be measured to this precision in the RGB star, and 19 in the TRGB star. For a fixed element the precision is roughly two times better for the TRGB star than the RGB star and another two times better than the MSTO star. 

These differences are expected since the absorption features of hot sub-giants are significantly weaker than for cool giants. This is especially true for elements like C, N, and O, which are measured primarily from molecular features that are pronounced in TRGB stars but practically non-existent in MSTO stars. Similarly, Fe, Si, Mg, Al, and other elements whose abundances affect a star's atmospheric structure leave a larger signature in cool, low surface gravity stars than hot, high surface gravity stars.

Recovering $T_\text{eff}$ and $\log(g)$, on the other hand, can be done more precisely in MSTO stars, due to the strong dependence of the Paschen lines on the star's atmospheric parameters.

\subsubsection{CRLBs with Priors} \label{sec:d1200g_priors}
For stars with secure distances (as members of external galaxies typically are), photometry can be used to constrain $T_\text{eff}$ and $\log(g)$ to roughly $\pm100$ K and $\pm0.15$ dex respectively \citep{kirby:2009, casagrande:2011, heiter:2015}. Knowledge of $\log(g)$ and Equation \ref{eq:logg_vmicro} can also constrain $v_{micro}$ to roughly $\pm0.25$ km/s \citep{holtzman:2015}.
We can incorporate these photometric estimates as priors on our spectroscopically determined labels as shown in \S\ref{sec:including_priors}. To do so we adopt Gaussian priors on these parameter with standard deviations equal to their photometric uncertainties. We once more consider a 1 hour observation of our fiducial $\log(Z)=-1.5$ RGB star at 50, 100, and 200 kpc.

\begin{figure*}[ht!]
	\includegraphics[width=\textwidth]{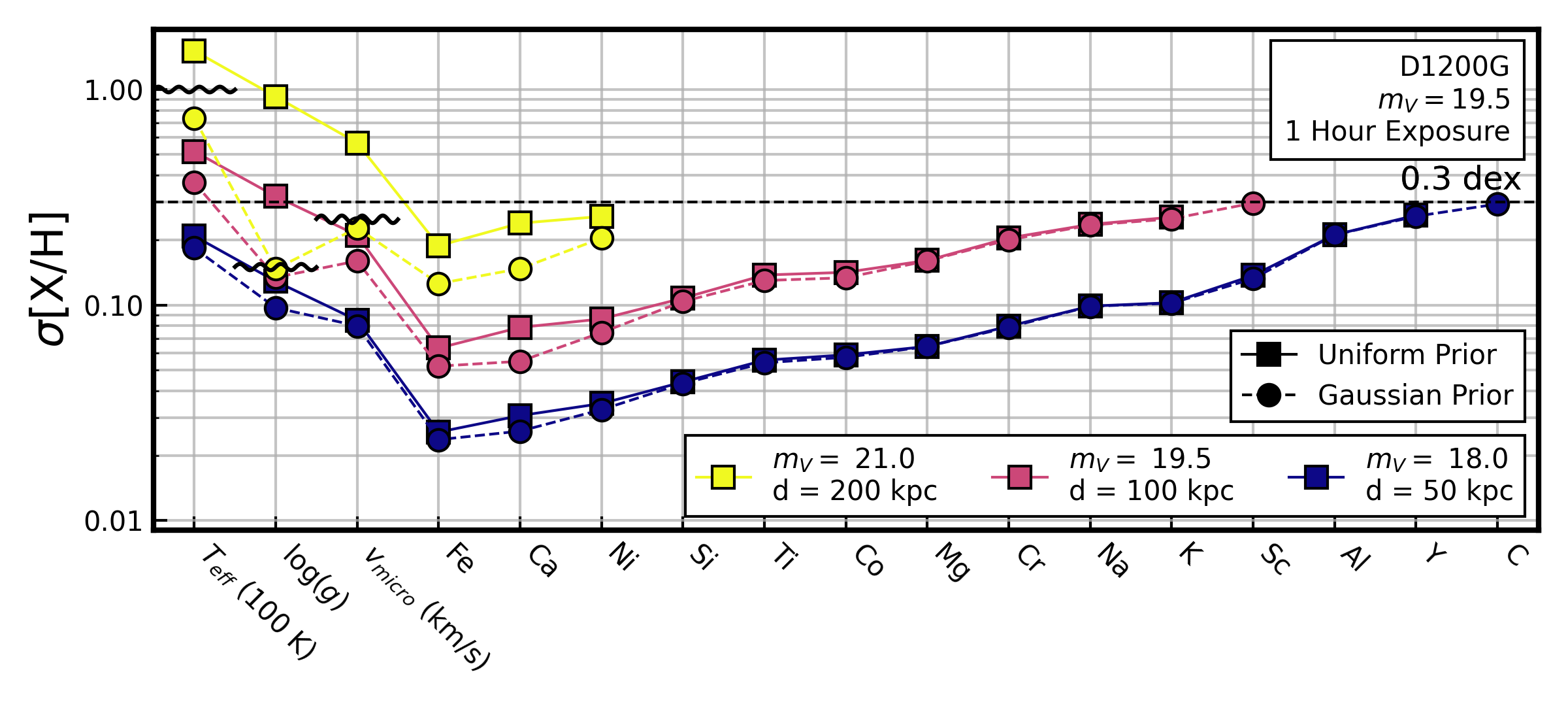}
    	\caption{Same as the top panel of Figure \ref{fig:crlb_d1200g} but also including the Bayesian CRLBs assuming $\sigma_{T_\text{eff},\text{prior}}=100$ K, $\sigma_{\log(g),\text{prior}}=0.15$ dex and $\sigma_{v_\text{micro},\text{prior}}=0.25$ km/s (dashed lines). The black wavy lines mark the priors on $T_\text{eff}$ and $\log(g)$. In addition to better constrained $T_\text{eff}$ and $\log(g)$, the inclusion of priors also improves the precision of abundance determinations, particularly at lower S/N.}
   	 \label{fig:crlb_priors}
\end{figure*}

Figure \ref{fig:crlb_priors} shows the results of the CRLBs assuming Gaussian priors. For references, we include the CRLBs from Figure \ref{fig:crlb_d1200g} (top), which assume uniform priors.

For the highest S/N case (at 50 kpc; $\text{S/N}\sim75$ pixel$^{-1}$), the precision on $T_\text{eff}$ and $\log(g)$ from D1200G spectroscopy alone is significantly better than the priors. The priors therefor contribute negligible additional information, and the CRLBs only minimally improve. 

However, in the lowest S/N case (at 200 kpc; S/N$\sim$10 pixel$^{-1}$), $T_\text{eff}$, $\log(g)$, and $v_\text{micro}$ are substantially less constrained by the spectroscopy compared to the priors and so nearly all of the information about these stellar labels are coming from the prior. As a result, use of these priors improve the precision of $T_\text{eff}$, $\log(g)$, and $v_\text{micro}$ by factors of 2-6 compared to the uniform prior case. 

In addition, because spectral gradients of $T_\text{eff}$ and $\log(g)$ are covariant with the spectral gradients of elements like Fe, Ca, and Ni, priors that better constrain $T_\text{eff}$ and $\log(g)$ also lead to improved precision on these chemical abundances. For example, in the case of our faintest star, the Fe, Ca, and Ni abundance precision improves by $\sim$50\% when Gaussian priors on $T_\text{eff}$ and $\log(g)$ are included. We expect the inclusion of photometric priors to have more impact when the spectral gradients of different labels are more covariant (i.e., for low-resolution spectra with heavily blended lines and spectra with very limited wavelength coverage and few absorption lines).

\subsection{Low and Medium Resolution MOS} \label{sec:lowres}
All other things being equal, high-resolution spectra would be preferable for abundance measurements, as fewer lines are blended which results in fewer coupled abundance determinations. Unfortunately, as described in \S \ref{sec:highres}, high-resolution spectrographs are typically limited to the brightest extragalactic stars due to their high spectral dispersion, relatively low throughput, and limited multiplexing capabilities. As a result, it is not possible at present to efficiently observe large numbers of extragalactic resolved stars with broad wavelength coverage and $R>10000$ spectroscopy.

Low- and medium-resolution multi-object spectrographs (MOS), on the other hand, provide high multiplexing capabilities, increased throughput, and broad wavelength coverage, enabling them to achieve modest S/N of many faint stars simultaneously in distant systems. Furthermore, as we will show, wavelengths bluer than $\sim$5000 \AA---even at low resolution---are incredibly rich in absorption features, especially for the cool low-mass giants typically observed outside the MW.
 
Historically, low- and moderate-resolution blue-optical spectra have not been favored for abundance determinations due to the challenge in identifying the continuum and substantial blending of lines \citep{ting:2017}. However, in recent years, advances in spectral fitting techniques have lead to large improvements in abundance recovery from  low-resolution blue-optical spectra. Notably, \citet{ting:2017b} and \citet{xiang:2019} have shown that it is possible to measure 16+ elements of $\sim$6 million MW stars from $R\sim1800$ LAMOST spectroscopy with a wavelength coverage of $3700$-$9000$ \AA. While the small aperture of LAMOST (1.75 m) precludes it from abundance measurements of most stars outside the MW, there are a handful of MOS already in commission that provide similar resolving power and wavelength coverage on 6+ meter telescopes (e.g., Keck/LRIS, LBT/MODS, and MMT/Hectospec). In the following sections, we quantify the potential of these facilities for chemical abundance measurements outside the MW.

\subsubsection{Blue-Optical MOS on Keck} \label{sec:blue_keck}
On the Keck/DEIMOS spectrograph there are several options that provide access to wavelengths bluer than 5000 \AA. As listed in Table \ref{tab:instruments}, the 900ZD, 600ZD, and 1200B gratings all provide bluer wavelength coverage, but slightly lower resolution, compared to the D1200G setup.
These gratings have already enabled  abundance determinations not possible from red-optical spectroscopy, such as the measurement of $\alpha$ elements in the M31 halo \citep{escala:2019a} and Ba in several dwarf galaxies \citep{duggan:2018}. The 1200B grating is a recent addition to DEIMOS's grating collection and has not been used to measure stellar abundances at the time of this paper's writing.

In addition to DEIMOS, the Keck telescopes also host the LRIS multi-object spectrograph, which operates using separate red and blue channels. The 600/4000 grism on the blue arm boasts impressive blue throughput compared to DEIMOS gratings\footnote{25\% at 4500 \AA\ compared to 13\% for DEIMOS 1200B and 4\% for DEIMOS 1200G}, while the 1200/7500 grating on the red arm provides coverage around the Ca II triplet (Table \ref{tab:instruments}). While LRIS has only ever been used for very limited stellar abundance determinations \citep{shetrone:2009, lai:2011}, it is nonetheless a promising instrument, particularly given the demonstrated success of LAMOST.

To quantify the information content accessible in the blue optical by these instrumental setups, we calculate their CRLBs given a 1 hour exposure of our fiducial $\log(Z)=-1.5$ RGB star at 100 kpc and the relevant ETC configurations for each instrument from Table \ref{tab:ETC}. 

\begin{figure*}[ht!] 
	\includegraphics[width=\textwidth]{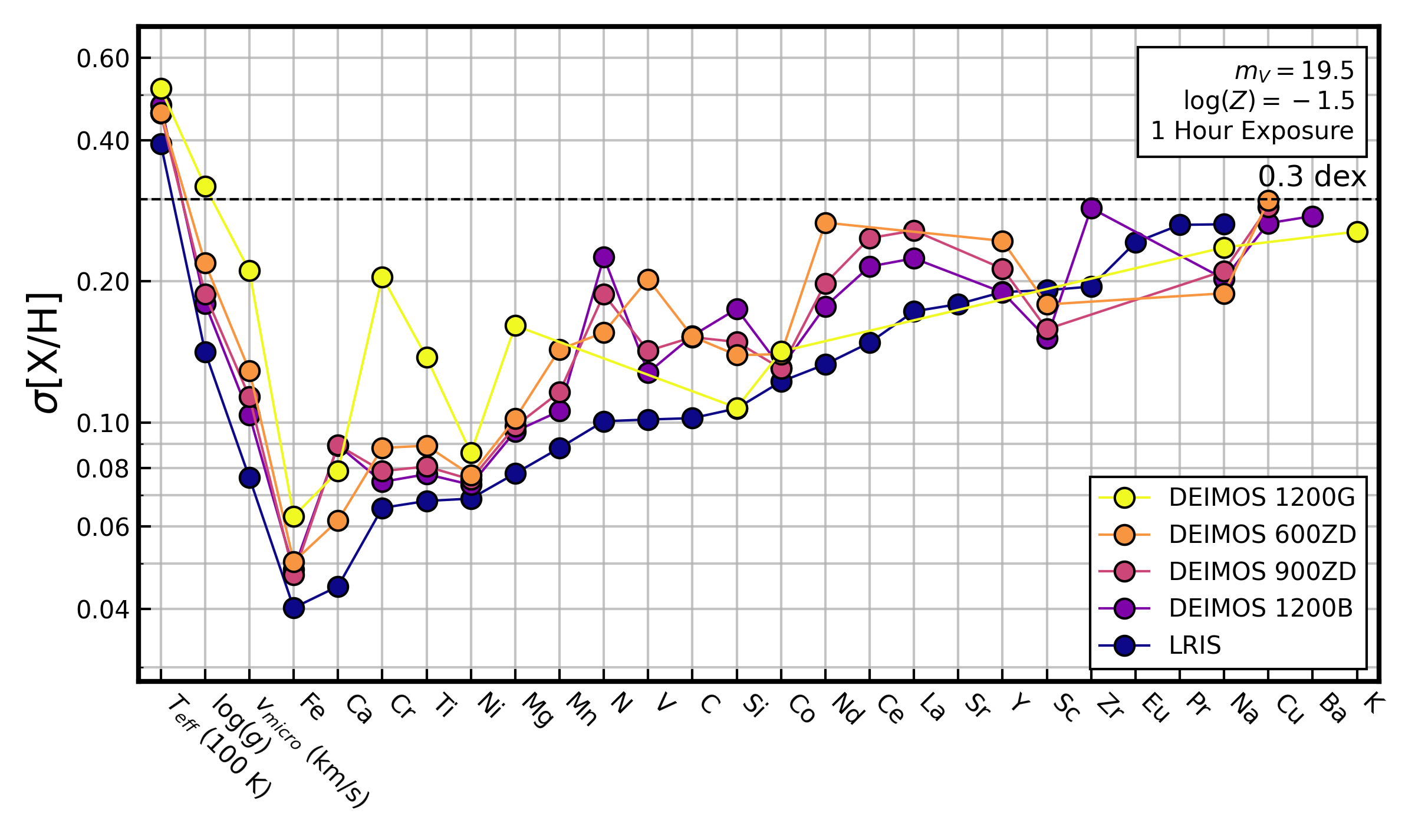}
    \caption{Comparison of CRLBs for several multi-object spectroscopic setups on Keck/DEIMOS and Keck/LRIS assuming a 1 hour exposure of a $\log(Z)=-1.5$, $M_V=-0.5$ RGB star at 100 kpc. The LRIS setup includes the spectral coverage of both its blue and red channels. The elements are ordered by decreasing precision as forecasted for LRIS up to 0.3 dex. The CRLB for D1200G is the same as shown previously in Figures \ref{fig:crlb_d1200g} (top), \ref{fig:crlb_feh}, and \ref{fig:crlb_priors}.}
    \label{fig:crlb_keck}
\end{figure*}

The forecasted abundance precision for each element is present in Figure \ref{fig:crlb_keck}. Despite their lower resolving powers, instruments with bluer wavelength coverage provide more precise measurements of more elements than D1200G. For example, the 1200B grating on DEIMOS and the 600/4000+1200/7500 LRIS setup, enable the recovery of 21 and 22 elements respectively to better than 0.3 dex---about twice that from comparable red-optical spectroscopy at fixed integration time and stellar type.
This includes eight r- and s- process elements (Y, Ce, La, Zr, Ba, Sr, Pr, and Eu), which have most, if not all of their absorption features at wavelengths shorter than 5000 \AA\ and are thus largely inaccessible to D1200G and other longer wavelength spectrographs.  Information about C and N comes primarily from C$_2$, CH, and CN absorption bands between 4000 and 5000 \AA\ and to a lesser extent from CN bands between 7000 and 9000 \AA. 

D1200G does provides comparable or better precision for Fe, Ni, Si, and Co, which have many lines at wavelengths longer than $\sim$6500 \AA, as well as for Ca, Na, and K, which have strong features in the red-optical\footnote{The Ca II triplet at $\lambda\lambda8498,8542,8662$ \AA, the Na I doublet at $\lambda\lambda8185,8197$ \AA, and the K I doublet at $\lambda\lambda7667,7701$ \AA\ respectively.}. 

LRIS's improved precision is due to a combination of its exceptional throughput down to 3900 \AA\ and the additional wavelength coverage provided by its red arm\footnote{Though LRIS does lose considerable information for Sc, Na, Cu, Ba and K in the gap between its red and blue coverage. This can be mitigated to a degree by carefully choosing the dichroic and grating angle employed.}. However, it is important to remember that LRIS has roughly half the field of view and half the multiplexing as DEIMOS (Table \ref{tab:instruments2}). Meaning that it may ultimately be less efficient for some elements, when the number of stars is included in the calculation.

\edit1{\added{
As a reminder, the DEIMOS 600ZD and 900ZD gratings and the LRIS 1200/7500 grating all over-sample their spectra with 5 pixels/FWHM. If the pixels in these spectra are not completely independent as we assume here, the CRLBs we present may be slightly more precise than would be expected in practice (see \S\ref{sec:oversampling}).
}}

\subsubsection{Blue-Optical MOS on other Telescopes} \label{sec:other_mos}
We now turn our attention to blue-sensitive instruments on facilities other than the Keck Telescopes, which include MODS on the LBT, MUSE on the VLT, and Hectospec and Binospec on the MMT.

MODS, like LRIS, operates at low resolution ($R\sim2000$) across the optical spectrum with a red and a blue arm, and modest multiplexing (Tables \ref{tab:instruments} and \ref{tab:instruments2}). Other than a recent study on a chemically peculiar ultra metal-poor star in the dwarf galaxy Canes Venatici I \citep{yoon:2019}, MODS has not been utilized for stellar chemical abundance measurements. 

While MUSE is not technically a MOS but rather an integral field unit (IFU), it can nonetheless be used effectively for low-resolution resolved star spectroscopy of many stars at the same time. MUSE has already been used to conduct several campaigns for both stellar radial velocity and chemical abundance measurements in globular clusters \citep[e.g.,][]{husser:2016, kamann:2016, kamann:2018, latour:2019} and in dwarf galaxies \citep[e.g.,][]{voggel:2016, evans:2019, alfaro-cuello:2019}. 

Hectospec, in comparison to MODS, MUSE, and the spectrographs on Keck, has a very large field of view ($1^{\circ}\times1^{\circ}$), which makes it a powerful instrument for spectroscopic observations of very extended stellar populations.
For example, \citet{carlin:2009} used Hectospec to measure the kinematics and bulk metallicity of stars in the disrupted MW dwarf galaxy Bo\"otes III.
Binospec is a new, complimentary MOS to Hectospec with very high throughput, but a significantly smaller field of view and a more limited multiplexing capability (Table \ref{tab:instruments2}). Both Hectospec and Binospec have a number of gratings that allow for a range in wavelength coverage and resolving power. We examine a few setups we consider to be most applicable to extragalactic stellar spectroscopy (see Table \ref{tab:instruments} for specifics).

\begin{figure*}[ht!]
	\includegraphics[width=\textwidth]{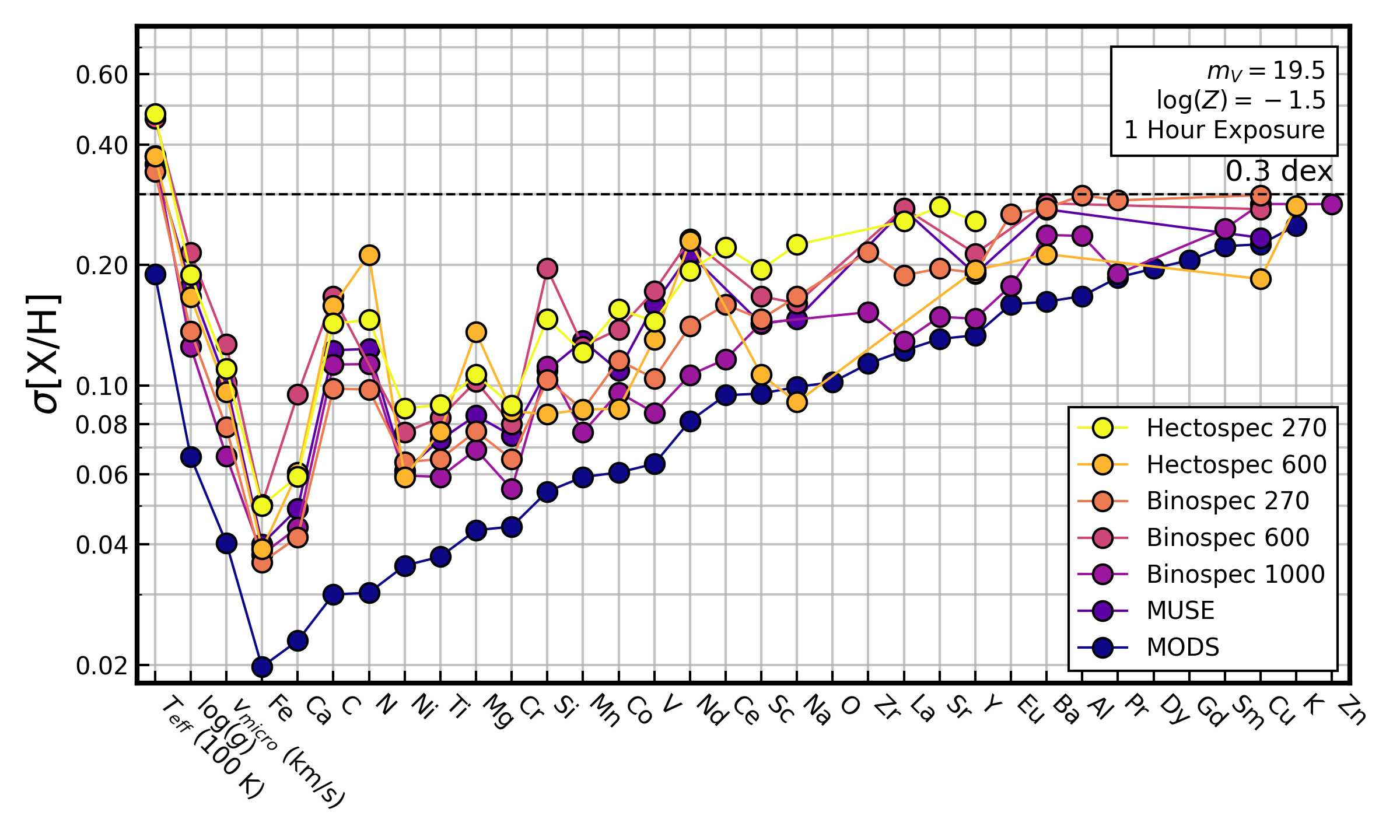}
    \caption{Same as Figure \ref{fig:crlb_keck} but for LBT/MODS, MMT/Hectospec, and MMT/Binospec. Elements are ordered by the precision forecasted for LBT/MODS up to 0.3 dex.} 
    \label{fig:crlb_other}
\end{figure*}

Figure \ref{fig:crlb_other} shows the CRLBs for our fiducial RGB star ($\log(Z)=-1.5$, $m_V=19.5$) and a 1 hour exposure.
For these observing conditions, MODS is forecasted to recover up to \edit1{\deleted{26}\added{30}} individual elements to better than 0.3 dex. MODS's precision can be attributed to two key factors: its large, nearly 12-m effective aperture and its throughput below 4000 \AA, which together achieve S/N of $>$\edit1{\deleted{25}\added{40}} pixel$^{-1}$ down to 4000 \AA\ and $>$\edit1{\deleted{20}\added{10}} pixel$^{-1}$ down to 3500 \AA. As discussed in \S \ref{sec:blue_keck}, these regions become increasingly information rich due to the high densities and and strengths of absorption features of many elements.

There are a few specific elements that are worth examining in more detail. Just as with the blue-optimized spectrographs on Keck, the constraints on C and N abundances come predominantly from absorption bands at wavelengths bluer than 5000 \AA\ and (to a lesser extent) between 8000 \AA\ and 1 $\mu$m. 
MODS's sensitivity across both of these ranges leads exceptional recovery of C and N compared to the other instruments analyzed here. MUSE and the 600 gratings of Hectospec and Binospec do not push nearly as blue (or red) and thus recover C and N abundances less precisely or not at all. 
\edit1{
\deleted{
While  the  270  grating  on  Hectospec  and  the  270  and 1000 gratings on Binospec do include most of the blue carbon features (and a small portion of the blue nitro-gen features),  they only achieve a S/N of $\sim$10 pixel$^{-1}$ in this region and thus also do not recover C and N as precisely as MODS.
}\added{
While the 270 grating on Hectospec and the 270 and 1000 gratings on Binospec do include most of the blue carbon features, they miss most of the blue nitrogen features and (with the exception of the 270 grating on Binospec) achieve a S/N in this region roughly half that of MODS. As a result they also do not recover C and N as precisely as MODS.
}}

In addition to C and N, MODS is also able to recover O to better than 0.2 dex because of strong OH absorption features below 3500 \AA\ and the important role of O in the CNO molecular network \citep{ting:2018}.

Again, it is worth highlighting the precision capable of these blue-optimized spectrographs for heavy r- and s-process elements \edit1{\deleted{Nd, Ce, Zr, La, Y, Sr, Ba, Eu, Pr, Dy, and Gd}\added{Nd, Ce, Zr, La, Sr, Y, Eu, Ba, Pr, Dy, Gd, Sm}} (in order of decreasing precision for MODS). In addition to those seen in Figure \ref{fig:crlb_keck}, the ability to recover \edit1{\deleted{Nd, Dy, and Gd}\added{Nd, Dy, Gd, and Sm}} is the direct result of MODS blue sensitivity (discussed further in \S \ref{sec:blue}). A few of these are recoverable by \edit1{\deleted{MUSE or the 1000 grating on Binospec}\added{MUSE, Hectospec, or Binospec}}, but measurement is made more difficult due to lower S/N and smaller wavelength coverage.

Given the smaller light-collecting power of MMT, it is reasonable that Hectospec and Binospec are forecasted to recover fewer elemental abundances and at larger uncertainties. It is nonetheless still interesting to look at them in greater detail and compare the various Hectospec and Binospec settings. Generally Binospec's higher throughput leads to higher precision measurements, but this of course comes with a diminished field of view and fewer fibers for stars. 

Similarly, the increased abundance precision of MODS, MUSE, and other Keck spectrographs is also modulated by much reduced fields of view. The choice between these instruments then ultimately comes down to weighing the importance of detailed abundance patterns versus the importance of a large sample size to the desired science.

\edit1{\added{
We remind the reader, that the Hectospec configurations over-sample their spectra with 5 pixels/FWHM. If the pixels in these spectra are not completely independent as we assume here, the CRLBs we present may be slightly more precise than would be expected in practice (see \S\ref{sec:oversampling}).
}}

\subsection{Low S/N, High Resolution Spectroscopy} \label{sec:highres}
In this section, we consider two classes of high-resolution spectrographs: single-slit echelle spectrographs and multiplexed single-order spectrographs.

\subsubsection{High Resolution, Single-Slit} \label{sec:single_slit}
High-resolution spectroscopic observations of stars provide precise radial velocities and are the gold standard for chemical abundance determinations.
Because high-resolution spectroscopy provides spectra with fewer blended absorption features, spectral abundance determinations preferentially use clean, isolated lines that can be fit with EW methods over blended lines, which require spectral synthesis techniques). By not fitting the star's entire spectrum simultaneously, some of the spectrum's chemical information goes un-utilized.
By calculating the CRLBs for several high-resolution spectrographs, we illustrate the chemical information that can be accessed through full-spectrum fitting techniques.

In the context of extragalactic studies, two commonly used single-slit echelle spectrographs are Magellan/MIKE and Keck/HIRES. Both instruments provide high-resolution spectra across the entire optical regime, and have been used extensively for abundance measurements in MW globular clusters \citep[e.g.,][]{boesgaard:2000, venn:2001, boesgaard:2005, koch:2010} and in nearby dwarf galaxies \citep[e.g.,][]{shetrone:1998, koch:2014, frebel:2014, frebel:2016, ji:2016b, ji:2016c, ji:2016a, ji:2019}.

\edit1{\deleted{
We also consider the single-slit spectrograph VLT/X-SHOOTER, which has also been used to measure abundances of bright stars in dwarf galaxies \citep{starkenburg:2013, spite:2018}. X-SHOOTER provides slightly lower resolution than MIKE and HIRES but significantly higher throughput and broader wavelength coverage\footnote{X-SHOOTER's NIR arm extends wavelength coverage to 2.48 $\mu$m, but due to the limitations of our linelist we only consider wavelengths shorter than 1.8 $\mu$m.}.
}\added{
We also consider two spectrographs on the VLT: FLAMES-UVES and X-SHOOTER. FLAMES-UVES is a high resolution spectrograph with a more limited wavelength coverage (only 4800-6800 \AA), but is capable of observing up to 8 stars at a time thanks to the FLAMES fiber feed\footnote{In this way, it straddles the boundary of the single-slit, multi-order spectrographs discussed in this section and the highly multiplexed, single-order spectrographs discussed in \S\ref{sec:single_order}.}. It has been used to observe RGB stars in MW globular clusters \citep[e.g.,][]{alves-brito:2006} and in nearby dwarf galaxies \citep[e.g.,][]{shetrone:2003, letarte:2006, hill:2019, lucchesi:2020}. X-SHOOTER has also been used to measure abundances of bright stars in dwarf galaxies \citep{starkenburg:2013, spite:2018} and provides slightly lower resolution than MIKE, HIRES, and UVES but significantly higher throughput and broader wavelength coverage\footnote{X-SHOOTER's NIR arm extends wavelength coverage to 2.48 $\mu$m, but due to the limitations of our linelist we only consider wavelengths shorter than 1.8 $\mu$m.}.
}}

As discussed in \S \ref{sec:snr}, a 1 hour exposure of a $m_V=19.5$ RGB star is typically insufficient for high-resolution spectrographs to overcome the read-noise limited regime of faint object spectroscopy. Instead we consider a more realistic 6 hours ($\sim$1 night) of integration, which yields $\text{S/N}>15$ (10) pixel$^{-1}$ at 4500 \AA\ and $\text{S/N}>20$ (20) pixel$^{-1}$ at 7500 \AA\ for HIRES (MIKE) when adopting the ETC configurations in Table \ref{tab:ETC}. 

\begin{figure*}[ht!]
	\includegraphics[width=\textwidth]{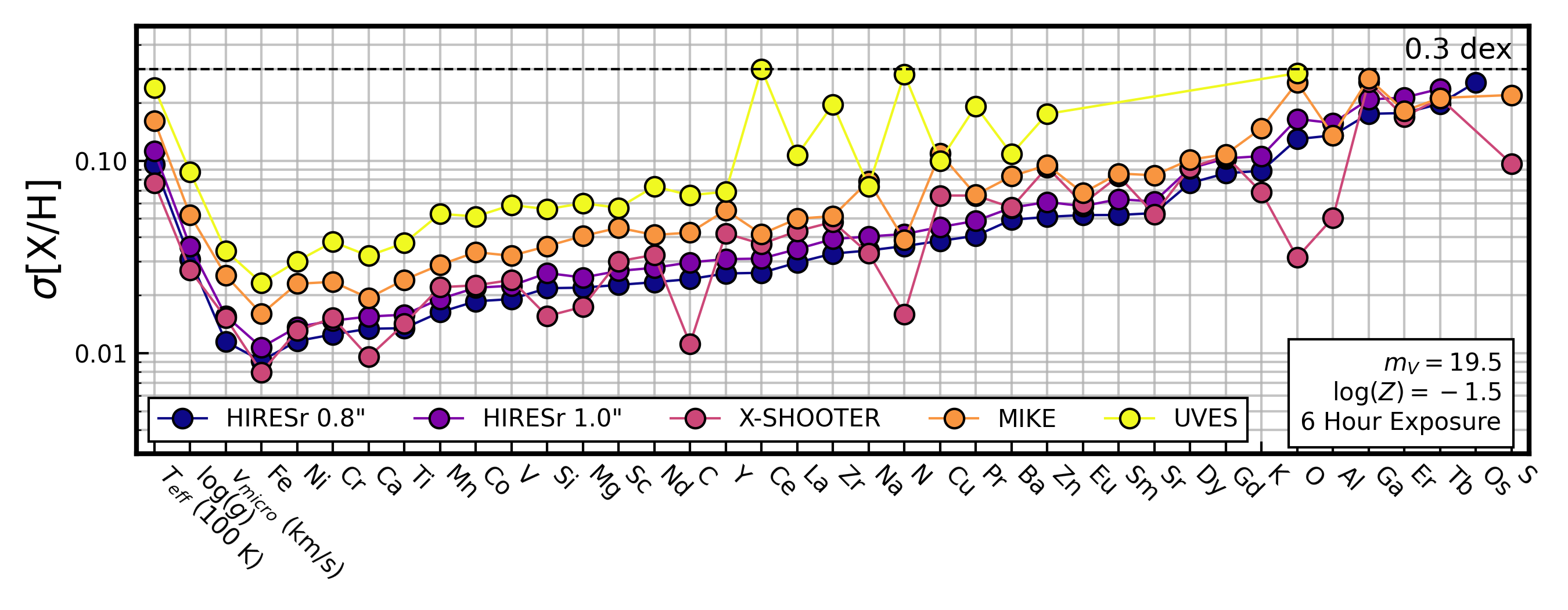}
    \caption{Comparison of CRLBs for high-resolution single-slit echelle spectrographs Keck/HIRES, Magellan/MIKE, and VLT/X-SHOOTER assuming a 6 hour exposure of a $\log(Z)=-1.5$, $M_V=-0.5$ RGB star at 100 kpc. The elements are ordered by decreasing precision as forecasted for HIRES up to 0.3 dex. The CRLBs suggest that even at low S/N ($\sim$15-20), the chemical information content of high-resolution spectra is considerable.} 
    \label{fig:crlb_highres1}
\end{figure*}

Figure \ref{fig:crlb_highres1} shows the CRLBs for HIRES, MIKE, \edit1{\added{FLAMES-UVES,}} and X-SHOOTER. As expected, high-resolution spectra provide very precise detailed chemical abundance patterns. \edit1{\deleted{All three spectrographs}\added{HIRES, MIKE, and X-SHOOTER}} are forecasted to measure a dozen elements to nearly 0.01 dex and over 30 elements to better than 0.3 dex. \edit1{\added{UVES, with its smaller wavelength coverage and lower S/N (5-10 pixel$^{-1}$), is still forecasted to recover over 20 elements.}}  This high precision is predicted despite the low S/N (\edit1{\deleted{$\sim$15-}\added{$<$}}20 red-ward of 4500 \AA) of these observations, demonstrating the potential power of full spectrum fitting applied to high-resolution spectra. While at low S/N any given absorption feature might be only weakly informative, the ensemble of all spectral features still provide strong constraints on the chemical abundances of a star.

The chemical information for many of the elements in Figure \ref{fig:crlb_highres1} can be traced to the same large numbers of features below $\sim$5000 \AA\ as previously discussed in \S \ref{sec:lowres}. While these absorption features are still subject to blending, the higher resolution of these instruments increases the \textit{rms} depth of the absorption feature and alleviates degeneracy between elements. This results in increased abundance precision over low-resolution instruments at fixed wavelength coverage. We can see this effect when comparing the CRLBs of the two HIRES settings, which have the same wavelength coverage but different resolving powers---the CRLBs scale with resolving power $\sigma_{CRLB}\propto R^{-1/2}$ as expected for instruments with the same wavelength range\footnote{A factor of $R^{-1}$ from the scaling of the absorption feature \textit{rms} depth and a factor of $R^{1/2}$ from the scaling of S/N with dispersion. For these two HIRES settings, $R^{-1/2}\sim0.85$.}.

In addition to elements previously discussed in \S\S\ref{sec:d1200g} and \ref{sec:lowres}, HIRES can recover the abundances of neutron-capture elements Sm, Er, Tb, and Os to better than 0.3 dex. At $R\sim50000$ there are nearly 100 Sm lines with gradients $>5\%$/dex and over 30 lines with gradients of 10-30\%/dex in the HIRES wavelength range---all of which are below 4500 \AA. The same spectrum has $\sim$15 (5) absorption lines with gradients of $>$5\%/dex (10\%/dex) absorption lines for Er (Tb) blue-ward of 5000 \AA. Os can be recovered to $\sim$0.3 dex from no more than 5 absorption lines with $>$5\%/dex gradients. 

MIKE's bluer wavelength coverage is largely offset by its lower resolving power ($R\sim28000$) and very low S/N ($<$5 pixel$^{-1}$) \edit1{\added{below 5000 \AA}}. Nevertheless, MIKE achieves slightly better precision for Tb and \edit1{\deleted{Ho}\added{Er}}, which have 2-3 times more lines between 3500 and 3900 \AA\ than they do at wavelengths longer than 3900 \AA. MIKE's recovery of N is aided by strong molecular absorption bands at $\lambda$3550 and $\lambda$3800 \AA\ and another in the red at $\lambda$9150 \AA. Its \edit1{\deleted{comparatively}} higher precision for Al and S \edit1{\added{compared to HIRES}} is the result of additional atomic absorption lines beyond 8500 \AA\ and its higher S/N \edit1{\deleted{compared to HIRES}} in the red.

X-SHOOTER, despite its lower resolution ($R\sim10000$), recovers \edit1{\deleted{Fe, Ca, Si, N, Mg, C, Na, Al, K, S, and O}\added{most elements}} as precisely as, if not better than, MIKE and HIRES. For C, N, and O, X-SHOOTER can achieve precisions 2-3 times better than MIKE and HIRES as a result of its larger wavelength coverage. It is sensitive to both the CNO molecular bands in the blue optical and the NIR molecular features beyond 1 $\mu$m. \edit1{\added{Si, Mg,}} Na, Al, K, and S also have a handful of absorption features in the NIR, enabling 1-2 times higher precision with X-Shooter than MIKE and HIRES. \edit1{\deleted{X-SHOOTER's achieves comparable precision for Si and Mg from a combination of NIR absorption lines and their indirect effects on other absorption features in the NIR.}}
Furthermore, since the NIR is generally less dense with absorption features, the gradients for these elements are less degenerate with other stellar labels and can thus be more precisely recovered.

\edit1{\added{
The comparatively lower precision of FLAMES-UVES can be attributed to it shorter (and redder) wavelength coverage, which does not include nearly as much of the high-information density spectral regions as the other spectrographs considered here. Furthermore, the S/N is roughly 2-3 times lower than that of MIKE or HIRES. Depending on the desired science, however, the multiplexing capabilities of UVES may more than make up for its lower throughput and wavelength coverage.
}}

At low S/N (e.g., 5 pixel$^{-1}$), there may be a concern that the assumptions of Gaussianity, which underlies the CRLB may not be valid. However, we show in Appendix \ref{app:validation} that the CRLBs are robust to the level of $\sim$0.01 dex down to $\text{S/N}\sim5$ pixel$^{-1}$.
\edit1{\added{
Thus we believe non-Gaussianity to have a minimal impact on the CRLBs, especially compared to other practical limitations (e.g., model fidelity) that make it difficult to fully realize the precision forecasted by the CRLBs.
}}

\edit1{\added{
UVES and the UVB arm of X-SHOOTER over-sample their spectra with 5 pixels/FWHM. If the pixels in these spectra are not completely independent as we assume, the CRLBs may not be as precise as we present here (see \S\ref{sec:oversampling}).
}}

\subsubsection{High-Resolution, Single-Order} \label{sec:single_order}
Another approach to high-resolution spectroscopy involves using order-blocking filters that block all but one order of the echelle spectrum. Doing so allows for improved multiplexing, but limits the observed wavelength to a small window of 50-300 \AA. Historically, the primary application of these instruments for extragalactic archaeology has been the efficient measurement of precise radial velocities in dwarf galaxies \citep[e.g.,][]{walker:2007, walker:2009a}, but these spectra clearly contain chemical information as well.

We consider three such high-resolution, single-order, fiber-fed MOS: VLT/FLAMES-GIRAFFE, MMT/Hectochelle, and Magellan/M2FS. Due to the nature of order blocking in these instruments, there is great flexibility in deciding what small portion of spectrum to observe. In this work, we will only look at spectral regions targeted by existing observations and save a detailed analysis of the optimal wavelength windows for a future paper. For M2FS, this includes a ``HiRes" and a ``MedRes" setting around the Mg I b triplet ($\lambda\lambda$5183, 5172, 5167 \AA), which have been used for membership determination and [Fe/H] measurement in several MW satellites \citep[e.g.,][]{walker:2007, walker:2009a, walker:2015_m2fs, walker:2016}. The RV31 order-blocking filter was used on Hectochelle for similar purposes \citep[e.g.,][]{walker:2009a, walker:2015_hectochelle, spencer:2017} and is also utilized by the H3 MW halo survey \citep{conroy:2019a, conroy:2019b}. On FLAMES-GIRAFFE, five setting have been used by the DART (Dwarf Abundances and Radial Velocities Team) program to measure various abundances and radial velocities in Local Group dwarf galaxies: LR8, HR10, HR13, HR14A, and HR15 \citep[e.g.,][]{hill:2019, theler:2019}. Details for all of these instruments and settings can be found in Table \ref{tab:instruments}.

Just as with the previous high-resolution CRLBs, we consider 6 hours of integration on our $\log(Z)=-1.5$ RGB star at 100 kpc and the the ETC configurations in Table \ref{tab:ETC}.

\begin{figure*}[ht!]
	\includegraphics[width=\textwidth]{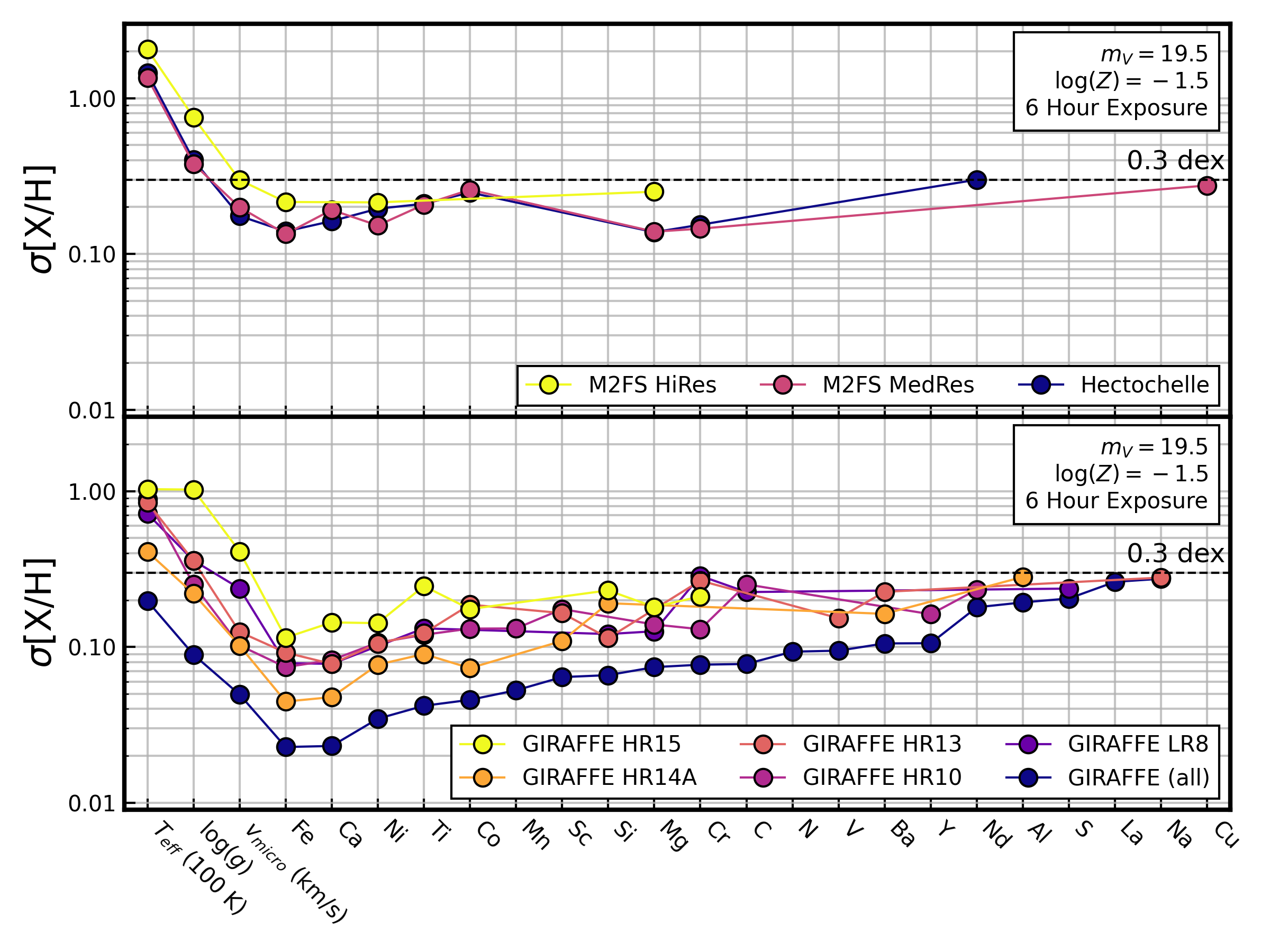}
    \caption{Same as Figure \ref{fig:crlb_highres1} but for multiplexed, single-order echelle spectrographs. CRLBs for Magellan/M2FS and MMT/Hectochelle are included in the top panel, and CRLBs for various VLT/FLAMES-GIRAFFE orders are included in the bottom panel. Elements are ordered by the precision forecasted for a combined analysis of all 5 GIRAFFE orders shown. The CRLBs suggest that even very small regions of spectrum, when well chosen, may contain non-negligible chemical information.} 
    \label{fig:crlb_highres2}
\end{figure*}

Figure \ref{fig:crlb_highres2} shows the forecasted precision for these single-order echelle spectrographs. As expected, the limited wavelength coverage of these setups severely reduces their chemical abundance recovery compared to the full-optical high-resolution spectrographs presented in Figure \ref{fig:crlb_highres1}. Even most low-resolution spectrographs can achieve comparable or better abundance recovery in a fraction of the time as presented in Figures \ref{fig:crlb_keck} and \ref{fig:crlb_other}. This is because the information content scales proportionally with the square root of the number of absorption features. A smaller wavelength range means fewer lines for a given element and worse precision.

Nevertheless, given the narrow wavelength range covered by these orders and the low S/N ($\sim$15-30 pixel$^{-1}$), it is promising that more than a handful of elements beyond Fe can be recovered to better than 0.3 dex. We first consider the abundance precision for M2FS and Hectochelle (Figure \ref{fig:crlb_highres2}; top), which cover 5100-5300 \AA. This narrow region of the spectrum contains numerous absorption lines of Fe, and to a lesser extent also of Ni, Ti, Co, Cr, and Nd, which enable their recovery. All three filters were designed to include the Mg I b triplet and as a result Mg can also be measured. 
\edit1{\deleted{
Also included in this portion of the spectrum are a few ($<$5) strong ($\sim$10-20\%/dex at $R\sim20000$) lines each for Ca, Y, Cu, and Sc, which enable Hectochelle to recover their abundances.
}\added{
There are also a few ($<$5) strong ($\sim$10-20\%/dex at $R\sim20000$) lines each for Ca and Cu in this wavelength range that enable the M2FS MedRes configuration with its broader wavelength coverage to recover these elements. Hectochelle's wavelength range excludes two Cu lines between 5100 \AA\ and 5160 \AA\ and thus recovers only Ca and not Cu to better than 0.3 dex.
}}
Because M2FS's \edit1{\deleted{filters have}\added{HiRes filter has}} a more limited wavelength range, \edit1{\deleted{they miss}\added{it misses}} a considerable fraction of these lines and thus cannot measure these abundances as precisely.

\edit1{\added{
 Hectochelle, however, does over-sample its spectra with 6 pixels/FWHM. If the pixels in these spectra are not completely independent the precision we present here may be slightly overestimated (see \S\ref{sec:oversampling}).
}}

Next we consider GIRAFFE, which has several orders that span the entire optical spectrum.
Fe, Ca, Ni, Ti, and Co all have numerous strong lines ($>$10\%/dex at $R\gtrsim20000$) below 7000 \AA, enabling their recovery by all the high-resolution order-blocking filters. Mn, however, has the majority of its strongest lines between 5300 and 5600 \AA\ and is thus \edit1{\deleted{best}\added{only}} recovered by HR10. The same is approximately true for Y \edit1{\added{and Nd}}. Ba has two moderate absorption features ($>$10\%/dex at $R\gtrsim20000$) at $\lambda$6143 \AA\ and $\lambda$6499 \AA\ in the HR13 and HR14A filters respectively, but is \edit1{\deleted{only recoverable}\added{better recovered}} in HR14A because of the filter's higher S/N and resolving power. 
\edit1{\deleted{
Overall, the combination of throughput and resolution enables HR14A to recover more elements to higher precision than the other individual filters despite its redder wavelength coverage.
}\added{
The combination of throughput and resolution enables HR14A to achieve higher precision for its recoverable elements than the other individual filters, though its redder wavelength coverage precludes it from measuring elements whose lines reside primarily at wavelengths bluer than 6000 \AA.
}}

For reference, we also include the CRLB for the combined analysis of all five GIRAFFE orders as was done in \cite{hill:2019} (Figure \ref{fig:crlb_highres2}; bottom). It is clear that by combining the many information-carrying absorption features across all orders provides a significant improvement in the possible stellar label precision \edit1{\added{and enables the measurement of elements that no individual filter alone could recover (e.g., N and La)}}.
However, to achieve the S/N and abundance precision found here, would require 6 hours of integration on each of the five GIRAFFE orders for a total of 30 hours of integration. Still, it is useful to compare this precision to that of low-resolution MOS and high-resolution single-slit echelle spectrographs. While low-resolution blue-optical spectroscopy can achieve similar precision abundance determinations for a similar number of stars in a small fraction of the time, the kinematic information in these observations is limited---at $R\sim2000$, the precision of radial velocity measurements is only $\sigma_{RV}\sim150$ km/s, which is good enough for membership determination, but not for detailed kinematic studies. In contrast, $R\sim20000$ spectra yield $\sigma_{RV}\sim5$ km/s, which are precise enough for stellar multiplicity determinations, orbit reconstruction, and dark matter mass measurements. Furthermore, these high resolution observations will be less prone to systematics incurred by model imperfections in blended lines.

A drawback to high-resolution single-slit echelle spectrographs is the amount of time required to build up large samples of stars.
In 30 hours of integration time, assuming 6 hours per pointing and ignoring overheads, HIRES, MIKE, and X-SHOOTER could observe 5 stars, while 5 echelle orders (6 hours each) could be acquired by GIRAFFE for $\sim$100 stars. Ultimately, the choice of instrument and observing strategy is highly dependent on the science case and whether higher abundance precision or a larger sample size is most valuable and whether precise radial velocities are needed. However, in the specific case of chemo-dynamical studies of dwarf galaxies, where both chemical and kinematic information is desired for a large number of stars, it may be worth trading in full optical coverage for \edit1{\added{specific wavelength regions and}} higher multiplexing.

\section{Forecasted Precision of Future Instruments} \label{sec:future}
In this section, we forecast the precision achievable by instruments currently in their construction or design stages. 
Our lengthy, but incomplete, list includes JWST/NIRSpec, 30-m class ELTs, and several planned survey facilities (e.g., MSE, FOBOS).
Because many of these instruments are still undergoing conceptual and practical revisions, the specifications we adopt in this section are estimates based on the best currently available information.

\subsection{JWST/NIRSpec} \label{sec:jwst}
The unprecedented angular resolution of the Near-Infrared Spectrograph (NIRSpec) on JWST opens up a new domain of crowded-field extragalactic stellar spectroscopy that is currently at or beyond  the limits of the most powerful ground-based telescopes (e.g., faint stars in the disk of M31 or beyond the Local Group).

In this analysis, we consider 4 of the 9 NIRSpec MOS disperser-filter combinations whose details can be found in Table \ref{tab:instruments}.
We consider 6 hours of integration and a $\log(Z)=-1.5$ TRGB star at a magnitude of $m_V=21$, which is similar to observing such a star in M31 or at the edge of the Local Group. 

\begin{figure*}[ht!]
	\includegraphics[width=\textwidth]{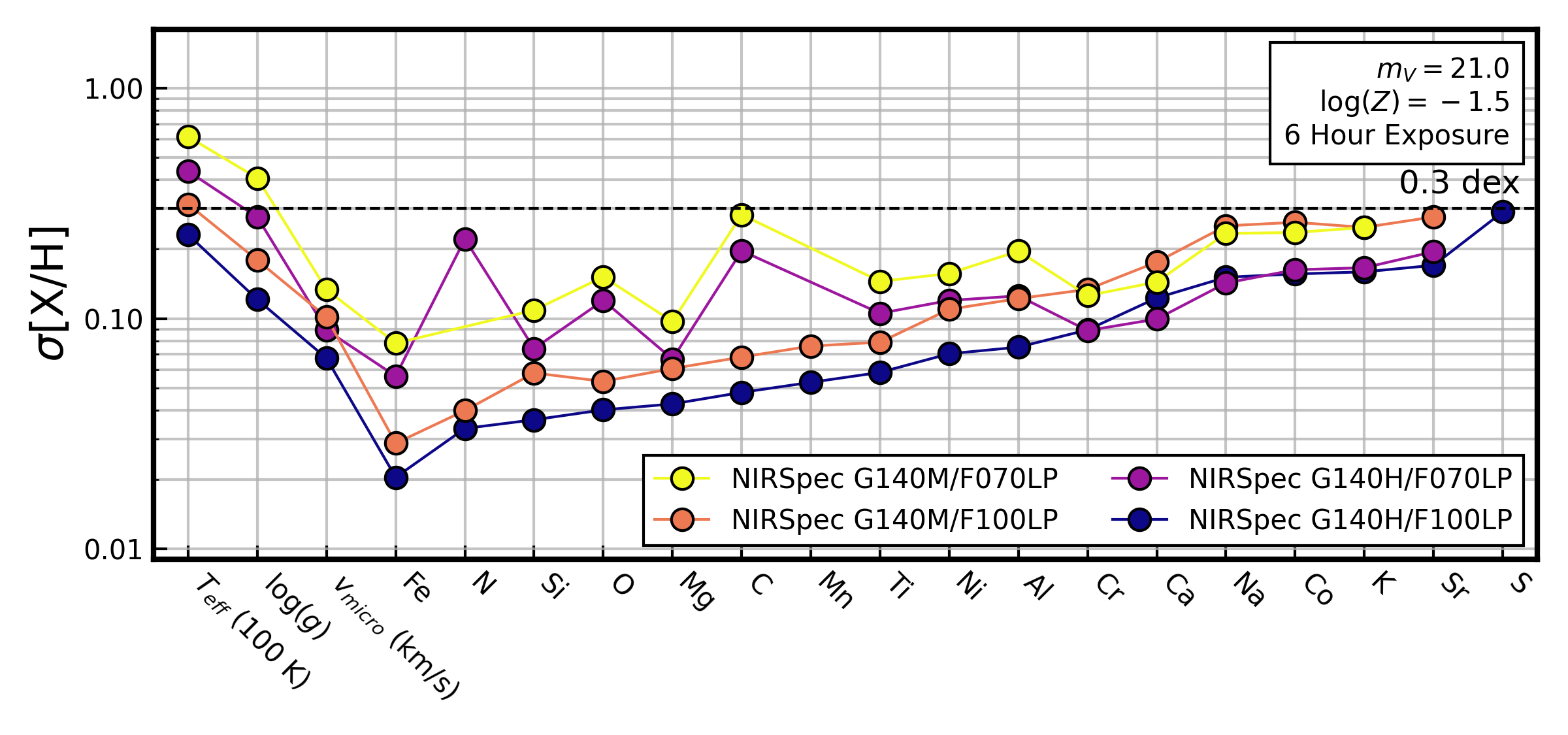}
    \caption{CRLBs for four gratings on JWST/NIRSpec assuming a 6 hour exposure of a $\log(Z)=-1.5$, $m_V=21$ TRGB star. The elements are ordered by decreasing precision as forecasted up to 0.3 dex. These CRLBs represent the abundance precision that can be measured for RGB stars in M31 or in dwarf galaxies at the edge of the Local Group.} 
    \label{fig:crlb_jwst}
\end{figure*}

Figure \ref{fig:crlb_jwst} shows the CRLBs for JWST/NIRSpec. We predict that NIRSpec can recover between 13 and 17 individual elemental abundances to better than 0.3 dex despite its low-resolution of these spectra ($R<3000$) and the faintness of the target star. This is quite promising for the future of extragalactic stellar spectroscopy as the field moves towards more distant and crowded extragalactic systems. For comparison, ground-based observations are presently limited to measuring only [Fe/H], bulk $\alpha$-element enhancements, and a few other elements in the M31's halo and satellites \citep[e.g.,][]{collins:2013, vargas:2014b, escala:2019a, gilbert:2019, kirby:2020}. 

Figure \ref{fig:crlb_jwst} also shows that for the same filter (i.e., wavelength coverage) the slightly higher resolution of the G140H grating provides an advantage in precision over the G140M grating despite the reduced S/N (100 pixel$^{-1}$ vs.\ 160 pixel$^{-1}$ at 1.2 $\mu$m). Just as in \S \ref{sec:highres}, this is consistent with the CRLBs scaling with $R^{-1/2}$ at fixed wavelength coverage.

Further, we see that the redder F100LP filter provides better abundance precision than the blue F070LP filter. This is due to a combination of factors including the F100LP's larger wavelength coverage and marginally higher S/N. Though it is true that blue optical wavelengths are rich in information, the situation changes in the red, where molecular bands in the NIR are more information rich than the red-optical.

In fact, the abundance precision benefits greatly from information contained at wavelengths longer than 1.4 $\mu$m provided by the F100LP filter. These redder wavelengths include numerous molecular features like the strong H$_2$O absorption lines that extend to 1.8 $\mu$m. Also included are bands of CN ($\lambda1.1$ $\mu$m), OH ($\lambda1.4$ $\mu$m), and CO ($\lambda1.5$ $\mu$m), features, which enable precise determinations of C, N, and O. In addition to Fe, Si, and Mg, which have absorption features somewhat uniformly distributed from 7000 \AA\ to 1.8 $\mu$m, the F100LP filter also enables precise recovery of Mn, which has $\sim$10 lines between 1.2 and 1.4 $\mu$m with strengths greater than 1\%/dex (at $R=2700$).

The redder wavelength coverage of the F100LP filter also allow for more precise recovery of $T_\text{eff}$ and $\log(g)$. This is the result of both Paschen lines at $\lambda\lambda$1.05, 1.09, and 1.28 $\mu$m and Brackett lines red-ward of 1.46. These lines are all sensitive to atmospheric parameters and thus provide strong constraints on $T_\text{eff}$ and $\log(g)$ (and to a lesser extent Fe, Si, Mg, and Al).

The bluer wavelength coverage of the F070LP filter does provide better recovery for Ti, Ca, Na, and Cr. Constraints on Ti abundance come from several TiO bands blueward of 1 $\mu$m and constraints on Cr come from roughly a dozen weak ($<$2\%/dex at $R=2700$) lines blueward of 1.2 $\mu$m. The precision of Ca and Na is a result of the Ca I triplet at $\lambda\lambda$8498, 8542, 8662 \AA\ and Na I doublet at $\lambda\lambda$8185, 8197 \AA\ as discussed previously in \S \ref{sec:blue_keck}.

We conclude by noting potential challenges in achieving the NIRSpec CRLBs. NIRSpec's elemental precision is strongly contingent on the information content of complicated molecular features. As a result, the abundances measured by NIRSpec may be quite sensitive to assumptions of the model atmosphere, molecular network, and linelists employed. Achieving the reported CRLBs and avoiding large systematics at $R<3000$ will require careful treatment of this portion of the spectrum.

In addition, due to the rigid nature of NIRSpec's mechanical slit mask, it will frequently be the case that stars will lie slightly off the center of their slit. In addition to a small cut in S/N to lost light, this introduces deviations to the expected LSF of the spectrum. Accounting for this effect will be important for abundance recovery to approach the forecasted precision and avoid systematics caused by variations in the LSF. Efforts to calibrate NIRSpec early in the lifetime of JWST should help to mitigate this issue.

\subsection{Extremely Large Telescopes} \label{sec:elt}
The advent of extremely large telescopes (ELTs) with apertures in excess of 30 meters have the potential to revolutionize extragalactic archaeology. Their higher angular resolution and increased light collecting power will enable the spectroscopic observation of resolved stars in some of the most distant and compact systems in and around the Local Group.
The Thirty Meter Telescope (TMT; 30-m aperture), the European-Extremely Large Telescope (E-ELT; 39-m aperture), and the Giant Magellan Telescope (GMT; 24.5-m aperture) all have plans for a highly multiplexed spectrographs---TMT/WFOS, E-ELT/MOSAIC, GMT/GMACS, and GMT/G-CLEF. 

\subsubsection{Low-Resolution ELT MOS} \label{sec:elt_lowres}
We first consider the three low-resolution spectrographs WFOS, MOSAIC, and GMACS, which all enable observations of $100+$ stars across the full optical spectrum at resolving powers between $R\sim1000$ and $R\sim5000$. The configurations we consider are listed in Table \ref{tab:instruments}. 
As in \S \ref{sec:lowres}, we assume a 1 hour observation of our fiducial $\log(Z)=-1.5$ RGB star with $m_V=19.5$ and the ETC configurations in Table \ref{tab:ETC}.

\begin{figure*}[ht!]
	\includegraphics[width=\textwidth]{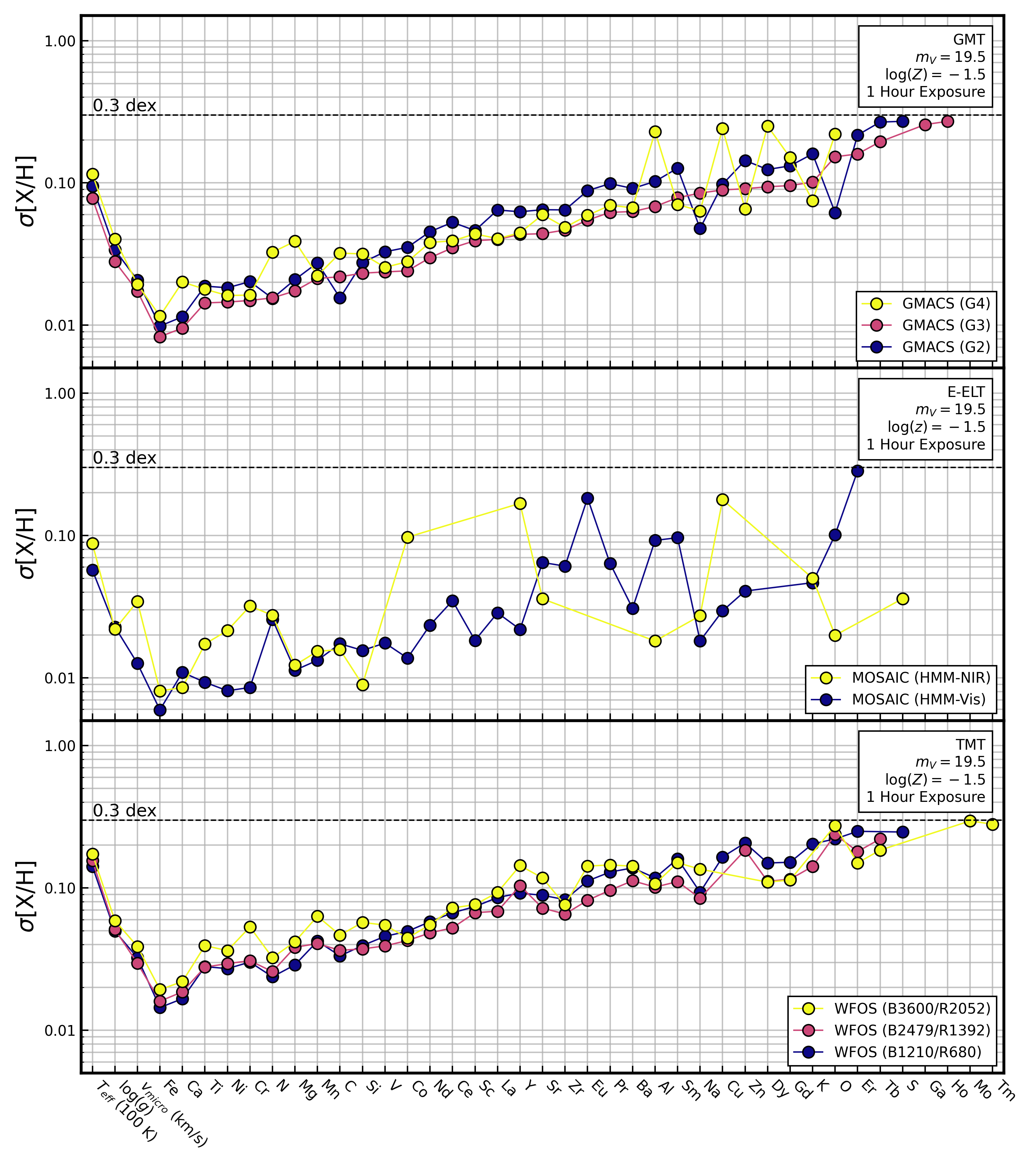}
    \caption{Same as Figure \ref{fig:crlb_keck} but for the low-resolution ELT spectrographs GMT/GMACS, E-ELT/MOSAIC, and TMT/WFOS.} 
    \label{fig:crlb_elt1}
\end{figure*}

Figure \ref{fig:crlb_elt1} presents the CRLBs for these ELT spectrographs. We predict that all three optical ELT spectrographs are capable of measuring 30 to 40 elemental abundances to better than 0.3 dex. In addition to all Fe-peak elements and most $\alpha$-elements, this includes 22 neutron-capture elements spanning all three r- and s-process peaks. Of these, \edit1{\deleted{10}\added{12}}, \edit1{\deleted{12}\added{9}}, and \edit1{\deleted{13}\added{8}} can be recovered to better than 0.1 dex by GMACS (G3), MOSAIC (HMM-VIS), and WFOS (B2479/R1392) respectively. 

Many of these elements have only weak features below 4000 \AA, which necessitate high S/N in the blue-optical and near-UV for their recovery. Tb and Tm, for example, have $\sim$20 absorption lines with 1-3\%/dex gradients at $R\sim3500$, but nearly all are found at wavelengths shorter than 4000 \AA. Similarly, Pd, Os, and Hf have fewer than 10 absorption lines of similar strengths, which are also predominantly located blue-ward of 4000 \AA. The strongest line of Th is at $\lambda$4019 \AA\ with a gradient of $\sim$1.5\%/dex, while $\sim$20 weaker (0.5-1.0\%/dex) features exist between 3100 and 4000 \AA. Despite the limited chemical information, spectrographs on ELTs are capable of measuring these elements because their large aperture telescopes and blue wavelength coverage can achieve S/N$\sim$100 at 4000 \AA. 

The informative power of blue-optical spectroscopy can be further seen in the comparatively poorer abundance recovery of MOSAIC's HMM-Vis and HMM-NIR settings. Because the optical arm only extends to 4500 \AA, it cannot capitalize on the information rich near-ultraviolet stellar spectrum. The NIR is expected to recover even fewer abundances than the optical arm due the lower information density beyond 8000 \AA. Nevertheless there are some elements (e.g., \edit1{\added{Ca, Si}}, Sr, O, Al, and S) whose absorption features are better observed in the NIR. CN absorption in the red and NIR also allow for recovery of C and N to a similar degree as can be done with spectra down to 4500\AA. We note, however, that because the JWST NIRSpec ETC was re-purposed to provide S/N in the NIR for MOSAIC, the S/N used here does not include the effects of troublesome NIR telluric features. As a result, we expect the abundance precision of MOSAIC's HMM-NIR spectra to be noticeably worse in practice. 

Figure \ref{fig:crlb_elt1} (top) illustrates the trade-offs in S/N, wavelength coverage, and resolution at fixed number of detector pixels for 3 different GMACS gratings.
As predicted by \citet{ting:2017}, the abundance precision of a detector with fixed pixel real estate under the assumption of the uniform distribution of chemical information is relatively invariant of the resolving power. Of course, there are slight differences in the expected precision of the gratings. For many elements, G2 ($R=1000$) performs more poorly than the higher resolution gratings, which is likely due to strongly blended lines at $R=1000$ and the resulting increased covariance between elements. It is also apparent that the chemical information is not uniformly distributed; there are several abundances (e.g., Cr, C, Ba, Al, Dy, Gd, and K) which the G4 grating recovers noticeably worse if not at all because the absorption features of these elements lie outside of its reduced wavelength coverage. These elements are predominantly those with few strong features that lie below 4200 \AA. \edit1{\added{Similar conclusions can be drawn from a comparison of the 3 WFOS grating combinations.}}

\subsubsection{High-Resolution ELT MOS}
Here, we consider G-CLEF, a GMT first-light fiber-fed echelle spectrograph. While it is primarily optimized for very high-resolution  ($R\sim100000$) single-slit spectroscopy across the optical, it will also feature a MOS mode that will combine modest multiplexing, Keck/HIRES-like spectra, and a 24.5-m aperture telescope that will dramatically increase the feasibility of high-resolution spectroscopy of stars beyond the immediate vicinity of the Local Group (see Tables \ref{tab:instruments} and \ref{tab:instruments2} for details). We calculate the S/N using the G-CLEF ETC given the same observational conditions used for the forecasting of existing high-resolution instruments (see Table \ref{tab:ETC}).

\begin{figure*}[ht!]
	\includegraphics[width=\textwidth]{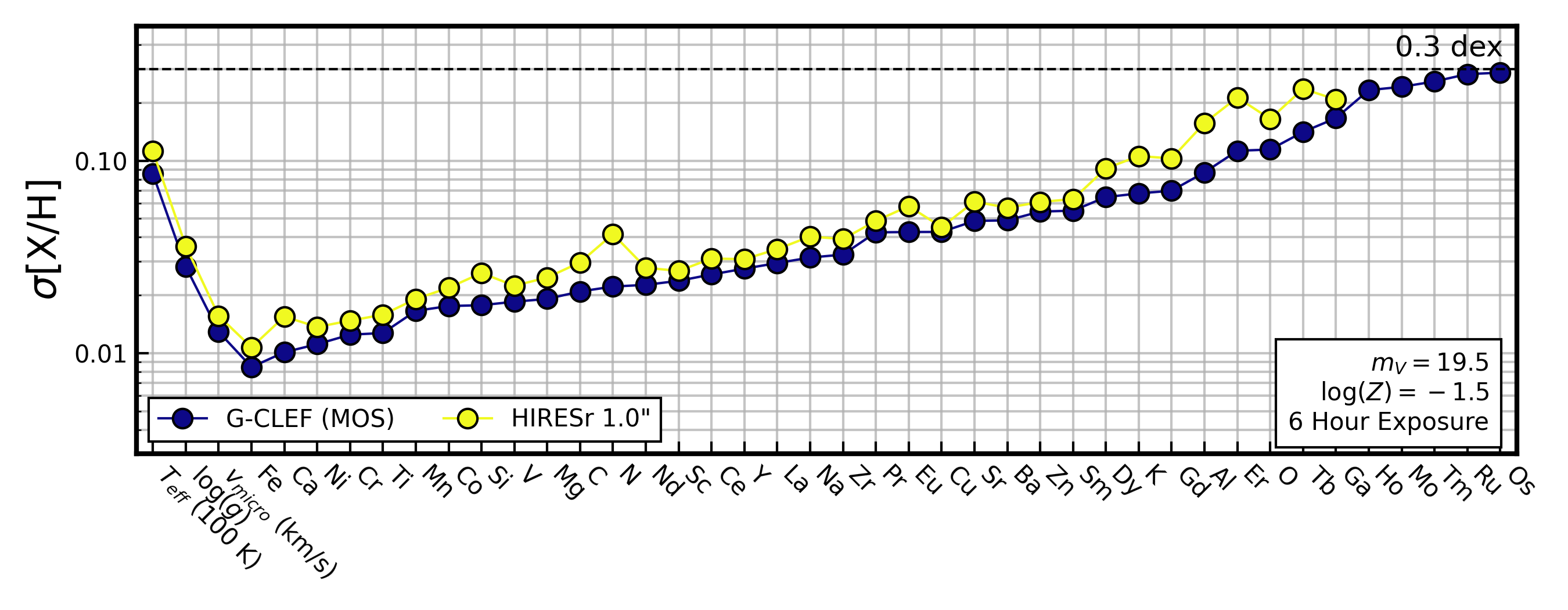}
    \caption{Same as Figure \ref{fig:crlb_highres1} but for high-resolution ELT spectrograph GMT/G-CLEF.} 
    \label{fig:crlb_elt2}
\end{figure*}

Figure \ref{fig:crlb_elt2} shows the CRLBs of G-CLEF with the HIRES 1".0 CRLBs for comparison. We forecast that G-CLEF observations will recover 30 elements to better than 0.1 dex (and nearly 40 to 0.3 dex) similar to HIRES and the other single-slit high-resolution spectrographs analyzed previously in \S\ref{sec:single_slit}. In addition to achieving HIRES-like abundance recovery, G-CLEF's multiplexing enables the simultaneous observation of up to 40 stars at a time. This dramatically increasing the feasibility of high-resolution studies of substantial numbers of stars in extragalactic systems (for both chemistry and kinematics).

The reason G-CLEF does not achieve substantially better abundance precision than its 10-m class analogues appears to be largely a consequence of G-CLEFs lower predicted throughput. Despite having a much larger light collecting power, G-CLEF acquires roughly the same S/N as Keck/HIRES at wavelengths shorter than 6000 \AA\ where most of the chemical information resides. G-CLEF achieves higher S/N ($\sim$35 pixel$^{-1}$ compared to $\sim$20 pixel$^{-1}$) at longer wavelengths, but this only yields small improvements in abundance precision. Furthermore, G-CLEF's bluer wavelength coverage is at $\text{S/N}<5$ pixel$^{-1}$ and thus provides little additional information.

\subsection{Spectroscopic Surveys} \label{sec:surveys}
Galactic archaeology in the MW has been revolutionized by several large-scale spectroscopic surveys (e.g., RAVE; \citealt{steinmetz:2006}, SEGUE; \citealt{yanny:2009}, LAMOST; \citealt{luo:2015}, GALAH; \citealt{desilva:2015}, APOGEE; \citealt{majewski:2017}, DESI)\footnote{\edit1{\deleted{In Appendix}\added{In Appendices \ref{app:lamost_compare} and}} \ref{app:desi}, we forecast the precision of \edit1{\added{the ongoing LAMOST MW survey and the}} recently begun DESI survey of MW halo stars. For forecasted precision of other MW surveys we refer the reader to \cite{ting:2017}.}; \citealt{desi:2016_sci}. These surveys have collected millions of stellar spectra from which detailed abundance patterns have been measured. The success of these surveys in the realm of stellar abundance measurements is in part due to the high-quality and homogeneity of the spectra collected. This has allowed for rigorous, self-consistent analyses, the implementation of data-driven approaches, and the refining of stellar models. However, similarly ambitious observing campaigns outside the MW are in their early stages, primarily because it requires a dedicated survey instrument on a 10-m class telescope. 

The next decades is poised to bring the field of extragalactic stellar spectroscopy its first large sample of homogeneously collected stellar spectra. For example, the Prime Focus Spectrograph (PFS) on Subaru will begin science observations in early 2020. PFS will dedicate $\sim$100 nights to surveying M31's disk and halo, making it the largest extragalactic stellar spectroscopic survey to date \citep{tamura:2018}. 

The MSE will replace the CFHT with an 11.25-m dedicated survey telescope, while FOBOS is a next-generation instrument proposed for the Keck telescopes with time dedicated for a stellar (extra-)galactic archaeology survey. Both MSE and FOBOS are much earlier in their conceptual design and plan to be on sky by $\sim$2030 \citep{MSE:2019, bundy:2019}.  

The details for these spectrographs can be found in Tables \ref{tab:instruments} and \ref{tab:instruments2}. For all three survey instruments we consider our standard 1 hour of integration time of our fiducial $\log(Z)=-1.5$, $m_V=19.5$ RGB star and the ETC configurations in Table \ref{tab:ETC}.

\begin{figure*}[ht!]
	\includegraphics[width=\textwidth]{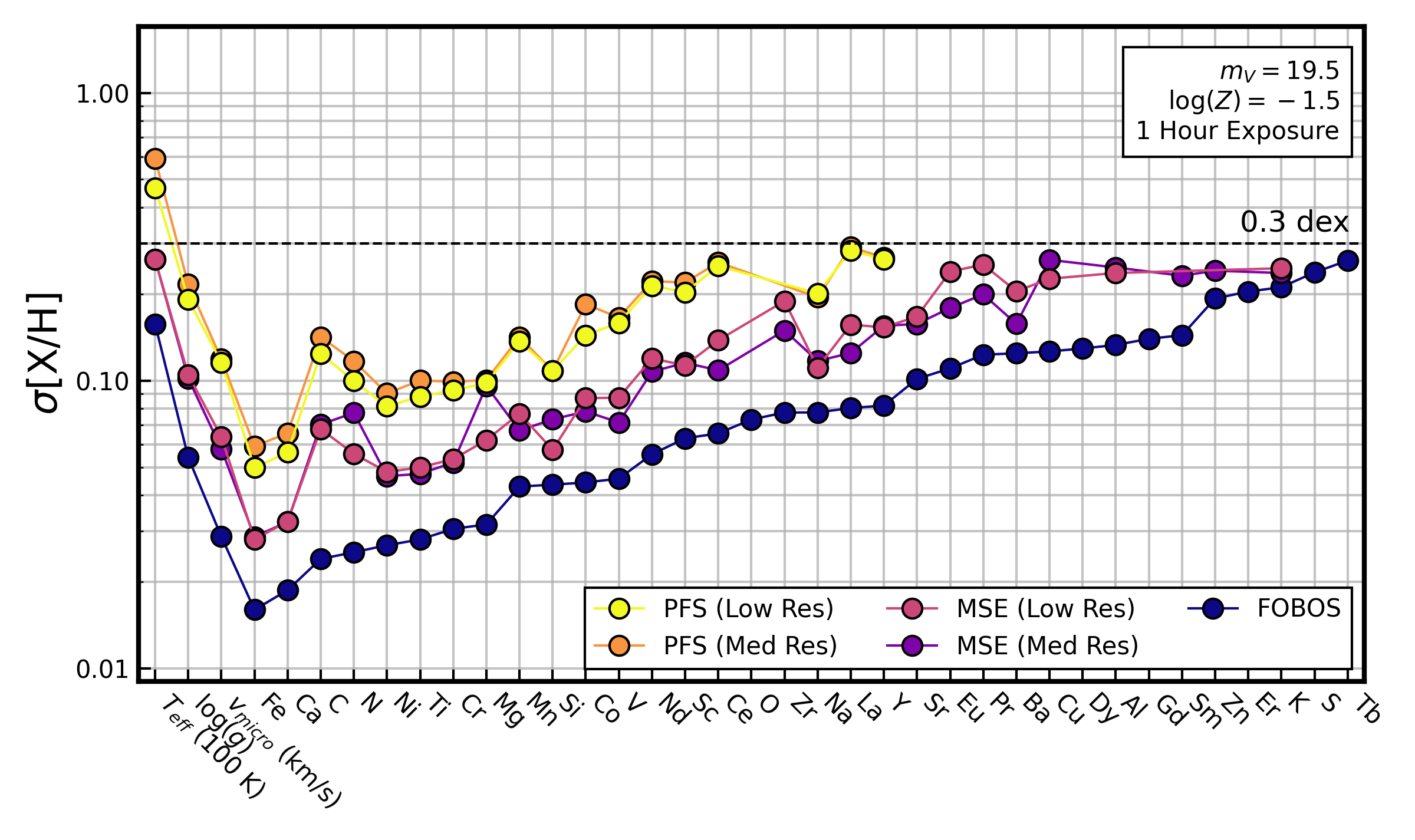}
    \caption{Same as Figure \ref{fig:crlb_keck} but for the survey instruments PFS, MSE, and FOBOS.} 
    \label{fig:crlb_survey}
\end{figure*}

We present the abundance precisions of PFS, MSE, and FOBOS for this observing scenario in Figure \ref{fig:crlb_survey}. All three spectrographs are capable of similar chemical abundance precision as blue-optimized spectrographs considered in \S\S\ref{sec:blue_keck} and \ref{sec:other_mos} (e.g., DEIMOS 1200B, LRIS, and MODS), recovering $>$20 elements to better than 0.3 dex. As seen in previous analyses, there are only minor differences between the low- and medium-resolution setting on PFS and MSE. The increase in resolution is roughly cancelled out by decreases in S/N and wavelength coverage. In this comparison, the additional wavelength coverage beyond 1 $\mu$m by the NIR and red arms of PFS and MSE (low-res) provide improved precision of Si and Al, but not C, N, and O which would require even redder spectra that extend past 1.4 $\mu$m.

\edit1{\deleted{
The lower precision of PFS compared to MSE and FOBOS is simply the result of lower S/N, which is mostly due to Subaru's smaller aperture, but to a lesser degree because the PFS S/N includes the effects of stellar absorption features and sky lines while the MSE and FOBOS S/N does not\footnote{The MSE ETC does include these effects, but we were unable to extract the detailed features from the ETC output.}.
}\added{
Despite the relatively similar specifications of these three survey spectrographs, there is a considerable spread in their forecasted abundance precision. This can be attributed to two predominant factors. The first and most important factor is the S/N of the observations. Throughout most of the optical, PFS achieves a S/N only 1/2 to 3/4 that of FOBOS and MSE. In addition, FOBOS's blue sensitivity enables a $\text{S/N}>10$ pixel$^{-1}$ down to 3500 \AA\ for these observations, while the S/N of MSE and PFS drop below a S/N of 10 pixel$^{-1}$ at $\sim$4000 \AA.
}}

\edit1{\added{
The second factor contribution to the higher precision predicted for FOBOS is its higher wavelength sampling (6 pixels/FWHM), which is nearly twice that of MSE and PFS. Even holding all other instrument specifications constant (e.g., wavelength coverage, resolving power, S/N), the higher sampling alone leads to a $\sqrt{2}$ improvement in the forecasted precision. Of course, oversampling the spectrum by this degree in practice would likely lead to increased correlations between adjacent pixels, resulting in a smaller improvement than our na\"ive scaling with $n^{-1/2}$ predicts (see Appendix \ref{app:sampling}).
}}
%
%
%
%

\section{Discussion} \label{sec:discussion}
\subsection{Information Rich Blue Spectra} \label{sec:blue}
In the context of extragalactic spectroscopy (i.e., at medium- and low-resolution), a key result of this paper is the importance of the blue-optical spectrum for measuring abundances. Spectral regions bluer than $\sim$4500 \AA\ are rich in absorption features of $\alpha$-elements and r- and s-process elements, and overall enable the recovery of more than double the number of elements than red-optical only wavelengths. \replaced{This finding echoes the importance of blue-optical spectra highlighted by \citet{ting:2017} and demonstrated by \cite{xiang:2019} with LAMOST spectra.}{This finding echoes the power of low-resolution blue-optical spectra highlighted in \citet{ting:2017} and demonstrated by \cite{xiang:2019} with LAMOST spectra.}

\begin{figure*}[ht!]
	\includegraphics[width=\textwidth]{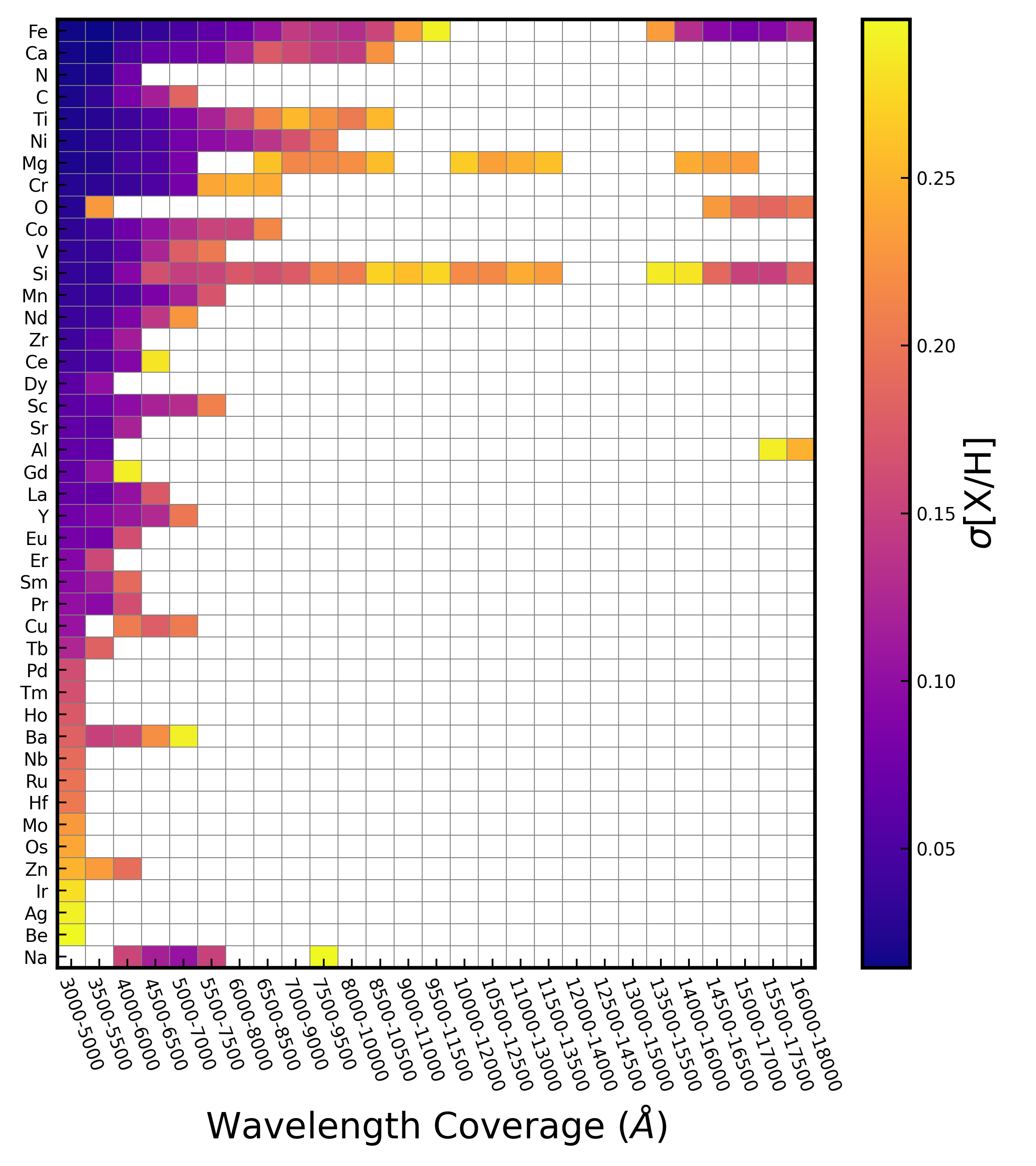}
    \caption{CRLBs for a $\log(Z)=-1.5$ RGB star observed in 2000 \AA\ wavelength regions from 3000 \AA\ to 1.8 $\mu$m, assuming $R=2000$, $R_\text{samp}=3$, constant throughput, a K2V stellar SED, and S/N$=100$ pixel$^{-1}$ at 6000\AA. This figure demonstrates the high density of chemical information found at wavelengths shorter than 4500\AA, especially for many neutron capture elements.} 
    \label{fig:windows_2000}
\end{figure*}

\begin{figure*}[ht!]
	\includegraphics[width=\textwidth]{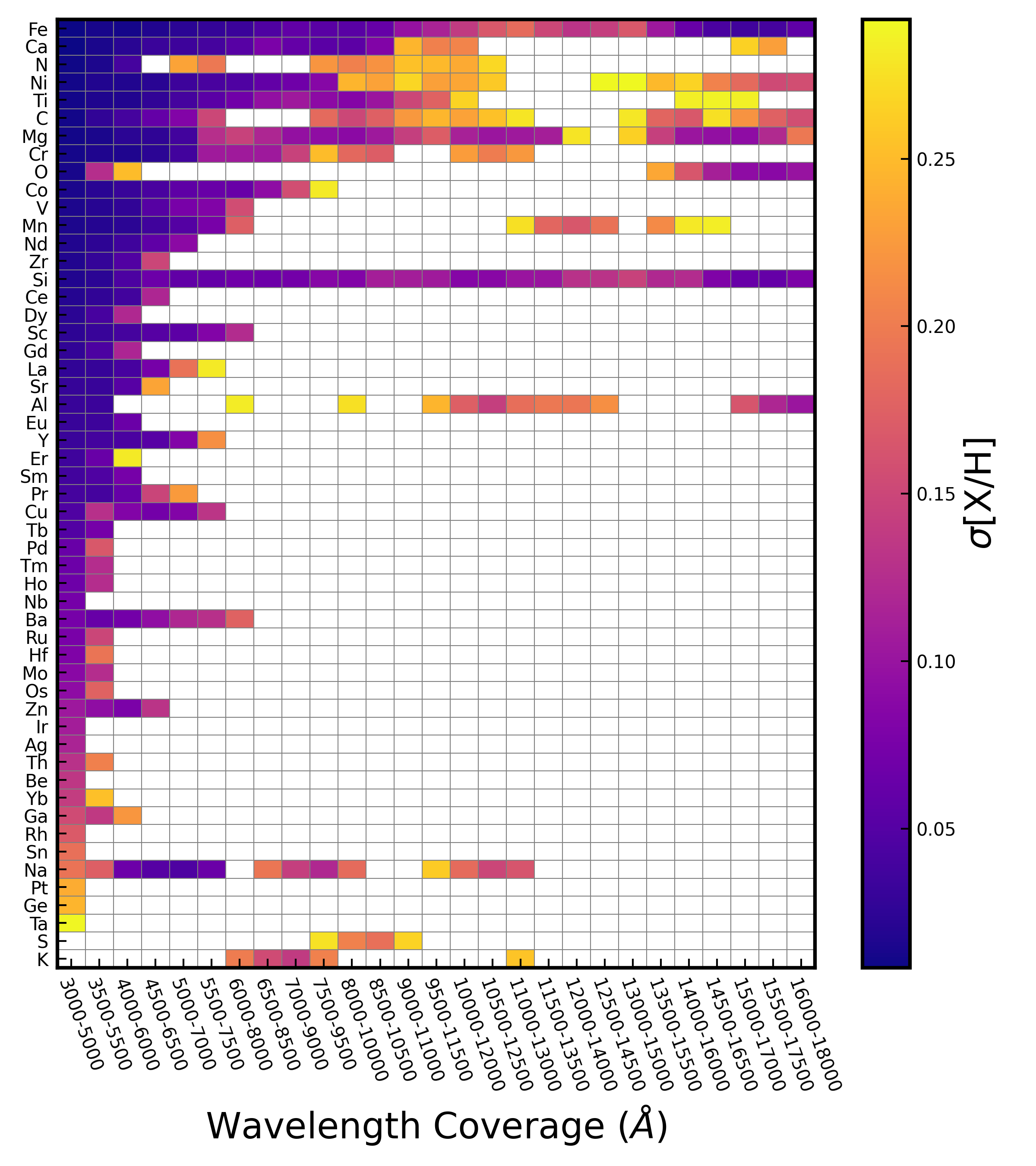}
    \caption{Same as Figure \ref{fig:windows_2000}, except for $R=5000$.} 
    \label{fig:windows_5000}
\end{figure*}

Figures \ref{fig:windows_2000} and \ref{fig:windows_5000} summarize the power of blue-optical spectroscopy for abundance recovery. To generate these figures, we have simulated a spectra with  $R\sim2000$ and 5000 respectively and a spectral sampling of 3 pixels per resolution element for a $\log(Z)=-1.5$ RGB star. We then computed the CRLB for each element for the 2000 \AA\ wavelength regions shown on the x-axis. We assume a K2V SED, constant throughput with wavelength, and a S/N of 100 pixel$^{-1}$ at 6000 \AA\ ($\sim$40 pixel$^{-1}$ at 3000 \AA; $\sim$55 pixel$^{-1}$ at 1.5 $\mu$m). Each cell is color-coded by the CRLB precision\footnote{To first order, the precision of a given element from a combination of two or more wavelengths windows can be found by taking the inverse square sum of the abundance's precision in the relevant wavelength ranges.}.

Figures \ref{fig:windows_2000} and \ref{fig:windows_5000} show that the largest number of elements can be recovered in the spectrum spanning 3000-5000 \AA. In this range, 42 (50) elements are recovered to a precision of $<$0.3 dex for $R=2000$ (5000). The number of elements available drops to 27 (31) in the 2000 \AA\ range between 4000 and 6000 \AA, indicating the rich information available below 4000 \AA.

In the 5000-7000 \AA\ range, 17 (20) elements can be recovered. As the wavelength coverage shifts redder, fewer elements are precisely measurable. At $R=2000$, no elements, including Fe, can be measured from 2000 \AA\ regions between 1.2 $\mu$m and 1.5 $\mu$m. This is because there are few absorption features for any elements---Fe with only $\sim$20 lines with gradients larger than 1\%/dex has the strongest of any element in this portion of the spectrum. The paucity of lines means there is little information to break the degeneracy between the poorly constrained $T_\text{eff}$ and $\log(g)$ ($\sigma_{T_\text{eff}}>300$ K and $\sigma_{\log(g)}>1.5$ dex) and the elemental abundances. Applying the same priors as in \S \ref{sec:d1200g_priors}, enables the recovery of Fe, Si, and Mn to better than 0.3 dex. As the wavelength coverage moves further into the near-IR (1.5-1.8 $\mu$m) the number of elements that can be recovered increases as a result of molecular features (e.g., H$_2$O and CO) and larger numbers of Fe, Si, Mg, and Al lines \citep[see APOGEE results:][]{ness:2015, garcia-perez:2016, ting:2019}.

Beyond increasing the number of elements that can be recovered, the blue-optical is rich in the absorption lines of neutron capture elements. For this reason, the blue-optical portion of the spectrum has long been targeted by high-resolution spectroscopy \citep[e.g.,][]{sneden:1983, cowan:2002, sneden:2003, hansen:2015}. 

\begin{figure*}[ht!]
	\includegraphics[width=\textwidth]{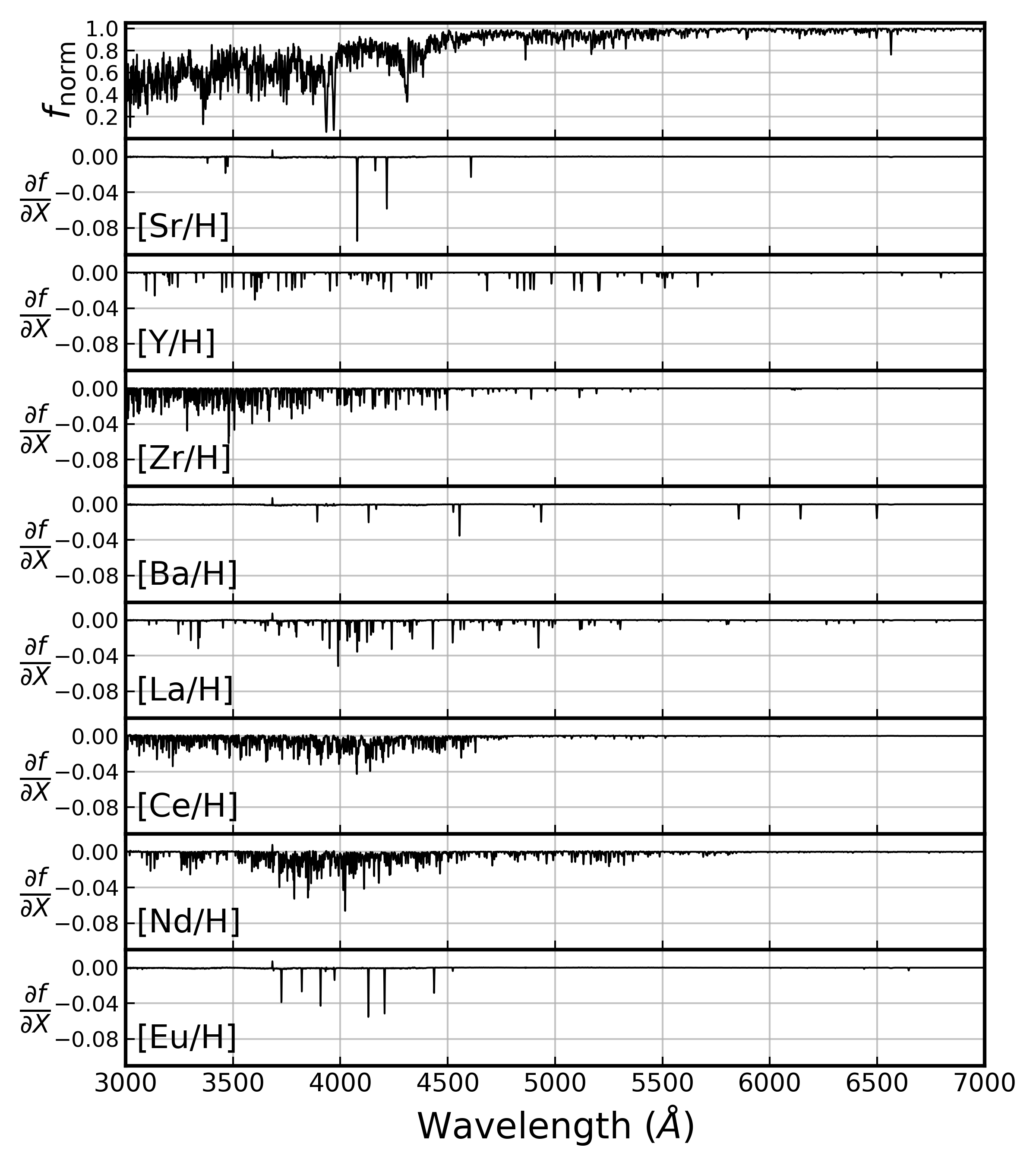}
    \caption{(Top) Spectrum of a $\log(Z)=-1.5$ RGB star convolved down to $R=2000$. (Below) Gradients of the spectrum with respect to r-/s-process elements recoverable by LBT/MODS given the setup in \S \ref{sec:other_mos}. Most of the information for these elements is at wavelengths shorter than 4500 \AA. Not shown in this figure are three modest Sr lines with gradients of 1\% dex$^{-1}$ between 1.0 and 1.1 $\mu$m and a handful of weak Y lines (all with gradients of $<$0.5\% dex$^{-1}$) that lie red-ward of 7000 \AA.} 
    \label{fig:rs_gradients}
\end{figure*}

However, as shown in Figure \ref{fig:rs_gradients}, these elements have strong gradients even at low-resolution ($R\sim2000$). Sr and Eu, for example, have a handful of absorption lines between 3500 and 4500 \AA\ with gradients of 4-8\%/dex. Other elements, like Zr, Ce, and Nd, have a forest of weaker ($\sim$2\%/dex) absorption lines that extend blue-ward of 4500 \AA. The results of Figures \ref{fig:windows_2000}, \ref{fig:windows_5000}, and \ref{fig:rs_gradients} together indicate that full spectral fitting methods have the potential to recover neutron capture elements outside the immediate vicinity of the MW.

The high information density of the blue-optical also introduces challenges to abundance recovery. For example, the large number of lines makes it challenging to define a continuum. Most spectral fitting routines operate on normalized spectra and the lack of a clearly defined continuum introduces additional sources of uncertainties into the fitting process.

A second challenge is the blending of absorption lines. The blending of spectral features is not inherently a problem for full spectral fitting, provided that all stellar labels are fit simultaneously to account for degeneracies. However, doing so requires a high degree of trust in the stellar atmosphere models, radiative transfer treatment, and linelists. When lines are resolved, individual lines that are imperfectly modelled (e.g., from non-LTE or 3D effects) can be isolated and ignored. But when lines are severely blended as they are in the blue-optical, identifying and masking (or calibrating) problematic lines becomes a non-trivial, but crucial, endeavor.

Finally, blue-optical spectra will typically have lower S/N than redder observations of the cool RGB stars we are considering---their flux peaks at $\sim$6100 \AA. To achieve the same S/N at 3000 \AA\ as at 6100\AA\ requires at least 50\% longer integration times in the blue\footnote{Assuming a constant throughput and a K2V stellar SED.}. We have attempted to take this into account by using ETCs with SEDs of cool stars to determine realistic S/N our the observing scenarios.

Taken together, the challenges of dealing with line blending and lower S/N, has meant that medium- and low-resolution blue optical spectroscopy has seldom been used for extragalactic stellar chemical abundance measurements.

These difficulties, however, do not invalidate the enormity of the information content contained in the near-UV and blue portions of a star's spectrum. Given the current designs of upcoming instruments and surveys, we will soon be awash in low-resolution blue stellar spectroscopy and the potential for major advances in abundance determinations. Fully taking advantage of this dataset will not be trivial and will take significant investments in stellar models, instrumental calibrations, and spectral fitting techniques, but we believe that it will be well worth the investment.

\subsection{Stellar Chemistry Beyond 1 Mpc} \label{sec:how_far}
At present, a full night ($\sim$6 hours) of observing time on a 10-m telescope is necessary to measure [Fe/H], [$\alpha$/Fe], and a few individual elemental abundances in stars as faint as $m_{V}\sim23$ \citep[e.g.,][]{vargas:2014a, vargas:2014b, escala:2019b, escala:2019a, gilbert:2019, kirby:2020}. While this enables the measurement of stellar metallicities in the halo of M31 with current facilities, measuring elemental abundances in systems at greater distances and stellar densities is currently out of reach, due to long integration times, read noise limitations, and crowding. Outside the Local Group, stellar spectroscopy is not possible for resolved stars. 

However, both JWST/NIRSpec and the ELT spectrographs will excel in the observation of faint stars in crowded systems. They provide Hubble-like angular resolution ($\lesssim$0".2) for spectroscopy, can achieve reasonable S/N for faint stars in modest integration times, and are sensitive to the spectral features of many elements (see \S\S \ref{sec:jwst} and \ref{sec:elt_lowres}). 

\begin{figure*}[ht!]
	\includegraphics[width=\textwidth]{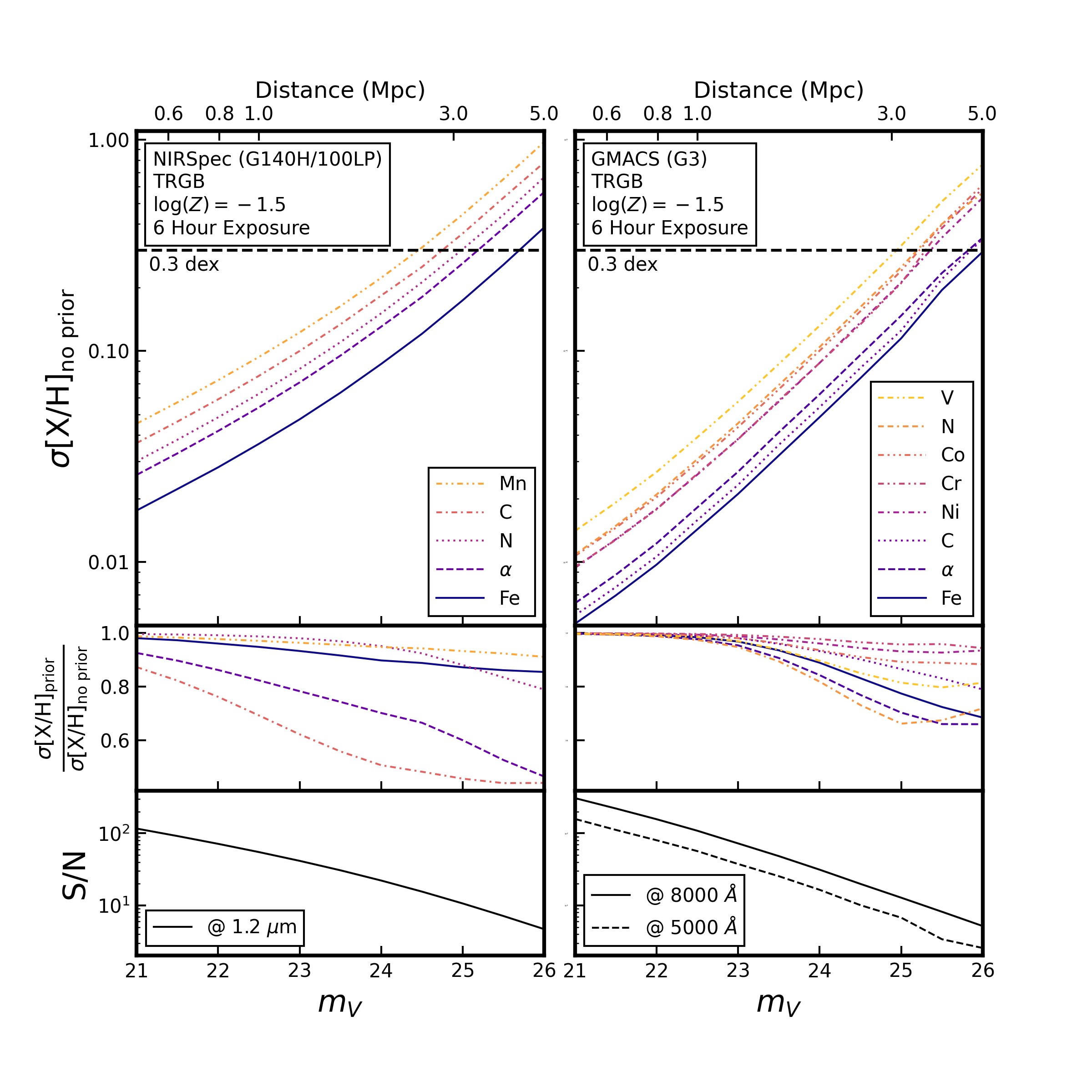}
    \caption{CRLBs for the JWST/NIRSpec G140H/100LP (left) and the GMT/GMACS G3 (right) setups given a 6-hour observation of a $\log(Z)=-1.5$ TRGB star as a function of apparent magnitude and distance. The middle panels show how the CRLBs improve when assuming Gaussian priors of $\sigma_{T_\text{eff},\text{prior}}=100$ K, $\sigma_{\log(g),\text{prior}} = 0.15$ dex, and $\sigma_{v_\text{micro},\text{prior}}=0.25$ km/s. The S/N at a characteristic wavelength is plotted in the bottom panels for each instrument. Small wiggles in the G3 S/N at 5000 \AA\ (and CRLBs) are due to interpolation errors in the extraction of data from the GMACS ETC at low S/N. JWST and ELTs will enable the recovery of Fe and $\alpha$ to better than 0.3 dex beyond 4 Mpc, and out to $\sim$3 Mpc for a handful of other elements.} 
    \label{fig:how_far}
\end{figure*}

Figure \ref{fig:how_far} illustrates the potential of JWST and the ELTs for resolved star spectroscopy in and beyond the Local Group. Here we plot the CRLB for several elements as a function of distance for two telescope configurations: JWST/NIRSpec (G140H/100LP) and GMT/GMACS (G3) (see Table \ref{tab:instruments}). For these calculations, we assume 6-hour observations of a $\log(Z)=-1.5$ TRGB star (see Table \ref{tab:ref_stars}) and replace the CRLBs of individual $\alpha$-elements (O, Ne, Mg, Si, S, Ar, Ca, and Ti) with a CRLB for [$\alpha$/H]\footnote{The gradients for $\alpha$ were calculated as in \S \ref{sec:gradients} except that offsets were applied to all $\alpha$-element abundances in lockstep instead of individually.}.
The CRLBs indicate that JWST and GMACS will be able to measure the Fe abundance to 0.3 dex in individual stars out to 4.4 and 5.0 Mpc respectively\footnote{We note that the S/N for both instruments is quite low beyond 4 Mpc; $<$10 pixel$^{-1}$ for NIRSpec and  $<$5 ($<$10)  pixel$^{-1}$ for GMACS at 5000 (8000) \AA.}. GMACS is capable of recovering $\alpha$ abundances, primarily through Ca features and to a lesser extent from Ti, Mg, and Si features, out to 4.5 Mpc. For NIRSpec, $\alpha$ is recovered through a combination of Si, O, and Mg features (in order of decreasing importance) out 3.5 Mpc. 

The small wiggles in the \edit1{\deleted{G2}\added{G3}} S/N at 5000 \AA\ (and CRLBs) seen beyond 25 Mpc are the result of interpolation errors in the extraction of data from the GMACS ETC at low S/N.

 We also calculate the Bayesian CRLB using the same Gaussian priors as in \S\ref{sec:d1200g_priors} ($\sigma_{T_\text{eff},\text{prior}}=100$ K, $\sigma_{\log(g),\text{prior}} = 0.15$ dex, and $\sigma_{v_\text{micro},\text{prior}}=0.25$ km/s). The middle panel of Figure \ref{fig:how_far} illustrates that these priors can improve the precision of C and $\alpha$ (N, Fe, and $\alpha$) by up to a factor of 2 (1.5) for JWST (GMACS) observations of faint stars.

In addition to Fe and $\alpha$, NIRSpec and GMACS are capable of recovering a handful of other individual abundances at a distance of $\sim$3 Mpc---N, C, and Mn for NIRSpec and C, Ni, Cr, Co, N, and V for GMACS. These elements can all be measured to better than 0.2 dex at 2 Mpc and 0.1 dex at 1 Mpc. Other elements not shown that can also be recovered to 0.3 dex out to 1 Mpc include Mn, Nd, Sc, Ce, La, Zr, Y, Pr, Sm, Ba, Na, K, Al, Sr, Eu, Cu, Gd, Zn and Dy for GMACS and Ni, Al, and Cr for JWST. This would not only enable precise chemical abundance measurements of stars in M31 and its satellites, but also enable detailed chemical enrichment studies of galaxies at the periphery of the Local Group and beyond, including potential new faint galaxy discoveries by LSST.

Though we didn't explicitly compute the CRLBs as a function of distance for TMT/WFOS and E-ELT/MOSAIC, we expect that each of these powerful facilities have similar abundance recovery potential for stars outside the Local Group.

\subsection{Planning Observations} \label{sec:planning}
For stellar abundance work, selecting the appropriate spectrograph, setup, and exposure time for a specific science case can be daunting given the large number of facilities and instrumental configurations. This can often lead to inefficiencies in observational strategies.

As illustrated in \S\S \ref{sec:existing} and \ref{sec:future}, The CRLB provides a useful and quantitative way to evaluate abundance recovery for a given spectroscopic set up. As an example, consider the comparison of Keck spectrographs and gratings in Figure \ref{fig:crlb_keck}, which displays the numerous trade-offs of each setup on an element-by-element basis. LRIS generally provides the most chemically informative spectra, but if high-multiplexing is a priority, the 1200B grating on DEIMOS is likely the better choice. However if a specific element is of interest (e.g., Ca), one of the lower resolution DEIMOS grating might be more valuable than the 1200B grating.

Given the simplicity in its computation, we suggest that CRLB should be standardized as part of observational planning for resolved star spectroscopic abundance measurements as a logical extension of the standard ETC usage. An ETC determines the S/N of a spectrum based on the integration time and observing conditions, and the CRLB in turn relates that S/N into an expected abundance precision. Figure \ref{fig:crlb_d1200g} provides a clear example of how calculating CRLBs for an instrument can inform an observing strategy. If the intended science goals necessitate simply measuring Fe and an alpha-element out beyond 100 kpc, an hour long exposure with the D1200G grating will likely suffice, allowing for a handful of fields to be observed in a night. However, if the science requires measuring specifically the alpha-element, magnesium, an integration time of 3 or more hours is necessary per field and a different observing strategy is required.

\subsection{Caveats and Assumptions} \label{sec:caveats}
\edit1{\deleted{
In this section we discuss the assumptions inherent to the calculation of CRLBs, namely that: 1) the model spectra perfectly reproduce real stellar spectra, 2) the likelihood and noise properties are Gaussian.
}\added{
In this section we discuss in more detail the assumptions adopted in our calculation of CRLBs, namely that: 1) the model spectra perfectly reproduce real stellar spectra, 2) the likelihood and noise properties are Gaussian, and 3) that adjacent pixels are uncorrelated. We save a more technical discussion of the CRLB for a biased estimator for Appendix \ref{app:biased}.
}}

\subsubsection{Model Fidelity} \label{sec:model_fidelity}
Model fidelity is a fundamental assumption inherent in all problems of parameter estimation. The CRLB of stellar spectra is no exception to this as the gradient spectra used in the above calculations are strongly dependent on the physical assumptions and spectral linelists that underpin any spectral synthesis model. It is important to keep in mind that the CRLB makes no claims about the accuracy of stellar label measurements, merely the possible precision. Nevertheless, incomplete or incorrect linelists will leave out or misplace spectral information, while models that assume 1-D atmospheres in local thermodynamic equilibrium (LTE) may incorrectly predict the spectral response to varying stellar labels for non-LTE lines. It is thus important to strive for consistency and consider the CRLBs calculated using the models relevant to the spectral fitting that will be conducted. While comparing CRLBs of different models is a valuable exercise to evaluate systematics in the predicted CRLBs, this should not be done to pass judgment on model quality.

A common practice in full-spectrum fitting is the masking of spectral regions that are known to be poorly fit by the spectral model to avoid introducing potential systematics into the analysis. Often the poor fit is due to non-LTE effects, but may also be the result of 3-D effects, poorly calibrated oscillator strengths, or an incomplete (or incorrect) linelist (see \citealt{nissen:2018} and references therein). When these regions are masked, so too is the information that it holds. In such a case the appropriate CRLB should be calculated with gradient spectra masked in the same regions \edit1{\added{(as we do in \S\ref{sec:comparison})}}, resulting in a higher uncertainty for the stellar labels. We note, however, that because information adds in quadrature, masking 90\% of the lines only worsens the CRLB by a factor of $\sim$3. For a more thorough analysis of the CRLBs dependence on masked regions see \citet{ting:2017}.

Another underlying challenge for our CRLBs is the assumption that the continuum can be perfectly determined. In the red-optical and near-infrared region of the spectrum, lines are sufficiently sparse that even at $R\sim2000$ identifying the continuum and dividing it out is routine. Unfortunately, the many absorption features in the blue-optical and UV, make it challenging to define a stellar continuum. Instead, a pseudo-continuum is defined using a polynomial function (or some smoothing kernel) and divided out, potentially introducing systematics \edit1{\added{or additional uncertainty in the normalized flux}} that will worsen the precision. By similarly normalizing the model spectra (instead of using the true continuum), any systematics introduced through imperfect normalization can be minimized.

Knowledge of the instrumental LSF is necessary to fit observed spectra with model spectra at the same resolving power. In this work, we have assumed a constant LSF. However, in practice, the LSF is not always known to great precision and can vary from object to object depending on where in the field of view the star lies. Use of the wrong LSF is thus another means by which systematics may be introduced into the fitting of stellar labels. \citet{ting:2017} showed that at least at moderate resolution ($R\sim6000$) and high S/N ($>$200), mismatched LSFs only bias stellar label recovery for differences in broadening greater than 10 km/s and is unlikely to effect the measurement precision. Spectral fitting at lower resolving powers should be even less sensitive to mismatches in LSF.

In addition, when using rest-frame synthetic spectra, it is necessary to properly determine and correct for the radial velocity of stars.
As with the continuum normalization and LSF, we have not quantified the uncertainty in stellar labels that is introduced when the radial velocity is fit simultaneously with other stellar labels. We expect any changes in the CRLBs to be small given that radial velocity is unlikely to correlate with other stellar labels. We will pursue this analysis in a future study.

Even with perfect spectral models, continuum normalization, and instrument characterization, fully extracting the chemical information content of a spectrum requires fitting the full wavelength range (as opposed to measuring EWs) for all stellar labels simultaneously. This is particularly important at low- and moderate-resolution to account for the degeneracies between labels introduced by blended spectral features. In practice, this can be computationally challenging owing to the high dimensionality of stellar label space and the large runtimes needed to generate even 1D LTE stellar atmospheres.

Despite these challenges, the future of extragalactic stellar spectroscopy looks bright as steady progress is being made in all of the aforementioned areas.
Attempts to incorporated non-LTE and 3D effects into stellar atmosphere and radiative transfer models have been undertaken by a number of groups \citep[e.g.,][]{caffau:2011, bergemann:2012, amarsi:2016}.
Several groups have committed to further refining linelists through the identification of unknown (or misplaced) lines in stellar spectra \citep[e.g.,][]{shetrone:2015, andreasen:2016} and the improved calibration of transition oscillator strengths \citep[e.g.,][]{pickering:2001, aldenius:2007, pehlivan-rhodin:2017, laverick:2018}.
Lastly, full spectrum fitting techniques have made major strides with spectral ``emulators" trained through data-driven (e.g., the Cannon; \citealt{ness:2015}), \textit{ab initio} (e.g., the Payne; \citealt{ting:2019}), or combined (e.g., the DD-Payne; \citealt{xiang:2019}) methods, which bypass the computationally expensive stellar atmosphere and radiative transfer calculations.

The above challenges to achieving the precision predicted by the CRLBs should not dissuade the use of CRLBs. Instead, the precision forecasted by the CRLBs provide strong motivation for the continued efforts towards understanding stars, their atmospheres, and their spectra.

\subsubsection{Assumptions of Gaussian Posteriors}
Implicit in the derivations of Equations \ref{eq:loglike}, \ref{eq:FIM_like}, and \ref{eq:FIM_grad_long} was that of Gaussian likelihoods and uncertainties. When these conditions are not met, the CRLB will inaccurately predict measurement errors and the degeneracies between stellar labels. In such situations, a more accurate estimate of the achievable precision can be found using Bayesian sampling techniques.
A comparison of the CRLB and the precision predicted by Hamiltonian Monte-Carlo sampling in the low S/N limit is performed in Appendix \ref{app:validation} and we find it robust down to a S/N of 5 in the case of D1200G (assuming a constant S/N with wavelength).

\subsubsection{Pixel-to-pixel Correlation}\label{sec:oversampling}
\edit1{\added{
Throughout this study we simplify our analysis by setting the correlation between adjacent pixels to zero when calculating the CRLBs\footnote{A similar simplification is employed nearly ubiquitously in the measurement of chemical abundances from stellar spectroscopy.}. In practice, however, most spectrographs are designed to over-sample their spectra such that the number of pixels per resolution element is larger than the Nyquist sampling ($\sim$2 pixels/FWHM)\footnote{For most instrumental LSFs the Nyquist sampling is somewhat larger than 2 pixels/FWHM (see \citealt{robertson:2017}).}. As a result, adjacent pixels will show some correlation and not be truly independent as we have assumed.
}}

\edit1{\added{
While this is unlikely to make a large difference for most spectrographs, which only slightly over-sample their spectra (3-4 pixels/FWHM), the pixel-to-pixel correlation of spectrographs that more highly over-sample (e.g., Hectospec, Hectochelle, FLAMES-UVES, FOBOS, and some DEIMOS and LRIS gratings) may be non-negligible in practice. If instead we believe that only 2 pixels per resolution element are informative then the CRLBs should be a factor of $\sqrt{2}$ ($\sqrt{3}$) larger than presented for spectrographs with a sampling of 4 (6) pixels/FWHM since the CRLBs scale as $n^{-1/2}$. More realistically, additional sampling beyond the Nyquist limit will yield pixels that are still informative, just less so than wholly independent pixels. Thus, we expect the increase in the CRLB to be considerably less than a factor of $\sqrt{2}$ ($\sqrt{3}$) when the correlation of adjacent pixels are taken into account.
In Appendix \ref{app:sampling}, we present an illustrative example of the impact of wavelength sampling and pixel-to-pixel correlation on the CRLBs.
}}

\section{Chem-I-Calc} \label{sec:chemicalc}
Forecasting stellar label recovery for spectroscopic observations is crucial to planning realistic observational campaigns and for validating the reported precision of spectral fitting analyses. However, there are far more combinations of instruments, observational conditions, and stellar targets than can be presented in a single paper. To make the calculation of stellar CRLBs convenient to the astronomical community, we have developed the open-source python package, \texttt{Chem-I-Calc}---the Chemical Information Calculator\footnote{\url{https://github.com/NathanSandford/Chem-I-Calc}}.

The \texttt{Chem-I-Calc} python package provides all the tools necessary to perform all of the computational work presented in this paper, excluding the generation of high-resolution spectra. All of this paper's calculations are included in a Jupyter Notebook on the \texttt{Chem-I-Calc} Github repository along with several other helpful tutorials and instructions for downloading the synthetic spectra described in \S\ref{sec:methods}. The code base is designed to be easy to modify for users that need more flexibility in their CRLB calculations (e.g., for incorporating wavelength-dependent resolution, alternative stellar models, or masking of specific wavelength regions).

While \texttt{Chem-I-Calc} is ready to be used in its current state, it is still under active development. Over time we expect to add additional commonly used spectrographs as presets and include a larger range of stellar types and metallicities as reference stars. We gratefully welcome community feedback and contributions to the python package.

\section{Summary and Conclusions} \label{sec:conclusion}
Current and future generations of powerful, highly-multiplexed spectrographs on large-aperture telescopes make accessible an enormous wealth of chemical information in the spectra of stars outside the MW. Already these instruments have observed the spectra of tens of thousands of individual stars in extragalactic systems, enabling the measurement of their abundance patterns \citep[e.g.,][and references therein]{suda:2017}. With the advent of large-scale extragalactic spectroscopic surveys and ELTs, the number of stars outside the MW with observed spectra will increase by at least an order of magnitude \citep{takada:2014, MSE:2019, bundy:2019}.

The majority of these spectra will be acquired at low- and moderate-resolution ($R<10000$) and feature heavy blending of spectral lines, necessitating that the entire spectrum be fit for all stellar labels simultaneously. Recently, novel full-spectral fitting techniques (e.g., The Cannon; \citealt{ness:2015}, The Payne; \citealt{ting:2019}, and The DD-Payne; \citealt{xiang:2019}) applied to stellar spectra from MW surveys have proven capable of measuring dozens of elemental abundances from low-resolution spectra.

With the field of extragalactic stellar spectroscopy poised for substantial growth, it is imperative that we understand the chemical information content of the spectra we collect and the precision to which it enables the recovery of elemental abundances. To that end, we have employed CRLBs to quantify the information content of extragalactic stellar spectra and forecast chemical abundance precision for 41 existing, future, and proposed spectrograph configurations on 14 telescopes. 
Here we summarize our findings.
\begin{itemize}
    \item The CRLB is an efficient method for computing the expected precision of stellar labels determined via full spectral fitting. We find that the \edit1{\added{precision of}} literature abundances for the commonly used DEIMOS 1200G grating \edit1{\added{and the LAMOST MW survey}} are within a factor of 2 of our CRLBs.
    \item Low- and moderate-resolution spectroscopy at blue-optical wavelengths ($\lambda\lesssim4500$ \AA) are incredibly information rich, enabling the recovery of 2-4 times as many elemental abundances as red-optical spectroscopy ($5000\lesssim\lambda\lesssim10000$ \AA) at similar resolutions. Further, We low-resolution, blue-optical spectroscopy is capable of constraining the abundances of several neutron capture elements (e.g., Sr, Ba, La, Eu).
    \item High-resolution ($R\gtrsim20000$) spectra contain substantial chemical information even at low S/N ($\sim$10 pixel$^{-1}$). Maximizing the precision of abundance recovery from high-resolution spectra benefits from full spectral fitting over equivalent width techniques.
    \item Even small ($\sim$100-500 \AA) windows of low S/N, high-resolution spectra can constrain [Fe/H] and a handful of other elements to better than 0.3 dex.
    \item JWST/NIRSpec and ELTs can recover 10-30 elements for red giant stars throughout the Local Group and [Fe/H] and [$\alpha$/Fe] for resolved stars in galaxies out to several Mpc with 6 hours ($\sim$1 night) of integration time.
    \item Our analysis strictly concerns the precision, not accuracy, of chemical abundance measurements. In practice, imperfect stellar models, linelists, and data reduction can introduce systematics that can bias abundance measurements and hinder attainment of near-CRLB precision. Further investment in the development of stellar models and spectral analysis are necessary to maximally use the chemical information content of the spectra collected.
    \item CRLBs, like ETCs should be used when planning stellar spectroscopic observations or developing spectroscopic instrumentation. To facilitate the calculation of CRLBs, we present Chem-I-Calc, an open-source python package for calculating CRLBs of arbitrary spectrograph configurations.
\end{itemize}
%
%
%
%

\acknowledgments
{We thank the anonymous referee for constructive comments that helped improve the paper.
We thank Bob Kurucz for developing and maintaining programs and databases without which this work would not be possible, as well as all those who have contributed to the design and ongoing utility of the various ETCs used throughout this work. Specifically, we thank Brad Holden, Luke Schmidt, Nicolas Flagey, and Kyle Westfall for providing additional information about the Keck, GMACS, MSE, and FOBOS/WFOS ETCs respectively, as well as Maosheng Xiang for providing example S/N curves of LAMOST spectra.
We would also like to thank Hans-Walter Rix, Kim Venn, Alex Ji, Douglas Finkbeiner, Julianne Delcanton, Josh Speagle, and Gregory Green for insightful discussions.

DRW acknowledges support from an Alfred P. Sloan Foundation Fellowship, an Alexander von Humboldt Fellowship, and a Hellman Faculty Fellowship.  DRW and NRS are grateful to the Max Planck Institute for Astronomie for their hospitality during the writing of this paper. Support for this work was provided by NASA through grants HST-GO-15901, HST-GO-15902, and JWST-DD-ERS-1334 from the Space Telescope Science Institute, which is operated by AURA, Inc., under NASA contract NAS5-26555.
Y.S.T. is grateful to be supported by the NASA Hubble Fellowship grant HST-HF2-51425.001 awarded by the Space Telescope Science Institute.
The computations in this paper were partially run on the Savio computational cluster resource provided by the Berkeley Research Computing Program at the University of California, Berkeley.}


\software{iPython \citep{perez:2007},
          Matplotlib \citep{hunter:2007},
          pandas \citep{mckinney:2010},
          NumPy \citep{walt:2011},
          Astropy \citep{astropy-collaboration:2013, astropy-collaboration:2018},
          PyMC3 \citep{salvatier:2016},
          PyTorch \citep{paszke:2019},
          SciPy \citep{virtanen:2019}
          }

\clearpage  
\bibliography{crlb}{}

\begin{thebibliography}{}
\expandafter\ifx\csname natexlab\endcsname\relax\def\natexlab#1{#1}\fi
\providecommand{\url}[1]{\href{#1}{#1}}
\providecommand{\dodoi}[1]{doi:~\href{http://doi.org/#1}{\nolinkurl{#1}}}
\providecommand{\doeprint}[1]{\href{http://ascl.net/#1}{\nolinkurl{http://ascl.net/#1}}}
\providecommand{\doarXiv}[1]{\href{https://arxiv.org/abs/#1}{\nolinkurl{https://arxiv.org/abs/#1}}}

\bibitem[{Adshead \& Easther(2008)}]{adshead:2008}
Adshead, P., \& Easther, R. 2008, Journal of Cosmology and Astroparticle
  Physics, 10, 047, \dodoi{10.1088/1475-7516/2008/10/047}

\bibitem[{Albrecht {et~al.}(2006)Albrecht, Bernstein, Cahn, Freedman, Hewitt,
  Hu, Huth, Kamionkowski, Kolb, Knox, Mather, Staggs, \&
  Suntzeff}]{albrecht:2006}
Albrecht, A., Bernstein, G., Cahn, R., {et~al.} 2006, arXiv:astro-ph/0609591.
\newblock \url{http://arxiv.org/abs/astro-ph/0609591}

\bibitem[{Aldenius {et~al.}(2007)Aldenius, Tanner, Johansson, Lundberg, \&
  Ryan}]{aldenius:2007}
Aldenius, M., Tanner, J.~D., Johansson, S., Lundberg, H., \& Ryan, S.~G. 2007,
  Astronomy and Astrophysics, 461, 767, \dodoi{10.1051/0004-6361:20066266}

\bibitem[{Alfaro-Cuello {et~al.}(2019)Alfaro-Cuello, Kacharov, Neumayer,
  Luetzgendorf, Seth, Boeker, Kamann, Leaman, van~de Ven, Bianchini, Watkins,
  \& Lyubenova}]{alfaro-cuello:2019}
Alfaro-Cuello, M., Kacharov, N., Neumayer, N., {et~al.} 2019, arXiv e-prints,
  arXiv:1909.10529.
\newblock \url{https://ui.adsabs.harvard.edu/2019arXiv190910529A/abstract}

\bibitem[{Allende~Prieto(2016)}]{allende-prieto:2016}
Allende~Prieto, C. 2016, Living Reviews in Solar Physics, 13, 1,
  \dodoi{10.1007/s41116-016-0001-6}

\bibitem[{Aller(1942)}]{aller:1942}
Aller, L.~H. 1942, The Astrophysical Journal, 96, 321, \dodoi{10.1086/144468}

\bibitem[{Aller(1946)}]{aller:1946}
---. 1946, The Astrophysical Journal, 104, 347, \dodoi{10.1086/144864}

\bibitem[{Alves-Brito {et~al.}(2006)Alves-Brito, Barbuy, Zoccali, Minniti,
  Ortolani, Hill, Renzini, Pasquini, Bica, Rich, Mel{\'e}ndez, \&
  Momany}]{alves-brito:2006}
Alves-Brito, A., Barbuy, B., Zoccali, M., {et~al.} 2006, Astronomy and
  Astrophysics, 460, 269, \dodoi{10.1051/0004-6361:20065488}

\bibitem[{Amarsi {et~al.}(2016)Amarsi, Lind, Asplund, Barklem, \&
  Collet}]{amarsi:2016}
Amarsi, A.~M., Lind, K., Asplund, M., Barklem, P.~S., \& Collet, R. 2016,
  Monthly Notices of the Royal Astronomical Society, 463, 1518,
  \dodoi{10.1093/mnras/stw2077}

\bibitem[{Andreasen {et~al.}(2016)Andreasen, Sousa, Delgado~Mena, Santos,
  Tsantaki, Rojas-Ayala, \& Neves}]{andreasen:2016}
Andreasen, D.~T., Sousa, S.~G., Delgado~Mena, E., {et~al.} 2016, Astronomy and
  Astrophysics, 585, A143, \dodoi{10.1051/0004-6361/201527308}

\bibitem[{Aoki {et~al.}(2009)Aoki, Arimoto, Sadakane, Tolstoy, Battaglia,
  Jablonka, Shetrone, Letarte, Irwin, Hill, Francois, Venn, Primas, Helmi,
  Kaufer, Tafelmeyer, Szeifert, \& Babusiaux}]{aoki:2009}
Aoki, W., Arimoto, N., Sadakane, K., {et~al.} 2009, Astronomy and Astrophysics,
  502, 569, \dodoi{10.1051/0004-6361/200911959}

\bibitem[{Asplund {et~al.}(2009)Asplund, Grevesse, Sauval, \&
  Scott}]{asplund:2009}
Asplund, M., Grevesse, N., Sauval, A.~J., \& Scott, P. 2009, Annual Review of
  Astronomy and Astrophysics, 47, 481,
  \dodoi{10.1146/annurev.astro.46.060407.145222}

\bibitem[{{Astropy Collaboration} {et~al.}(2013){Astropy Collaboration},
  Robitaille, Tollerud, Greenfield, Droettboom, Bray, Aldcroft, Davis,
  Ginsburg, Price-Whelan, Kerzendorf, Conley, Crighton, Barbary, Muna,
  Ferguson, Grollier, Parikh, Nair, Unther, Deil, Woillez, Conseil, Kramer,
  Turner, Singer, Fox, Weaver, Zabalza, Edwards, Azalee~Bostroem, Burke, Casey,
  Crawford, Dencheva, Ely, Jenness, Labrie, Lim, Pierfederici, Pontzen, Ptak,
  Refsdal, Servillat, \& Streicher}]{astropy-collaboration:2013}
{Astropy Collaboration}, Robitaille, T.~P., Tollerud, E.~J., {et~al.} 2013,
  Astronomy and Astrophysics, 558, A33, \dodoi{10.1051/0004-6361/201322068}

\bibitem[{{Astropy Collaboration} {et~al.}(2018){Astropy Collaboration},
  Price-Whelan, Sip{\H o}cz, G{\"u}nther, Lim, Crawford, Conseil, Shupe, Craig,
  Dencheva, Ginsburg, VanderPlas, Bradley, P{\'e}rez-Su{\'a}rez, de~Val-Borro,
  Aldcroft, Cruz, Robitaille, Tollerud, Ardelean, Babej, Bach, Bachetti,
  Bakanov, Bamford, Barentsen, Barmby, Baumbach, Berry, Biscani, Boquien,
  Bostroem, Bouma, Brammer, Bray, Breytenbach, Buddelmeijer, Burke, Calderone,
  Cano~Rodr{\'\i}guez, Cara, Cardoso, Cheedella, Copin, Corrales, Crichton,
  D'Avella, Deil, Depagne, Dietrich, Donath, Droettboom, Earl, Erben, Fabbro,
  Ferreira, Finethy, Fox, Garrison, Gibbons, Goldstein, Gommers, Greco,
  Greenfield, Groener, Grollier, Hagen, Hirst, Homeier, Horton, Hosseinzadeh,
  Hu, Hunkeler, Ivezi{\'c}, Jain, Jenness, Kanarek, Kendrew, Kern, Kerzendorf,
  Khvalko, King, Kirkby, Kulkarni, Kumar, Lee, Lenz, Littlefair, Ma, Macleod,
  Mastropietro, McCully, Montagnac, Morris, Mueller, Mumford, Muna, Murphy,
  Nelson, Nguyen, Ninan, N{\"o}the, Ogaz, Oh, Parejko, Parley, Pascual, Patil,
  Patil, Plunkett, Prochaska, Rastogi, Reddy~Janga, Sabater, Sakurikar,
  Seifert, Sherbert, Sherwood-Taylor, Shih, Sick, Silbiger, Singanamalla,
  Singer, Sladen, Sooley, Sornarajah, Streicher, Teuben, Thomas, Tremblay,
  Turner, Terr{\'o}n, van Kerkwijk, de~la Vega, Watkins, Weaver, Whitmore,
  Woillez, Zabalza, \& {Astropy Contributors}}]{astropy-collaboration:2018}
{Astropy Collaboration}, Price-Whelan, A.~M., Sip{\H o}cz, B.~M., {et~al.}
  2018, The Astronomical Journal, 156, 123, \dodoi{10.3847/1538-3881/aabc4f}

\bibitem[{Bacon {et~al.}(2010)Bacon, Accardo, Adjali, Anwand, Bauer, Biswas,
  Blaizot, Boudon, Brau-Nogue, Brinchmann, Caillier, Capoani, Carollo, Contini,
  Couderc, Daguis{\'e}, Deiries, Delabre, Dreizler, Dubois, Dupieux, Dupuy,
  Emsellem, Fechner, Fleischmann, Fran{\c c}ois, Gallou, Gharsa, Glindemann,
  Gojak, Guiderdoni, Hansali, Hahn, Jarno, Kelz, Koehler, Kosmalski, Laurent,
  Floch, Lilly, Lizon, Loupias, Manescau, Monstein, Nicklas, Olaya, Pares,
  Pasquini, P{\'e}contal-Rousset, Pell{\'o}, Petit, Popow, Reiss, Remillieux,
  Renault, Roth, Rupprecht, Serre, Schaye, Soucail, Steinmetz, Streicher,
  Stuik, H, Vernet, Weilbacher, Wisotzki, \& Yerle}]{bacon:2010}
Bacon, R., Accardo, M., Adjali, L., {et~al.} 2010, in Ground-based and
  {Airborne} {Instrumentation} for {Astronomy} {III}, Vol. 7735 (International
  Society for Optics and Photonics), 773508, \dodoi{10.1117/12.856027}

\bibitem[{Bagnasco {et~al.}(2007)Bagnasco, Kolm, Ferruit, Honnen, Koehler,
  Lemke, Maschmann, Melf, Noyer, Rumler, Salvignol, Strada, \&
  Te~Plate}]{bagnasco:2007}
Bagnasco, G., Kolm, M., Ferruit, P., {et~al.} 2007, Cryogenic Optical Systems
  and Instruments XII, 6692, 66920M, \dodoi{10.1117/12.735602}

\bibitem[{Bailer-Jones(2000)}]{bailer-jones:2000}
Bailer-Jones, C. a.~L. 2000, Astronomy and Astrophysics, 357, 197.
\newblock
  \url{https://ui.adsabs.harvard.edu/abs/2000A%26A...357..197B/abstract}

\bibitem[{Baschek(1959)}]{baschek:1959}
Baschek, B. 1959, Zeitschrift fur Astrophysik, 48, 95.
\newblock \url{http://adsabs.harvard.edu/abs/1959ZA.....48...95B}

\bibitem[{Bastian \& Lardo(2018)}]{bastian:2018}
Bastian, N., \& Lardo, C. 2018, Annual Review of Astronomy and Astrophysics,
  56, 83, \dodoi{10.1146/annurev-astro-081817-051839}

\bibitem[{Battaglia {et~al.}(2008)Battaglia, Helmi, Tolstoy, Irwin, Hill, \&
  Jablonka}]{battaglia:2008}
Battaglia, G., Helmi, A., Tolstoy, E., {et~al.} 2008, The Astrophysical Journal
  Letters, 681, L13, \dodoi{10.1086/590179}

\bibitem[{Battaglia {et~al.}(2011)Battaglia, Tolstoy, Helmi, Irwin, Parisi,
  Hill, \& Jablonka}]{battaglia:2011}
Battaglia, G., Tolstoy, E., Helmi, A., {et~al.} 2011, Monthly Notices of the
  Royal Astronomical Society, 411, 1013,
  \dodoi{10.1111/j.1365-2966.2010.17745.x}

\bibitem[{Battaglia {et~al.}(2006)Battaglia, Tolstoy, Helmi, Irwin, Letarte,
  Jablonka, Hill, Venn, Shetrone, Arimoto, Primas, Kaufer, Francois, Szeifert,
  Abel, \& Sadakane}]{battaglia:2006}
---. 2006, Astronomy and Astrophysics, 459, 423,
  \dodoi{10.1051/0004-6361:20065720}

\bibitem[{Becker {et~al.}(2012)Becker, Huterer, \& Kadota}]{becker:2012}
Becker, A., Huterer, D., \& Kadota, K. 2012, Journal of Cosmology and
  Astroparticle Physics, 12, 034, \dodoi{10.1088/1475-7516/2012/12/034}

\bibitem[{Bedell {et~al.}(2014)Bedell, Mel{\'e}ndez, Bean, Ram{\'\i}rez, Leite,
  \& Asplund}]{bedell:2014}
Bedell, M., Mel{\'e}ndez, J., Bean, J.~L., {et~al.} 2014, The Astrophysical
  Journal, 795, 23, \dodoi{10.1088/0004-637X/795/1/23}

\bibitem[{Bell(1970)}]{bell:1970}
Bell, R.~A. 1970, Monthly Notices of the Royal Astronomical Society, 148, 25,
  \dodoi{10.1093/mnras/148.1.25}

\bibitem[{Bell \& Branch(1976)}]{bell:1976}
Bell, R.~A., \& Branch, D. 1976, Monthly Notices of the Royal Astronomical
  Society, 175, 25, \dodoi{10.1093/mnras/175.1.25}

\bibitem[{Bergemann {et~al.}(2012)Bergemann, Lind, Collet, Magic, \&
  Asplund}]{bergemann:2012}
Bergemann, M., Lind, K., Collet, R., Magic, Z., \& Asplund, M. 2012, Monthly
  Notices of the Royal Astronomical Society, 427, 27,
  \dodoi{10.1111/j.1365-2966.2012.21687.x}

\bibitem[{Bernstein {et~al.}(2003)Bernstein, Shectman, Gunnels, Mochnacki, \&
  Athey}]{bernstein:2003}
Bernstein, R., Shectman, S.~A., Gunnels, S.~M., Mochnacki, S., \& Athey, A.~E.
  2003, in Instrument {Design} and {Performance} for {Optical}/{Infrared}
  {Ground}-based {Telescopes}, Vol. 4841 (International Society for Optics and
  Photonics), 1694--1704, \dodoi{10.1117/12.461502}

\bibitem[{Betoule {et~al.}(2014)Betoule, Kessler, Guy, Mosher, Hardin, Biswas,
  Astier, El-Hage, Konig, Kuhlmann, Marriner, Pain, Regnault, Balland, Bassett,
  Brown, Campbell, Carlberg, Cellier-Holzem, Cinabro, Conley, D'Andrea, DePoy,
  Doi, Ellis, Fabbro, Filippenko, Foley, Frieman, Fouchez, Galbany, Goobar,
  Gupta, Hill, Hlozek, Hogan, Hook, Howell, Jha, Le~Guillou, Leloudas, Lidman,
  Marshall, M{\"o}ller, Mour{\~a}o, Neveu, Nichol, Olmstead,
  Palanque-Delabrouille, Perlmutter, Prieto, Pritchet, Richmond, Riess,
  Ruhlmann-Kleider, Sako, Schahmaneche, Schneider, Smith, Sollerman, Sullivan,
  Walton, \& Wheeler}]{betoule:2014}
Betoule, M., Kessler, R., Guy, J., {et~al.} 2014, Astronomy and Astrophysics,
  568, A22, \dodoi{10.1051/0004-6361/201423413}

\bibitem[{Blanco-Cuaresma(2019)}]{blanco-cuaresma:2019}
Blanco-Cuaresma, S. 2019, Monthly Notices of the Royal Astronomical Society,
  486, 2075, \dodoi{10.1093/mnras/stz549}

\bibitem[{Boesgaard {et~al.}(2005)Boesgaard, King, Cody, Stephens, \&
  Deliyannis}]{boesgaard:2005}
Boesgaard, A.~M., King, J.~R., Cody, A.~M., Stephens, A., \& Deliyannis, C.~P.
  2005, The Astrophysical Journal, 629, 832, \dodoi{10.1086/431645}

\bibitem[{Boesgaard {et~al.}(2000)Boesgaard, Stephens, King, \&
  Deliyannis}]{boesgaard:2000}
Boesgaard, A.~M., Stephens, A., King, J.~R., \& Deliyannis, C.~P. 2000, in
  Discoveries and {Research} {Prospects} from 8- to 10-{Meter}-{Class}
  {Telescopes}, Vol. 4005 (International Society for Optics and Photonics),
  274--284, \dodoi{10.1117/12.390137}

\bibitem[{Bundy {et~al.}(2019)Bundy, Westfall, MacDonald, Kupke, Savage,
  Poppett, Alabi, Becker, Burchett, Capak, Coil, Cooper, Cowley, Deich, Dillon,
  Edelstein, Guhathakurta, Hennawi, Kassis, Lee, Masters, Miller, Newman,
  O'Meara, Prochaska, Rau, Rhodes, Rich, Rockosi, Romanowsky, Schafer,
  Schlegel, Shapley, Siana, Ting, Weisz, White, Williams, Wilson, Wilson, \&
  Yan}]{bundy:2019}
Bundy, K., Westfall, K., MacDonald, N., {et~al.} 2019, Bulletin of the American
  Astronomical Society, 51, 198.
\newblock \url{https://ui.adsabs.harvard.edu/abs/2019BAAS...51g.198B/abstract}

\bibitem[{Caffau {et~al.}(2011)Caffau, Ludwig, Steffen, Freytag, \&
  Bonifacio}]{caffau:2011}
Caffau, E., Ludwig, H.-G., Steffen, M., Freytag, B., \& Bonifacio, P. 2011,
  Solar Physics, 268, 255, \dodoi{10.1007/s11207-010-9541-4}

\bibitem[{Caffau {et~al.}(2013)Caffau, Koch, Sbordone, Sartoretti, Hansen,
  Royer, Leclerc, Bonifacio, Christlieb, Ludwig, Grebel, de~Jong, Chiappini,
  Walcher, Mignot, Feltzing, Cohen, Minchev, Helmi, Piffl, Depagne, \&
  Schnurr}]{caffau:2013}
Caffau, E., Koch, A., Sbordone, L., {et~al.} 2013, Astronomische Nachrichten,
  334, 197, \dodoi{10.1002/asna.201211814}

\bibitem[{Carbon {et~al.}(1982)Carbon, Langer, Butler, Kraft, Suntzeff, Kemper,
  Trefzger, \& Romanishin}]{carbon:1982}
Carbon, D.~F., Langer, G.~E., Butler, D., {et~al.} 1982, The Astrophysical
  Journal Supplement Series, 49, 207, \dodoi{10.1086/190796}

\bibitem[{Carlin {et~al.}(2009)Carlin, Grillmair, Mu{\~n}oz, Nidever, \&
  Majewski}]{carlin:2009}
Carlin, J.~L., Grillmair, C.~J., Mu{\~n}oz, R.~R., Nidever, D.~L., \& Majewski,
  S.~R. 2009, The Astrophysical Journal, 702, L9,
  \dodoi{10.1088/0004-637X/702/1/L9}

\bibitem[{Carrera {et~al.}(2013)Carrera, Pancino, Gallart, \& del
  Pino}]{carrera:2013}
Carrera, R., Pancino, E., Gallart, C., \& del Pino, A. 2013, Monthly Notices of
  the Royal Astronomical Society, 434, 1681, \dodoi{10.1093/mnras/stt1126}

\bibitem[{Casagrande {et~al.}(2011)Casagrande, Sch{\"o}nrich, Asplund, Cassisi,
  Ram{\'\i}rez, Mel{\'e}ndez, Bensby, \& Feltzing}]{casagrande:2011}
Casagrande, L., Sch{\"o}nrich, R., Asplund, M., {et~al.} 2011, Astronomy and
  Astrophysics, 530, A138, \dodoi{10.1051/0004-6361/201016276}

\bibitem[{Cenarro {et~al.}(2001{\natexlab{a}})Cenarro, Cardiel, Gorgas,
  Peletier, Vazdekis, \& Prada}]{cenarro:2001a}
Cenarro, A.~J., Cardiel, N., Gorgas, J., {et~al.} 2001{\natexlab{a}}, Monthly
  Notices of the Royal Astronomical Society, 326, 959,
  \dodoi{10.1046/j.1365-8711.2001.04688.x}

\bibitem[{Cenarro {et~al.}(2001{\natexlab{b}})Cenarro, Gorgas, Cardiel, Pedraz,
  Peletier, \& Vazdekis}]{cenarro:2001b}
Cenarro, A.~J., Gorgas, J., Cardiel, N., {et~al.} 2001{\natexlab{b}}, Monthly
  Notices of the Royal Astronomical Society, 326, 981,
  \dodoi{10.1046/j.1365-8711.2001.04689.x}

\bibitem[{Cenarro {et~al.}(2002)Cenarro, Gorgas, Cardiel, Vazdekis, \&
  Peletier}]{cenarro:2002}
Cenarro, A.~J., Gorgas, J., Cardiel, N., Vazdekis, A., \& Peletier, R.~F. 2002,
  Monthly Notices of the Royal Astronomical Society, 329, 863,
  \dodoi{10.1046/j.1365-8711.2002.05029.x}

\bibitem[{Chapman {et~al.}(2005)Chapman, Ibata, Lewis, Ferguson, Irwin,
  McConnachie, \& Tanvir}]{chapman:2005}
Chapman, S.~C., Ibata, R., Lewis, G.~F., {et~al.} 2005, The Astrophysical
  Journal, 632, L87, \dodoi{10.1086/497686}

\bibitem[{Choi {et~al.}(2016)Choi, Dotter, Conroy, Cantiello, Paxton, \&
  Johnson}]{choi:2016}
Choi, J., Dotter, A., Conroy, C., {et~al.} 2016, The Astrophysical Journal,
  823, 102, \dodoi{10.3847/0004-637X/823/2/102}

\bibitem[{Cohen \& Huang(2009)}]{cohen:2009}
Cohen, J.~G., \& Huang, W. 2009, The Astrophysical Journal, 701, 1053,
  \dodoi{10.1088/0004-637X/701/2/1053}

\bibitem[{Collins {et~al.}(2013)Collins, Chapman, Rich, Ibata, Martin, Irwin,
  Bate, Lewis, Pe{\~n}arrubia, Arimoto, Casey, Ferguson, Koch, McConnachie, \&
  Tanvir}]{collins:2013}
Collins, M. L.~M., Chapman, S.~C., Rich, R.~M., {et~al.} 2013, The
  Astrophysical Journal, 768, 172, \dodoi{10.1088/0004-637X/768/2/172}

\bibitem[{Conroy {et~al.}(2019{\natexlab{a}})Conroy, Naidu, Zaritsky, Bonaca,
  Cargile, Johnson, \& Caldwell}]{conroy:2019b}
Conroy, C., Naidu, R.~P., Zaritsky, D., {et~al.} 2019{\natexlab{a}}, arXiv
  e-prints, arXiv:1909.02007.
\newblock \url{https://ui.adsabs.harvard.edu/abs/2019arXiv190902007C/abstract}

\bibitem[{Conroy {et~al.}(2019{\natexlab{b}})Conroy, Bonaca, Cargile, Johnson,
  Caldwell, Naidu, Zaritsky, Fabricant, Moran, Rhee, Szentgyorgyi, Berlind,
  Calkins, Kattner, \& Ly}]{conroy:2019a}
Conroy, C., Bonaca, A., Cargile, P., {et~al.} 2019{\natexlab{b}}, The
  Astrophysical Journal, 883, 107, \dodoi{10.3847/1538-4357/ab38b8}

\bibitem[{Cowan {et~al.}(2002)Cowan, Sneden, Burles, Ivans, Beers, Truran,
  Lawler, Primas, Fuller, Pfeiffer, \& Kratz}]{cowan:2002}
Cowan, J.~J., Sneden, C., Burles, S., {et~al.} 2002, The Astrophysical Journal,
  572, 861, \dodoi{10.1086/340347}

\bibitem[{Cramer(1946)}]{cramer:1946}
Cramer, H. 1946, Mathematical methods of statistics, Princeton mathematical
  series No.~9 (Princeton: Princeton University Press)

\bibitem[{Cui {et~al.}(2012)Cui, Zhao, Chu, Li, Li, Zhang, Su, Yao, Wang, Xing,
  Li, Zhu, Wang, Gu, Luo, Xu, Zhang, Liu, Zhang, Yang, Cao, Chen, Chen, Chen,
  Chen, Chu, Feng, Gong, Hou, Hu, Hu, Hu, Jia, Jiang, Jiang, Jiang, Jin, Li,
  Li, Li, Liu, Liu, Lu, Mao, Men, Qi, Qi, Shi, Tang, Tao, Wang, Wang, Wang,
  Wang, Wang, Wang, Wang, Wang, Wang, Wang, Wang, Wang, Xu, Xu, Yang, Yu, Yuan,
  Yuan, Zhai, Zhang, Zhang, Zhang, Zhao, Zhou, Zhou, Zhu, \& Zou}]{cui:2012}
Cui, X.-Q., Zhao, Y.-H., Chu, Y.-Q., {et~al.} 2012, Research in Astronomy and
  Astrophysics, 12, 1197, \dodoi{10.1088/1674-4527/12/9/003}

\bibitem[{Czekala {et~al.}(2015)Czekala, Andrews, Mandel, Hogg, \&
  Green}]{czekala:2015}
Czekala, I., Andrews, S.~M., Mandel, K.~S., Hogg, D.~W., \& Green, G.~M. 2015,
  The Astrophysical Journal, 812, 128, \dodoi{10.1088/0004-637X/812/2/128}

\bibitem[{Darmois(1945)}]{darmois:1945}
Darmois, G. 1945, Revue de l'Institut International de Statistique / Review of
  the International Statistical Institute, 13, 9, \dodoi{10.2307/1400974}

\bibitem[{De~Silva {et~al.}(2015)De~Silva, Freeman, Bland-Hawthorn, Martell,
  de~Boer, Asplund, Keller, Sharma, Zucker, Zwitter, Anguiano, Bacigalupo,
  Bayliss, Beavis, Bergemann, Campbell, Cannon, Carollo, Casagrande, Casey,
  Da~Costa, D'Orazi, Dotter, Duong, Heger, Ireland, Kafle, Kos, Lattanzio,
  Lewis, Lin, Lind, Munari, Nataf, O'Toole, Parker, Reid, Schlesinger, Sheinis,
  Simpson, Stello, Ting, Traven, Watson, Wittenmyer, Yong, \& {\v
  Z}erjal}]{desilva:2015}
De~Silva, G.~M., Freeman, K.~C., Bland-Hawthorn, J., {et~al.} 2015, Monthly
  Notices of the Royal Astronomical Society, 449, 2604,
  \dodoi{10.1093/mnras/stv327}

\bibitem[{Dekker {et~al.}(2000)Dekker, D'Odorico, Kaufer, Delabre, \&
  Kotzlowski}]{dekker:2000}
Dekker, H., D'Odorico, S., Kaufer, A., Delabre, B., \& Kotzlowski, H. 2000, in
  Optical and IR Telescope Instrumentation and Detectors, ed. M.~Iye \&
  A.~F.~M. Moorwood, Vol. 4008, International Society for Optics and Photonics
  (SPIE), 534 -- 545, \dodoi{10.1117/12.395512}

\bibitem[{DePoy {et~al.}(2012)DePoy, Allen, Barkhouser, Boster, Carona,
  Harding, Hammond, Marshall, Orndorff, Papovich, Prochaska, Prochaska,
  Rheault, Smee, Shectman, \& Jr}]{depoy:2012}
DePoy, D.~L., Allen, R., Barkhouser, R., {et~al.} 2012, in Ground-based and
  {Airborne} {Instrumentation} for {Astronomy} {IV}, Vol. 8446 (International
  Society for Optics and Photonics), 84461N, \dodoi{10.1117/12.926186}

\bibitem[{{DESI Collaboration} {et~al.}(2016{\natexlab{a}}){DESI
  Collaboration}, Aghamousa, Aguilar, Ahlen, Alam, Allen, Allende~Prieto,
  Annis, Bailey, Balland, Ballester, Baltay, Beaufore, Bebek, Beers, Bell,
  Bernal, Besuner, Beutler, Blake, Bleuler, Blomqvist, Blum, Bolton, Briceno,
  Brooks, Brownstein, Buckley-Geer, Burden, Burtin, Busca, Cahn, Cai,
  Cardiel-Sas, Carlberg, Carton, Casas, Castander, Cervantes-Cota, Claybaugh,
  Close, Coker, Cole, Comparat, Cooper, Cousinou, Crocce, Cuby, Cunningham,
  Davis, Dawson, de~la Macorra, De~Vicente, Delubac, Derwent, Dey, Dhungana,
  Ding, Doel, Duan, Ealet, Edelstein, Eftekharzadeh, Eisenstein, Elliott,
  Escoffier, Evatt, Fagrelius, Fan, Fanning, Farahi, Farihi, Favole, Feng,
  Fernandez, Findlay, Finkbeiner, Fitzpatrick, Flaugher, Flender, Font-Ribera,
  Forero-Romero, Fosalba, Frenk, Fumagalli, Gaensicke, Gallo, Garcia-Bellido,
  Gaztanaga, Pietro Gentile~Fusillo, Gerard, Gershkovich, Giannantonio, Gillet,
  Gonzalez-de Rivera, Gonzalez-Perez, Gott, Graur, Gutierrez, Guy, Habib,
  Heetderks, Heetderks, Heitmann, Hellwing, Herrera, Ho, Holland, Honscheid,
  Huff, Hutchinson, Huterer, Hwang, Illa~Laguna, Ishikawa, Jacobs, Jeffrey,
  Jelinsky, Jennings, Jiang, Jimenez, Johnson, Joyce, Jullo, Juneau, Kama,
  Karcher, Karkar, Kehoe, Kennamer, Kent, Kilbinger, Kim, Kirkby, Kisner,
  Kitanidis, Kneib, Koposov, Kovacs, Koyama, Kremin, Kron, Kronig,
  Kueter-Young, Lacey, Lafever, Lahav, Lambert, Lampton, Landriau, Lang, Lauer,
  Le~Goff, Le~Guillou, Le~Van~Suu, Lee, Lee, Leitner, Lesser, Levi, L'Huillier,
  Li, Liang, Lin, Linder, Loebman, Luki{\'c}, Ma, MacCrann, Magneville,
  Makarem, Manera, Manser, Marshall, Martini, Massey, Matheson, McCauley,
  McDonald, McGreer, Meisner, Metcalfe, Miller, Miquel, Moustakas, Myers, Naik,
  Newman, Nichol, Nicola, Nicolati~da Costa, Nie, Niz, Norberg, Nord, Norman,
  Nugent, O'Brien, Oh, Olsen, Padilla, Padmanabhan, Padmanabhan,
  Palanque-Delabrouille, Palmese, Pappalardo, P{\^a}ris, Park, Patej, Peacock,
  Peiris, Peng, Percival, Perruchot, Pieri, Pogge, Pollack, Poppett, Prada,
  Prakash, Probst, Rabinowitz, Raichoor, Ree, Refregier, Regal, Reid, Reil,
  Rezaie, Rockosi, Roe, Ronayette, Roodman, Ross, Ross, Rossi, Rozo,
  Ruhlmann-Kleider, Rykoff, Sabiu, Samushia, Sanchez, Sanchez, Schlegel,
  Schneider, Schubnell, Secroun, Seljak, Seo, Serrano, Shafieloo, Shan,
  Sharples, Sholl, Shourt, Silber, Silva, Sirk, Slosar, Smith, Smoot, Som,
  Song, Sprayberry, Staten, Stefanik, Tarle, Sien~Tie, Tinker, Tojeiro, Valdes,
  Valenzuela, Valluri, Vargas-Magana, Verde, Walker, Wang, Wang, Weaver,
  Weaverdyck, Wechsler, Weinberg, White, Yang, Yeche, Zhang, Zhao, Zheng, Zhou,
  Zhou, Zhu, Zou, \& Zu}]{desi:2016_inst}
{DESI Collaboration}, Aghamousa, A., Aguilar, J., {et~al.} 2016{\natexlab{a}},
  arXiv e-prints, arXiv:1611.00037.
\newblock \url{https://ui.adsabs.harvard.edu/abs/2016arXiv161100037D/abstract}

\bibitem[{{DESI Collaboration} {et~al.}(2016{\natexlab{b}}){DESI
  Collaboration}, Aguilar, Ahlen, Alam, Allen, Allende~Prieto, Annis, Bailey,
  Balland, Ballester, Baltay, Beaufore, Bebek, Beers, Bell, Bernal, Besuner,
  Beutler, Blake, Bleuler, Blomqvist, Blum, Bolton, Briceno, Brooks,
  Brownstein, Buckley-Geer, Burden, Burtin, Busca, Cahn, Cai, Cardiel-Sas,
  Carlberg, Carton, Casas, Castander, Cervantes-Cota, Claybaugh, Close, Coker,
  Cole, Comparat, Cooper, Cousinou, Crocce, Cuby, Cunningham, Davis, Dawson,
  de~la Macorra, De~Vicente, Delubac, Derwent, Dey, Dhungana, Ding, Doel, Duan,
  Ealet, Edelstein, Eftekharzadeh, Eisenstein, Elliott, Escoffier, Evatt,
  Fagrelius, Fan, Fanning, Farahi, Farihi, Favole, Feng, Fernandez, Findlay,
  Finkbeiner, Fitzpatrick, Flaugher, Flender, Font-Ribera, Forero-Romero,
  Fosalba, Frenk, Fumagalli, Gaensicke, Gallo, Garcia-Bellido, Gaztanaga,
  Pietro Gentile~Fusillo, Gerard, Gershkovich, Giannantonio, Gillet,
  Gonzalez-de Rivera, Gonzalez-Perez, Gott, Graur, Gutierrez, Guy, Habib,
  Heetderks, Heetderks, Heitmann, Hellwing, Herrera, Ho, Holland, Honscheid,
  Huff, Hutchinson, Huterer, Hwang, Illa~Laguna, Ishikawa, Jacobs, Jeffrey,
  Jelinsky, Jennings, Jiang, Jimenez, Johnson, Joyce, Jullo, Juneau, Kama,
  Karcher, Karkar, Kehoe, Kennamer, Kent, Kilbinger, Kim, Kirkby, Kisner,
  Kitanidis, Kneib, Koposov, Kovacs, Koyama, Kremin, Kron, Kronig,
  Kueter-Young, Lacey, Lafever, Lahav, Lambert, Lampton, Landriau, Lang, Lauer,
  Le~Goff, Le~Guillou, Le~Van~Suu, Lee, Lee, Leitner, Lesser, Levi, L'Huillier,
  Li, Liang, Lin, Linder, Loebman, Luki{\'c}, Ma, MacCrann, Magneville,
  Makarem, Manera, Manser, Marshall, Martini, Massey, Matheson, McCauley,
  McDonald, McGreer, Meisner, Metcalfe, Miller, Miquel, Moustakas, Myers, Naik,
  Newman, Nichol, Nicola, Nicolati~da Costa, Nie, Niz, Norberg, Nord, Norman,
  Nugent, O'Brien, Oh, Olsen, Padilla, Padmanabhan, Padmanabhan,
  Palanque-Delabrouille, Palmese, Pappalardo, P{\^a}ris, Park, Patej, Peacock,
  Peiris, Peng, Percival, Perruchot, Pieri, Pogge, Pollack, Poppett, Prada,
  Prakash, Probst, Rabinowitz, Raichoor, Ree, Refregier, Regal, Reid, Reil,
  Rezaie, Rockosi, Roe, Ronayette, Roodman, Ross, Ross, Rossi, Rozo,
  Ruhlmann-Kleider, Rykoff, Sabiu, Samushia, Sanchez, Sanchez, Schlegel,
  Schneider, Schubnell, Secroun, Seljak, Seo, Serrano, Shafieloo, Shan,
  Sharples, Sholl, Shourt, Silber, Silva, Sirk, Slosar, Smith, Smoot, Som,
  Song, Sprayberry, Staten, Stefanik, Tarle, Sien~Tie, Tinker, Tojeiro, Valdes,
  Valenzuela, Valluri, Vargas-Magana, Verde, Walker, Wang, Wang, Weaver,
  Weaverdyck, Wechsler, Weinberg, White, Yang, Yeche, Zhang, Zhao, Zheng, Zhou,
  Zhou, Zhu, Zou, \& Zu}]{desi:2016_sci}
{DESI Collaboration}, Aguilar, J., Ahlen, S., {et~al.} 2016{\natexlab{b}},
  arXiv e-prints, arXiv:1611.00036.
\newblock \url{https://ui.adsabs.harvard.edu/abs/2016arXiv161100036D/abstract}

\bibitem[{Dotter(2016)}]{dotter:2016}
Dotter, A. 2016, The Astrophysical Journal Supplement Series, 222, 8,
  \dodoi{10.3847/0067-0049/222/1/8}

\bibitem[{Duane {et~al.}(1987)Duane, Kennedy, Pendleton, \&
  Roweth}]{duane:1987}
Duane, S., Kennedy, A.~D., Pendleton, B.~J., \& Roweth, D. 1987, Physics
  Letters B, 195, 216, \dodoi{10.1016/0370-2693(87)91197-X}

\bibitem[{Duggan {et~al.}(2018)Duggan, Kirby, Andrievsky, \&
  Korotin}]{duggan:2018}
Duggan, G.~E., Kirby, E.~N., Andrievsky, S.~M., \& Korotin, S.~A. 2018, The
  Astrophysical Journal, 869, 50, \dodoi{10.3847/1538-4357/aaeb8e}

\bibitem[{Echeverria {et~al.}(2016)Echeverria, Silva, Mendez, \&
  Orchard}]{echeverria:2016}
Echeverria, A., Silva, J.~F., Mendez, R.~A., \& Orchard, M. 2016, Astronomy and
  Astrophysics, 594, A111, \dodoi{10.1051/0004-6361/201628220}

\bibitem[{Eriksen \& Gazta{\~n}aga(2015)}]{eriksen:2015}
Eriksen, M., \& Gazta{\~n}aga, E. 2015, Monthly Notices of the Royal
  Astronomical Society, 452, 2168, \dodoi{10.1093/mnras/stv1075}

\bibitem[{Escala {et~al.}(2019{\natexlab{a}})Escala, Gilbert, Kirby, Wojno,
  Cunningham, \& Guhathakurta}]{escala:2019b}
Escala, I., Gilbert, K.~M., Kirby, E.~N., {et~al.} 2019{\natexlab{a}}, arXiv
  e-prints, arXiv:1909.00006.
\newblock \url{https://ui.adsabs.harvard.edu/abs/2019arXiv190900006E/abstract}

\bibitem[{Escala {et~al.}(2019{\natexlab{b}})Escala, Kirby, Gilbert,
  Cunningham, \& Wojno}]{escala:2019a}
Escala, I., Kirby, E.~N., Gilbert, K.~M., Cunningham, E.~C., \& Wojno, J.
  2019{\natexlab{b}}, The Astrophysical Journal, 878, 42,
  \dodoi{10.3847/1538-4357/ab1eac}

\bibitem[{Evans {et~al.}(2019)Evans, Castro, Gonzalez, Garcia, Bastian, Cioni,
  Clark, Davies, Ferguson, Kamann, Lennon, Patrick, Vink, \&
  Weisz}]{evans:2019}
Evans, C.~J., Castro, N., Gonzalez, O.~A., {et~al.} 2019, Astronomy and
  Astrophysics, 622, A129, \dodoi{10.1051/0004-6361/201834145}

\bibitem[{Faber {et~al.}(2003)Faber, Phillips, Kibrick, Alcott, Allen, Burrous,
  Cantrall, Clarke, Coil, Cowley, Davis, Deich, Dietsch, Gilmore, Harper,
  Hilyard, Lewis, McVeigh, Newman, Osborne, Schiavon, Stover, Tucker, Wallace,
  Wei, Wirth, \& Wright}]{faber:2003}
Faber, S.~M., Phillips, A.~C., Kibrick, R.~I., {et~al.} 2003, Instrument Design
  and Performance for Optical/Infrared Ground-based Telescopes, 4841, 1657,
  \dodoi{10.1117/12.460346}

\bibitem[{Fabricant {et~al.}(2005)Fabricant, Fata, Roll, Hertz, Caldwell,
  Gauron, Geary, McLeod, Szentgyorgyi, Zajac, Kurtz, Barberis, Bergner, Brown,
  Conroy, Eng, Geller, Goddard, Honsa, Mueller, Mink, Ordway, Tokarz, Woods,
  Wyatt, Epps, \& Dell'Antonio}]{fabricant:2005}
Fabricant, D., Fata, R., Roll, J., {et~al.} 2005, Publications of the
  Astronomical Society of the Pacific, 117, 1411, \dodoi{10.1086/497385}

\bibitem[{Fabricant {et~al.}(2019)Fabricant, Fata, Epps, Gauron, Mueller,
  Zajac, Amato, Barberis, Bergner, Brennan, Brown, Chilingarian, Geary,
  Kradinov, McLeod, Smith, \& Woods}]{fabricant:2019}
Fabricant, D., Fata, R., Epps, H., {et~al.} 2019, Publications of the
  Astronomical Society of the Pacific, 131, 075004,
  \dodoi{10.1088/1538-3873/ab1d78}

\bibitem[{Feeney {et~al.}(2019)Feeney, Wandelt, \& Ness}]{feeney:2019}
Feeney, S.~M., Wandelt, B.~D., \& Ness, M.~K. 2019, arXiv e-prints, 1912,
  arXiv:1912.09498.
\newblock \url{http://adsabs.harvard.edu/abs/2019arXiv191209498F}

\bibitem[{Fischel(1964)}]{fischel:1964}
Fischel, D. 1964, The Astrophysical Journal, 140, 221, \dodoi{10.1086/147909}

\bibitem[{Font-Ribera {et~al.}(2014)Font-Ribera, McDonald, Mostek, Reid, Seo,
  \& Slosar}]{font-ribera:2014}
Font-Ribera, A., McDonald, P., Mostek, N., {et~al.} 2014, Journal of Cosmology
  and Astroparticle Physics, 05, 023, \dodoi{10.1088/1475-7516/2014/05/023}

\bibitem[{Fraunhofer(1817)}]{fraunhofer:1817}
Fraunhofer, J. 1817, Annalen der Physik, 56, 264,
  \dodoi{10.1002/andp.18170560706}

\bibitem[{Frebel {et~al.}(2016)Frebel, Norris, Gilmore, \& Wyse}]{frebel:2016}
Frebel, A., Norris, J.~E., Gilmore, G., \& Wyse, R. F.~G. 2016, The
  Astrophysical Journal, 826, 110, \dodoi{10.3847/0004-637X/826/2/110}

\bibitem[{Frebel {et~al.}(2010)Frebel, Simon, Geha, \& Willman}]{frebel:2010}
Frebel, A., Simon, J.~D., Geha, M., \& Willman, B. 2010, The Astrophysical
  Journal, 708, 560, \dodoi{10.1088/0004-637X/708/1/560}

\bibitem[{Frebel {et~al.}(2014)Frebel, Simon, \& Kirby}]{frebel:2014}
Frebel, A., Simon, J.~D., \& Kirby, E.~N. 2014, The Astrophysical Journal, 786,
  74, \dodoi{10.1088/0004-637X/786/1/74}

\bibitem[{Fr{\'e}chet(1943)}]{frechet:1943}
Fr{\'e}chet, M. 1943, Revue de l'Institut International de Statistique / Review
  of the International Statistical Institute, 11, 182, \dodoi{10.2307/1401114}

\bibitem[{Fulbright {et~al.}(2004)Fulbright, Rich, \& Castro}]{fulbright:2004}
Fulbright, J.~P., Rich, R.~M., \& Castro, S. 2004, The Astrophysical Journal,
  612, 447, \dodoi{10.1086/421712}

\bibitem[{Garc{\'\i}a~P{\'e}rez {et~al.}(2016)Garc{\'\i}a~P{\'e}rez,
  Allende~Prieto, Holtzman, Shetrone, M{\'e}sz{\'a}ros, Bizyaev, Carrera,
  Cunha, Garc{\'\i}a-Hern{\'a}ndez, Johnson, Majewski, Nidever, Schiavon,
  Shane, Smith, Sobeck, Troup, Zamora, Weinberg, Bovy, Eisenstein, Feuillet,
  Frinchaboy, Hayden, Hearty, Nguyen, O'Connell, Pinsonneault, Wilson, \&
  Zasowski}]{garcia-perez:2016}
Garc{\'\i}a~P{\'e}rez, A.~E., Allende~Prieto, C., Holtzman, J.~A., {et~al.}
  2016, The Astronomical Journal, 151, 144, \dodoi{10.3847/0004-6256/151/6/144}

\bibitem[{Gilbert {et~al.}(2019)Gilbert, Kirby, Escala, Wojno, Kalirai, \&
  Guhathakurta}]{gilbert:2019}
Gilbert, K.~M., Kirby, E.~N., Escala, I., {et~al.} 2019, The Astrophysical
  Journal, 883, 128, \dodoi{10.3847/1538-4357/ab3807}

\bibitem[{Gingerich(1969)}]{cayrel:1969}
Gingerich, O., ed. 1969, Comparison of {Synthetic} {Spectra} with {Real}
  {Spectra}, ed. O.~Gingerich, Proceedings of the 3rd Harvard-Smithsonian
  Conference on Stellar Atmospheres, Massachusetts Institute of Technology,
  Cambridge, MA, 237.
\newblock \url{http://adsabs.harvard.edu/abs/1969tons.conf..237C}

\bibitem[{Greenstein(1948)}]{greenstein:1948}
Greenstein, J.~L. 1948, The Astrophysical Journal, 107, 151,
  \dodoi{10.1086/145002}

\bibitem[{Hansen {et~al.}(2015)Hansen, Ludwig, Seifert, Koch, Xu, Caffau,
  Christlieb, Korn, Lind, Sbordone, Ruchti, Feltzing, de~Jong, \&
  Barden}]{hansen:2015}
Hansen, C.~J., Ludwig, H.-G., Seifert, W., {et~al.} 2015, Astronomische
  Nachrichten, 336, 665, \dodoi{10.1002/asna.201512206}

\bibitem[{Hearnshaw(2010)}]{hearnshaw:2010}
Hearnshaw, J. 2010, Journal of Astronomical History and Heritage, 13, 90.
\newblock \url{http://adsabs.harvard.edu/abs/2010JAHH...13...90H}

\bibitem[{Heiter {et~al.}(2015)Heiter, Jofr{\'e}, Gustafsson, Korn, Soubiran,
  \& Th{\'e}venin}]{heiter:2015}
Heiter, U., Jofr{\'e}, P., Gustafsson, B., {et~al.} 2015, Astronomy and
  Astrophysics, 582, A49, \dodoi{10.1051/0004-6361/201526319}

\bibitem[{Hendricks {et~al.}(2014)Hendricks, Koch, Walker, Johnson,
  Pe{\~n}arrubia, \& Gilmore}]{hendricks:2014}
Hendricks, B., Koch, A., Walker, M., {et~al.} 2014, Astronomy and Astrophysics,
  572, A82, \dodoi{10.1051/0004-6361/201424645}

\bibitem[{Hill {et~al.}(2019)Hill, Sk{\'u}lad{\'o}ttir, Tolstoy, Venn,
  Shetrone, Jablonka, Primas, Battaglia, de~Boer, Fran{\c c}ois, Helmi, Kaufer,
  Letarte, Starkenburg, \& Spite}]{hill:2019}
Hill, V., Sk{\'u}lad{\'o}ttir, {\'A}., Tolstoy, E., {et~al.} 2019, Astronomy
  and Astrophysics, 626, A15, \dodoi{10.1051/0004-6361/201833950}

\bibitem[{Ho {et~al.}(2015)Ho, Geha, Tollerud, Zinn, Guhathakurta, \&
  Vargas}]{ho:2015}
Ho, N., Geha, M., Tollerud, E.~J., {et~al.} 2015, The Astrophysical Journal,
  798, 77, \dodoi{10.1088/0004-637X/798/2/77}

\bibitem[{Holtzman {et~al.}(2015)Holtzman, Shetrone, Johnson, Allende~Prieto,
  Anders, Andrews, Beers, Bizyaev, Blanton, Bovy, Carrera, Chojnowski, Cunha,
  Eisenstein, Feuillet, Frinchaboy, Galbraith-Frew, Garc{\'\i}a~P{\'e}rez,
  Garc{\'\i}a-Hern{\'a}ndez, Hasselquist, Hayden, Hearty, Ivans, Majewski,
  Martell, Meszaros, Muna, Nidever, Nguyen, O'Connell, Pan, Pinsonneault,
  Robin, Schiavon, Shane, Sobeck, Smith, Troup, Weinberg, Wilson, Wood-Vasey,
  Zamora, \& Zasowski}]{holtzman:2015}
Holtzman, J.~A., Shetrone, M., Johnson, J.~A., {et~al.} 2015, The Astronomical
  Journal, 150, 148, \dodoi{10.1088/0004-6256/150/5/148}

\bibitem[{Huggins \& Miller(1864)}]{huggins:1864}
Huggins, W., \& Miller, W.~A. 1864, Philosophical Transactions of the Royal
  Society of London Series I, 154, 413.
\newblock \url{http://adsabs.harvard.edu/abs/1864RSPT..154..413H}

\bibitem[{Hunter(2007)}]{hunter:2007}
Hunter, J.~D. 2007, Computing in Science \& Engineering, 9, 90,
  \dodoi{10.1109/MCSE.2007.55}

\bibitem[{Husser {et~al.}(2016)Husser, Kamann, Dreizler, Wendt, Wulff, Bacon,
  Wisotzki, Brinchmann, Weilbacher, Roth, \& Monreal-Ibero}]{husser:2016}
Husser, T.-O., Kamann, S., Dreizler, S., {et~al.} 2016, Astronomy and
  Astrophysics, 588, A148, \dodoi{10.1051/0004-6361/201526949}

\bibitem[{Ireland(2005)}]{ireland:2005}
Ireland, J. 2005, The Astrophysical Journal, 620, 1132, \dodoi{10.1086/427230}

\bibitem[{Jagourel {et~al.}(2018)Jagourel, Fitzsimons, Hammer, Frondat, Puech,
  Evans, Sanchez, Guinouard, Chemla, Frotin, Yang, Parr-Burman, Morris,
  Dubbeldam, Close, Middleton, Rousset, Gendron, Kelz, Janssen, Pragt, Navarro,
  Larrieu, Hadi, Dohlen, Dalton, Lewis, Rodrigues, Morris, Kaper, Barbuy, Cuby,
  \& F{\`e}vre}]{jagourel:2018}
Jagourel, P., Fitzsimons, E., Hammer, F., {et~al.} 2018, in Ground-based and
  {Airborne} {Instrumentation} for {Astronomy} {VII}, Vol. 10702 (International
  Society for Optics and Photonics), 10702A4, \dodoi{10.1117/12.2314135}

\bibitem[{Ji {et~al.}(2016{\natexlab{a}})Ji, Frebel, Ezzeddine, \&
  Casey}]{ji:2016b}
Ji, A.~P., Frebel, A., Ezzeddine, R., \& Casey, A.~R. 2016{\natexlab{a}}, The
  Astrophysical Journal, 832, L3, \dodoi{10.3847/2041-8205/832/1/L3}

\bibitem[{Ji {et~al.}(2016{\natexlab{b}})Ji, Frebel, Simon, \&
  Chiti}]{ji:2016c}
Ji, A.~P., Frebel, A., Simon, J.~D., \& Chiti, A. 2016{\natexlab{b}}, The
  Astrophysical Journal, 830, 93, \dodoi{10.3847/0004-637X/830/2/93}

\bibitem[{Ji {et~al.}(2016{\natexlab{c}})Ji, Frebel, Simon, \& Geha}]{ji:2016a}
Ji, A.~P., Frebel, A., Simon, J.~D., \& Geha, M. 2016{\natexlab{c}}, The
  Astrophysical Journal, 817, 41, \dodoi{10.3847/0004-637X/817/1/41}

\bibitem[{Ji {et~al.}(2019)Ji, Li, Simon, Marshall, Vivas, Pace, Bechtol,
  Drlica-Wagner, Koposov, Hansen, Allam, Gruendl, Johnson, McNanna, Noel,
  Tucker, Walker, \& {MagLiteS Collaboration}}]{ji:2019}
Ji, A.~P., Li, T.~S., Simon, J.~D., {et~al.} 2019, arXiv e-prints, 1912,
  arXiv:1912.04963.
\newblock \url{http://adsabs.harvard.edu/abs/2019arXiv191204963J}

\bibitem[{Jofr{\'e} {et~al.}(2019)Jofr{\'e}, Heiter, \& Soubiran}]{jofre:2019}
Jofr{\'e}, P., Heiter, U., \& Soubiran, C. 2019, Annual Review of Astronomy and
  Astrophysics, 57, 571, \dodoi{10.1146/annurev-astro-091918-104509}

\bibitem[{Jorgensen {et~al.}(1992)Jorgensen, Carlsson, \&
  Johnson}]{jorgensen:1992}
Jorgensen, U.~G., Carlsson, M., \& Johnson, H.~R. 1992, Astronomy and
  Astrophysics, 254, 258.
\newblock \url{http://adsabs.harvard.edu/abs/1992A%26A...254..258J}

\bibitem[{Kalirai {et~al.}(2010)Kalirai, Beaton, Geha, Gilbert, Guhathakurta,
  Kirby, Majewski, Ostheimer, Patterson, \& Wolf}]{kalirai:2010}
Kalirai, J.~S., Beaton, R.~L., Geha, M.~C., {et~al.} 2010, The Astrophysical
  Journal, 711, 671, \dodoi{10.1088/0004-637X/711/2/671}

\bibitem[{Kamann {et~al.}(2016)Kamann, Husser, Wendt, Bacon, Brinchmann,
  Dreizler, Emsellem, Krajnovi{\'c}, Monreal-Ibero, Roth, Weilbacher, \&
  Wisotzki}]{kamann:2016}
Kamann, S., Husser, T.-O., Wendt, M., {et~al.} 2016, The Messenger, 164, 18.
\newblock \url{https://ui.adsabs.harvard.edu/2016Msngr.164...18K/abstract}

\bibitem[{Kamann {et~al.}(2018)Kamann, Husser, Dreizler, Emsellem, Weilbacher,
  Martens, Bacon, den Brok, Giesers, Krajnovi{\'c}, Roth, Wendt, \&
  Wisotzki}]{kamann:2018}
Kamann, S., Husser, T.-O., Dreizler, S., {et~al.} 2018, Monthly Notices of the
  Royal Astronomical Society, 473, 5591, \dodoi{10.1093/mnras/stx2719}

\bibitem[{Kay(1993)}]{kay:1993}
Kay, S.~M. 1993, Fundamentals of {Statistical} {Signal} {Processing}, {Volume}
  {I}: {Estimation} {Theory}, 1st edn. (Englewood Cliffs, N.J: Prentice Hall)

\bibitem[{King {et~al.}(2014)King, Davis, Denney, Vestergaard, \&
  Watson}]{king:2014}
King, A.~L., Davis, T.~M., Denney, K.~D., Vestergaard, M., \& Watson, D. 2014,
  Monthly Notices of the Royal Astronomical Society, 441, 3454,
  \dodoi{10.1093/mnras/stu793}

\bibitem[{Kirby {et~al.}(2015{\natexlab{a}})Kirby, Cohen, Simon, \&
  Guhathakurta}]{kirby:2015c}
Kirby, E.~N., Cohen, J.~G., Simon, J.~D., \& Guhathakurta, P.
  2015{\natexlab{a}}, The Astrophysical Journal Letters, 814, L7,
  \dodoi{10.1088/2041-8205/814/1/L7}

\bibitem[{Kirby {et~al.}(2017{\natexlab{a}})Kirby, Cohen, Simon, Guhathakurta,
  Thygesen, \& Duggan}]{kirby:2017b}
Kirby, E.~N., Cohen, J.~G., Simon, J.~D., {et~al.} 2017{\natexlab{a}}, The
  Astrophysical Journal, 838, 83, \dodoi{10.3847/1538-4357/aa6570}

\bibitem[{Kirby {et~al.}(2020)Kirby, Gilbert, Escala, Wojno, Guhathakurta,
  Majewski, \& Beaton}]{kirby:2020}
Kirby, E.~N., Gilbert, K.~M., Escala, I., {et~al.} 2020, The Astronomical
  Journal, 159, 46, \dodoi{10.3847/1538-3881/ab5f0f}

\bibitem[{Kirby {et~al.}(2009)Kirby, Guhathakurta, Bolte, Sneden, \&
  Geha}]{kirby:2009}
Kirby, E.~N., Guhathakurta, P., Bolte, M., Sneden, C., \& Geha, M.~C. 2009, The
  Astrophysical Journal, 705, 328, \dodoi{10.1088/0004-637X/705/1/328}

\bibitem[{Kirby {et~al.}(2008)Kirby, Guhathakurta, \& Sneden}]{kirby:2008}
Kirby, E.~N., Guhathakurta, P., \& Sneden, C. 2008, The Astrophysical Journal,
  682, 1217, \dodoi{10.1086/589627}

\bibitem[{Kirby {et~al.}(2017{\natexlab{b}})Kirby, Rizzi, Held, Cohen, Cole,
  Manning, Skillman, \& Weisz}]{kirby:2017a}
Kirby, E.~N., Rizzi, L., Held, E.~V., {et~al.} 2017{\natexlab{b}}, The
  Astrophysical Journal, 834, 9, \dodoi{10.3847/1538-4357/834/1/9}

\bibitem[{Kirby {et~al.}(2015{\natexlab{b}})Kirby, Simon, \&
  Cohen}]{kirby:2015b}
Kirby, E.~N., Simon, J.~D., \& Cohen, J.~G. 2015{\natexlab{b}}, The
  Astrophysical Journal, 810, 56, \dodoi{10.1088/0004-637X/810/1/56}

\bibitem[{Kirby {et~al.}(2018)Kirby, Xie, Guo, Kovalev, \&
  Bergemann}]{kirby:2018}
Kirby, E.~N., Xie, J.~L., Guo, R., Kovalev, M., \& Bergemann, M. 2018, The
  Astrophysical Journal Supplement Series, 237, 18,
  \dodoi{10.3847/1538-4365/aac952}

\bibitem[{Kirby {et~al.}(2010)Kirby, Guhathakurta, Simon, Geha, Rockosi,
  Sneden, Cohen, Sohn, Majewski, \& Siegel}]{kirby:2010}
Kirby, E.~N., Guhathakurta, P., Simon, J.~D., {et~al.} 2010, The Astrophysical
  Journal Supplement Series, 191, 352, \dodoi{10.1088/0067-0049/191/2/352}

\bibitem[{Kirby {et~al.}(2015{\natexlab{c}})Kirby, Guo, Zhang, Deng, Cohen,
  Guhathakurta, Shetrone, Lee, \& Rizzi}]{kirby:2015a}
Kirby, E.~N., Guo, M., Zhang, A.~J., {et~al.} 2015{\natexlab{c}}, The
  Astrophysical Journal, 801, 125, \dodoi{10.1088/0004-637X/801/2/125}

\bibitem[{Kirchhoff(1860)}]{kirchhoff:1860}
Kirchhoff, G. 1860, Annalen der Physik, 185, 275,
  \dodoi{10.1002/andp.18601850205}

\bibitem[{Kirchhoff(1863)}]{kirchhoff:1863}
---. 1863, Annalen der Physik, 194, 94, \dodoi{10.1002/andp.18631940106}

\bibitem[{Kirchhoff \& Bunsen(1860)}]{kirchhoff_bunsen:1860}
Kirchhoff, G., \& Bunsen, R. 1860, Annalen der Physik, 186, 161,
  \dodoi{10.1002/andp.18601860602}

\bibitem[{Koch \& C{\^o}t{\'e}(2010)}]{koch:2010}
Koch, A., \& C{\^o}t{\'e}, P. 2010, Astronomy and Astrophysics, 517, A59,
  \dodoi{10.1051/0004-6361/201014155}

\bibitem[{Koch {et~al.}(2008{\natexlab{a}})Koch, Grebel, Gilmore, Wyse, Kleyna,
  Harbeck, Wilkinson, \& Evans}]{koch:2008a}
Koch, A., Grebel, E.~K., Gilmore, G.~F., {et~al.} 2008{\natexlab{a}}, The
  Astronomical Journal, 135, 1580, \dodoi{10.1088/0004-6256/135/4/1580}

\bibitem[{Koch {et~al.}(2007{\natexlab{a}})Koch, Grebel, Kleyna, Wilkinson,
  Harbeck, Gilmore, Wyse, \& Evans}]{koch:2007a}
Koch, A., Grebel, E.~K., Kleyna, J.~T., {et~al.} 2007{\natexlab{a}}, The
  Astronomical Journal, 133, 270, \dodoi{10.1086/509889}

\bibitem[{Koch {et~al.}(2008{\natexlab{b}})Koch, McWilliam, Grebel, Zucker, \&
  Belokurov}]{koch:2008b}
Koch, A., McWilliam, A., Grebel, E.~K., Zucker, D.~B., \& Belokurov, V.
  2008{\natexlab{b}}, The Astrophysical Journal Letters, 688, L13,
  \dodoi{10.1086/595001}

\bibitem[{Koch \& Rich(2014)}]{koch:2014}
Koch, A., \& Rich, R.~M. 2014, The Astrophysical Journal, 794, 89,
  \dodoi{10.1088/0004-637X/794/1/89}

\bibitem[{Koch {et~al.}(2007{\natexlab{b}})Koch, Wilkinson, Kleyna, Gilmore,
  Grebel, Mackey, Evans, \& Wyse}]{koch:2007b}
Koch, A., Wilkinson, M.~I., Kleyna, J.~T., {et~al.} 2007{\natexlab{b}}, The
  Astrophysical Journal, 657, 241, \dodoi{10.1086/510879}

\bibitem[{Koch {et~al.}(2009)Koch, Wilkinson, Kleyna, Irwin, Zucker, Belokurov,
  Gilmore, Fellhauer, \& Evans}]{koch:2009}
---. 2009, The Astrophysical Journal, 690, 453,
  \dodoi{10.1088/0004-637X/690/1/453}

\bibitem[{Kurucz(1970)}]{kurucz:1970}
Kurucz, R.~L. 1970, SAO Special Report, 309.
\newblock \url{https://ui.adsabs.harvard.edu/1970SAOSR.309.....K/abstract}

\bibitem[{Kurucz(1993)}]{kurucz:1993}
---. 1993, {SYNTHE} spectrum synthesis programs and line data, CD-ROM.
\newblock \url{http://adsabs.harvard.edu/abs/1993sssp.book.....K}

\bibitem[{Kurucz(2005)}]{kurucz:2005}
---. 2005, Memorie della Societa Astronomica Italiana Supplementi, 8, 14.
\newblock \url{https://ui.adsabs.harvard.edu/2005MSAIS...8...14K/abstract}

\bibitem[{Kurucz(2013)}]{kurucz:2013}
---. 2013, Astrophysics Source Code Library, ascl:1303.024.
\newblock \url{https://ui.adsabs.harvard.edu/2013ascl.soft03024K/abstract}

\bibitem[{Kurucz(2017)}]{kurucz:2017}
---. 2017, Astrophysics Source Code Library, ascl:1710.017.
\newblock \url{https://ui.adsabs.harvard.edu/2017ascl.soft10017K/abstract}

\bibitem[{Kurucz \& Avrett(1981)}]{kurucz:1981}
Kurucz, R.~L., \& Avrett, E.~H. 1981, SAO Special Report, 391.
\newblock \url{https://ui.adsabs.harvard.edu/1981SAOSR.391.....K/abstract}

\bibitem[{Lai {et~al.}(2011)Lai, Lee, Bolte, Lucatello, Beers, Johnson,
  Sivarani, \& Rockosi}]{lai:2011}
Lai, D.~K., Lee, Y.~S., Bolte, M., {et~al.} 2011, The Astrophysical Journal,
  738, 51, \dodoi{10.1088/0004-637X/738/1/51}

\bibitem[{Latour {et~al.}(2019)Latour, Husser, Giesers, Kamann, Goettgens,
  Dreizler, Brinchmann, Bastian, Wendt, Weilbacher, \& Molinski}]{latour:2019}
Latour, M., Husser, T.-O., Giesers, B., {et~al.} 2019, arXiv e-prints,
  arXiv:1909.04959.
\newblock \url{https://ui.adsabs.harvard.edu/abs/2019arXiv190904959L/abstract}

\bibitem[{Laverick {et~al.}(2018)Laverick, Lobel, Merle, Royer, Martayan,
  David, Hensberge, \& Thienpont}]{laverick:2018}
Laverick, M., Lobel, A., Merle, T., {et~al.} 2018, Astronomy and Astrophysics,
  612, A60, \dodoi{10.1051/0004-6361/201731933}

\bibitem[{Leaman {et~al.}(2009)Leaman, Cole, Venn, Tolstoy, Irwin, Szeifert,
  Skillman, \& McConnachie}]{leaman:2009}
Leaman, R., Cole, A.~A., Venn, K.~A., {et~al.} 2009, The Astrophysical Journal,
  699, 1, \dodoi{10.1088/0004-637X/699/1/1}

\bibitem[{Leep {et~al.}(1987)Leep, Oke, \& Wallerstein}]{leep:1987}
Leep, E.~M., Oke, J.~B., \& Wallerstein, G. 1987, The Astronomical Journal, 93,
  338, \dodoi{10.1086/114318}

\bibitem[{Leep {et~al.}(1986)Leep, Wallerstein, \& Oke}]{leep:1986}
Leep, E.~M., Wallerstein, G., \& Oke, J.~B. 1986, The Astronomical Journal, 91,
  1117, \dodoi{10.1086/114088}

\bibitem[{Letarte {et~al.}(2006)Letarte, Hill, Jablonka, Tolstoy, Fran{\c
  c}ois, \& Meylan}]{letarte:2006}
Letarte, B., Hill, V., Jablonka, P., {et~al.} 2006, Astronomy \& Astrophysics,
  453, 547, \dodoi{10.1051/0004-6361:20054439}

\bibitem[{Li {et~al.}(2017)Li, Simon, Drlica-Wagner, Bechtol, Wang,
  Garc{\'\i}a-Bellido, Frieman, Marshall, James, Strigari, Pace, Balbinot,
  Zhang, Abbott, Allam, Benoit-L{\'e}vy, Bernstein, Bertin, Brooks, Burke,
  Carnero~Rosell, Carrasco~Kind, Carretero, Cunha, D'Andrea, da~Costa, DePoy,
  Desai, Diehl, Eifler, Flaugher, Goldstein, Gruen, Gruendl, Gschwend,
  Gutierrez, Krause, Kuehn, Lin, Maia, March, Menanteau, Miquel, Plazas, Romer,
  Sanchez, Santiago, Schubnell, Sevilla-Noarbe, Smith, Sobreira, Suchyta,
  Tarle, Thomas, Tucker, Walker, Wechsler, Wester, Yanny, \& {DES
  Collaboration}}]{li:2017}
Li, T.~S., Simon, J.~D., Drlica-Wagner, A., {et~al.} 2017, The Astrophysical
  Journal, 838, 8, \dodoi{10.3847/1538-4357/aa6113}

\bibitem[{Longeard {et~al.}(2020)Longeard, Martin, Starkenburg, Ibata, Collins,
  Laevens, Mackey, Rich, Aguado, Arentsen, Jablonka,
  Gonz{\'a}lez~Hern{\'a}ndez, Navarro, \& S{\'a}nchez-Janssen}]{longeard:2020}
Longeard, N., Martin, N., Starkenburg, E., {et~al.} 2020, Monthly Notices of
  the Royal Astronomical Society, 491, 356, \dodoi{10.1093/mnras/stz2854}

\bibitem[{Lucchesi {et~al.}(2020)Lucchesi, Lardo, Primas, Jablonka, North,
  Battaglia, Starkenburg, Hill, Irwin, Francois, Shetrone, Tolstoy, \&
  Venn}]{lucchesi:2020}
Lucchesi, R., Lardo, C., Primas, F., {et~al.} 2020, arXiv e-prints, 2001,
  arXiv:2001.11033.
\newblock \url{http://adsabs.harvard.edu/abs/2020arXiv200111033L}

\bibitem[{Luo {et~al.}(2015)Luo, Zhao, Zhao, Deng, Liu, Jing, Wang, Zhang, Shi,
  Cui, Chu, Li, Bai, Wu, Cai, Cao, Cao, Carlin, Chen, Chen, Chen, Chen, Chen,
  Chen, Chen, Christlieb, Chu, Cui, Dong, Du, Fan, Feng, Fu, Gao, Gong, Gu,
  Guo, Han, He, Hou, Hou, Hou, Hu, Hu, Hu, Huo, Jia, Jiang, Jiang, Jiang, Jin,
  Kong, Kong, Lei, Li, Li, Li, Li, Li, Li, Li, Li, Li, Li, Li, Li, Liang, Lin,
  Liu, Liu, Liu, Liu, Lu, Luo, Mao, Newberg, Ni, Qi, Qi, Shen, Shi, Song, Song,
  Su, Su, Tang, Tao, Tian, Wang, Wang, Wang, Wang, Wang, Wang, Wang, Wang,
  Wang, Wang, Wang, Wang, Wang, Wang, Wang, Wang, Wang, Wang, Wang, Wang, Wei,
  Wei, Wu, Wu, Wu, Wu, Xing, Xu, Xu, Xu, Yan, Yang, Yang, Yang, Yang, Yao, Yu,
  Yuan, Yuan, Yuan, Yuan, Zhai, Zhang, Zhang, Zhang, Zhang, Zhang, Zhang,
  Zhang, Zhang, Zhao, Zhou, Zhou, Zhu, Zhu, Zou, \& Zuo}]{luo:2015}
Luo, A.-L., Zhao, Y.-H., Zhao, G., {et~al.} 2015, Research in Astronomy and
  Astrophysics, 15, 1095, \dodoi{10.1088/1674-4527/15/8/002}

\bibitem[{Majewski {et~al.}(2017)Majewski, Schiavon, Frinchaboy,
  Allende~Prieto, Barkhouser, Bizyaev, Blank, Brunner, Burton, Carrera,
  Chojnowski, Cunha, Epstein, Fitzgerald, Garc{\'\i}a~P{\'e}rez, Hearty,
  Henderson, Holtzman, Johnson, Lam, Lawler, Maseman, M{\'e}sz{\'a}ros, Nelson,
  Nguyen, Nidever, Pinsonneault, Shetrone, Smee, Smith, Stolberg, Skrutskie,
  Walker, Wilson, Zasowski, Anders, Basu, Beland, Blanton, Bovy, Brownstein,
  Carlberg, Chaplin, Chiappini, Eisenstein, Elsworth, Feuillet, Fleming,
  Galbraith-Frew, Garc{\'\i}a, Garc{\'\i}a-Hern{\'a}ndez, Gillespie, Girardi,
  Gunn, Hasselquist, Hayden, Hekker, Ivans, Kinemuchi, Klaene, Mahadevan,
  Mathur, Mosser, Muna, Munn, Nichol, O'Connell, Parejko, Robin, Rocha-Pinto,
  Schultheis, Serenelli, Shane, Silva~Aguirre, Sobeck, Thompson, Troup,
  Weinberg, \& Zamora}]{majewski:2017}
Majewski, S.~R., Schiavon, R.~P., Frinchaboy, P.~M., {et~al.} 2017, The
  Astronomical Journal, 154, 94, \dodoi{10.3847/1538-3881/aa784d}

\bibitem[{Martin {et~al.}(2007)Martin, Ibata, Chapman, Irwin, \&
  Lewis}]{martin:2007}
Martin, N.~F., Ibata, R.~A., Chapman, S.~C., Irwin, M., \& Lewis, G.~F. 2007,
  Monthly Notices of the Royal Astronomical Society, 380, 281,
  \dodoi{10.1111/j.1365-2966.2007.12055.x}

\bibitem[{Martin {et~al.}(2016{\natexlab{a}})Martin, Ibata, Collins, Rich,
  Bell, Ferguson, Laevens, Rix, Chapman, \& Koch}]{martin:2016a}
Martin, N.~F., Ibata, R.~A., Collins, M. L.~M., {et~al.} 2016{\natexlab{a}},
  The Astrophysical Journal, 818, 40, \dodoi{10.3847/0004-637X/818/1/40}

\bibitem[{Martin {et~al.}(2016{\natexlab{b}})Martin, Geha, Ibata, Collins,
  Laevens, Bell, Rix, Ferguson, Chambers, Wainscoat, \& Waters}]{martin:2016b}
Martin, N.~F., Geha, M., Ibata, R.~A., {et~al.} 2016{\natexlab{b}}, Monthly
  Notices of the Royal Astronomical Society, 458, L59,
  \dodoi{10.1093/mnrasl/slw013}

\bibitem[{Mashonkina {et~al.}(2007)Mashonkina, Korn, \&
  Przybilla}]{mashonkina:2007}
Mashonkina, L., Korn, A.~J., \& Przybilla, N. 2007, Astronomy and Astrophysics,
  461, 261, \dodoi{10.1051/0004-6361:20065999}

\bibitem[{Mateo {et~al.}(2012)Mateo, Iii, Crane, Shectman, Thompson, Roederer,
  Bigelow, \& Gunnels}]{mateo:2012}
Mateo, M., Iii, J. I.~B., Crane, J., {et~al.} 2012, in Ground-based and
  {Airborne} {Instrumentation} for {Astronomy} {IV}, Vol. 8446 (International
  Society for Optics and Photonics), 84464Y, \dodoi{10.1117/12.926448}

\bibitem[{McKinney(2010)}]{mckinney:2010}
McKinney, W. 2010, in Proceedings of the 9th Python in Science Conference, ed.
  S.~van~der Walt \& J.~Millman, 51 -- 56

\bibitem[{Minnaert(1934)}]{minnaert:1934}
Minnaert, M. 1934, The Observatory, 57, 328.
\newblock \url{http://adsabs.harvard.edu/abs/1934Obs....57..328M}

\bibitem[{Moore(1920)}]{moore:1920}
Moore, E. 1920, Bulletin of the American Mathematical Society, 26, 394,
  \dodoi{10.1090/S0002-9904-1920-03322-7.}

\bibitem[{{MSE Science Team} {et~al.}(2019){MSE Science Team}, Babusiaux,
  Bergemann, Burgasser, Ellison, Haggard, Huber, Kaplinghat, Li, Marshall,
  Martell, McConnachie, Percival, Robotham, Shen, Thirupathi, Tran, Yeche,
  Yong, Adibekyan, Silva~Aguirre, Angelou, Asplund, Balogh, Banerjee,
  Bannister, Barr{\'\i}a, Battaglia, Bayo, Bechtol, Beck, Beers, Bellinger,
  Berg, Bestenlehner, Bilicki, Bitsch, Bland-Hawthorn, Bolton, Boselli, Bovy,
  Bragaglia, Buzasi, Caffau, Cami, Carleton, Casagrande, Cassisi, Catelan,
  Chang, Cortese, Damjanov, Davies, de~Grijs, de~Rosa, Deason, di~Matteo,
  Drlica-Wagner, Erkal, Escorza, Ferrarese, Fleming, Font-Ribera, Freeman,
  G{\"a}nsicke, Gabdeev, Gallagher, Gandolfi, Garc{\'\i}a, Gaulme, Geha,
  Gennaro, Gieles, Gilbert, Gordon, Goswami, Greco, Grillmair, Guiglion,
  H{\'e}nault-Brunet, Hall, Handler, Hansen, Hathi, Hatzidimitriou, Haywood,
  Hern{\'a}ndez~Santisteban, Hillenbrand, Hopkins, Howlett, Hudson, Ibata,
  Ili{\'c}, Jablonka, Ji, Jiang, Juneau, Karakas, Karinkuzhi, Kim, Kong,
  Konstantopoulos, Krogager, Lagos, Lallement, Laporte, Lebreton, Lee, Lewis,
  Lianou, Liu, Lodieu, Loveday, M{\'e}sz{\'a}ros, Makler, Mao, Marchesini,
  Martin, Mateo, Melis, Merle, Miglio, Gohar~Mohammad, Molaverdikhani, Monier,
  Morel, Mosser, Nataf, Necib, Neilson, Newman, Nierenberg, Nord, Noterdaeme,
  O'Dea, Oshagh, Pace, Palanque-Delabrouille, Pandey, Parker, Pawlowski, Peter,
  Petitjean, Petric, Placco, Popovi{\'c}, Price-Whelan, Prsa, Ravindranath,
  Rich, Ruan, Rybizki, Sakari, Sanderson, Schiavon, Schimd, Serenelli, Siebert,
  Siudek, Smiljanic, Smith, Sobeck, Starkenburg, Stello, Szab{\'o}, Szabo,
  Taylor, Thanjavur, Thomas, Tollerud, Toonen, Tremblay, Tresse, Tsantaki,
  Valentini, Van~Eck, Variu, Venn, Villaver, Walker, Wang, Wang, Wilson,
  Wright, Xu, Yildiz, Zhang, Zwintz, Anguiano, Bedell, Chaplin, Collet,
  Cuillandre, Duc, Flagey, Hermes, Hill, Kamath, Laychak, Ma{\l}ek, Marley,
  Sheinis, Simons, Sousa, Szeto, Ting, Vegetti, Wells, Babas, Bauman, Bosselli,
  C{\^o}t{\'e}, Colless, Comparat, Courtois, Crampton, Croom, Davies, de~Grijs,
  Denny, Devost, di~Matteo, Driver, Fernandez-Lorenzo, Guhathakurta, Han,
  Higgs, Hill, Ho, Hopkins, Hudson, Ibata, Isani, Jarvis, Johnson, Jullo,
  Kaiser, Kneib, Koda, Koshy, Mignot, Murowinski, Newman, Nusser, Pancoast,
  Peng, Peroux, Pichon, Poggianti, Richard, Salmon, Seibert, Shastri, Smith,
  Sutaria, Tao, Taylor, Tully, van Waerbeke, Vermeulen, Walker, Willis, Willot,
  \& Withington}]{MSE:2019}
{MSE Science Team}, Babusiaux, C., Bergemann, M., {et~al.} 2019, arXiv
  e-prints, arXiv:1904.04907.
\newblock \url{https://ui.adsabs.harvard.edu/abs/2019arXiv190404907T/abstract}

\bibitem[{Mu{\~n}oz {et~al.}(2006)Mu{\~n}oz, Majewski, Zaggia, Kunkel,
  Frinchaboy, Nidever, Crnojevic, Patterson, Crane, Johnston, Sohn, Bernstein,
  \& Shectman}]{munoz:2006}
Mu{\~n}oz, R.~R., Majewski, S.~R., Zaggia, S., {et~al.} 2006, The Astrophysical
  Journal, 649, 201, \dodoi{10.1086/505620}

\bibitem[{Ness {et~al.}(2015)Ness, Hogg, Rix, Ho, \& Zasowski}]{ness:2015}
Ness, M., Hogg, D.~W., Rix, H.-W., Ho, A. Y.~Q., \& Zasowski, G. 2015, The
  Astrophysical Journal, 808, 16, \dodoi{10.1088/0004-637X/808/1/16}

\bibitem[{Nissen \& Gustafsson(2018)}]{nissen:2018}
Nissen, P.~E., \& Gustafsson, B. 2018, Astronomy and Astrophysics Review, 26,
  6, \dodoi{10.1007/s00159-018-0111-3}

\bibitem[{Norris {et~al.}(2008)Norris, Gilmore, Wyse, Wilkinson, Belokurov,
  Evans, \& Zucker}]{norris:2008}
Norris, J.~E., Gilmore, G., Wyse, R. F.~G., {et~al.} 2008, The Astrophysical
  Journal Letters, 689, L113, \dodoi{10.1086/595962}

\bibitem[{Oke {et~al.}(1995)Oke, Cohen, Carr, Cromer, Dingizian, Harris,
  Labrecque, Lucinio, Schaal, Epps, \& Miller}]{oke:1995}
Oke, J.~B., Cohen, J.~G., Carr, M., {et~al.} 1995, Publications of the
  Astronomical Society of the Pacific, 107, 375, \dodoi{10.1086/133562}

\bibitem[{Olszewski {et~al.}(1991)Olszewski, Schommer, Suntzeff, \&
  Harris}]{olszewski:1991}
Olszewski, E.~W., Schommer, R.~A., Suntzeff, N.~B., \& Harris, H.~C. 1991, The
  Astronomical Journal, 101, 515, \dodoi{10.1086/115701}

\bibitem[{Pasquini {et~al.}(2002)Pasquini, Avila, Blecha, Cacciari, Cayatte,
  Colless, Damiani, de~Propris, Dekker, di~Marcantonio, Farrell, Gillingham,
  Guinouard, Hammer, Kaufer, Hill, Marteaud, Modigliani, Mulas, North, Popovic,
  Rossetti, Royer, Santin, Schmutzer, Simond, Vola, Waller, \&
  Zoccali}]{pasquini:2002}
Pasquini, L., Avila, G., Blecha, A., {et~al.} 2002, The Messenger, 110, 1.
\newblock \url{https://ui.adsabs.harvard.edu/abs/2002Msngr.110....1P/abstract}

\bibitem[{Paszke {et~al.}(2019)Paszke, Gross, Massa, Lerer, Bradbury, Chanan,
  Killeen, Lin, Gimelshein, Antiga, Desmaison, K{\"o}pf, Yang, DeVito, Raison,
  Tejani, Chilamkurthy, Steiner, Fang, Bai, \& Chintala}]{paszke:2019}
Paszke, A., Gross, S., Massa, F., {et~al.} 2019, arXiv e-prints, 1912,
  arXiv:1912.01703.
\newblock \url{http://adsabs.harvard.edu/abs/2019arXiv191201703P}

\bibitem[{Paxton {et~al.}(2011)Paxton, Bildsten, Dotter, Herwig, Lesaffre, \&
  Timmes}]{paxton:2011}
Paxton, B., Bildsten, L., Dotter, A., {et~al.} 2011, The Astrophysical Journal
  Supplement Series, 192, 3, \dodoi{10.1088/0067-0049/192/1/3}

\bibitem[{Paxton {et~al.}(2013)Paxton, Cantiello, Arras, Bildsten, Brown,
  Dotter, Mankovich, Montgomery, Stello, Timmes, \& Townsend}]{paxton:2013}
Paxton, B., Cantiello, M., Arras, P., {et~al.} 2013, The Astrophysical Journal
  Supplement Series, 208, 4, \dodoi{10.1088/0067-0049/208/1/4}

\bibitem[{Paxton {et~al.}(2015)Paxton, Marchant, Schwab, Bauer, Bildsten,
  Cantiello, Dessart, Farmer, Hu, Langer, Townsend, Townsley, \&
  Timmes}]{paxton:2015}
Paxton, B., Marchant, P., Schwab, J., {et~al.} 2015, The Astrophysical Journal
  Supplement Series, 220, 15, \dodoi{10.1088/0067-0049/220/1/15}

\bibitem[{Payne(1925)}]{payne:1925}
Payne, C.~H. 1925, PhD thesis, Radcliffe College.
\newblock \url{http://adsabs.harvard.edu/abs/1925PhDT.........1P}

\bibitem[{Pazder {et~al.}(2006)Pazder, Roberts, Abraham, Anthony, Fletcher,
  Hardy, Loop, \& Sun}]{pazder:2006}
Pazder, J.~S., Roberts, S., Abraham, R., {et~al.} 2006, in Ground-based and
  {Airborne} {Instrumentation} for {Astronomy}, Vol. 6269 (International
  Society for Optics and Photonics), 62691X, \dodoi{10.1117/12.672712}

\bibitem[{Pehlivan~Rhodin {et~al.}(2017)Pehlivan~Rhodin, Hartman, Nilsson, \&
  J{\"o}nsson}]{pehlivan-rhodin:2017}
Pehlivan~Rhodin, A., Hartman, H., Nilsson, H., \& J{\"o}nsson, P. 2017,
  Astronomy and Astrophysics, 598, A102, \dodoi{10.1051/0004-6361/201629849}

\bibitem[{Penrose(1955)}]{penrose:1955}
Penrose, R. 1955, Mathematical Proceedings of the Cambridge Philosophical
  Society, 51, 406, \dodoi{10.1017/S0305004100030401}

\bibitem[{P{\'e}rez \& Granger(2007)}]{perez:2007}
P{\'e}rez, F., \& Granger, B.~E. 2007, Computing in Science \& Engineering, 9,
  21, \dodoi{10.1109/MCSE.2007.53}

\bibitem[{Pickering {et~al.}(2001)Pickering, Thorne, \& Perez}]{pickering:2001}
Pickering, J.~C., Thorne, A.~P., \& Perez, R. 2001, The Astrophysical Journal
  Supplement Series, 132, 403, \dodoi{10.1086/318958}

\bibitem[{Pogge {et~al.}(2010)Pogge, Atwood, Brewer, Byard, Derwent, Gonzalez,
  Martini, Mason, O'Brien, Osmer, Pappalardo, Steinbrecher, Teiga, \&
  Zhelem}]{pogge:2010}
Pogge, R.~W., Atwood, B., Brewer, D.~F., {et~al.} 2010, Ground-based and
  Airborne Instrumentation for Astronomy III, 7735, 77350A,
  \dodoi{10.1117/12.857215}

\bibitem[{Pont {et~al.}(2004)Pont, Zinn, Gallart, Hardy, \&
  Winnick}]{pont:2004}
Pont, F., Zinn, R., Gallart, C., Hardy, E., \& Winnick, R. 2004, The
  Astronomical Journal, 127, 840, \dodoi{10.1086/380608}

\bibitem[{Rao(1945)}]{rao:1945}
Rao, C.~R. 1945, Bulletin of the Calcutta Mathematical Society, 37, 81.
\newblock \url{http://bulletin.calmathsoc.org/article.php?ID=B.1945.37.14}

\bibitem[{Robertson(2017)}]{robertson:2017}
Robertson, J.~G. 2017, Publications of the Astronomical Society of Australia,
  34, e035, \dodoi{10.1017/pasa.2017.29}

\bibitem[{Roederer(2019)}]{roederer:2019}
Roederer, I. 2019, Bulletin of the American Astronomical Society, 51, 49.
\newblock \url{https://ui.adsabs.harvard.edu/2019BAAS...51c..49R/abstract}

\bibitem[{Ruchti {et~al.}(2016)Ruchti, Feltzing, Lind, Caffau, Korn, Schnurr,
  Hansen, Koch, Sbordone, \& de~Jong}]{ruchti:2016}
Ruchti, G.~R., Feltzing, S., Lind, K., {et~al.} 2016, Monthly Notices of the
  Royal Astronomical Society, 461, 2174, \dodoi{10.1093/mnras/stw1351}

\bibitem[{Russell(1929)}]{russell:1929}
Russell, H.~N. 1929, The Astrophysical Journal, 70, 11, \dodoi{10.1086/143197}

\bibitem[{Rutledge {et~al.}(1997)Rutledge, Hesser, \& Stetson}]{rutledge:1997}
Rutledge, G.~A., Hesser, J.~E., \& Stetson, P.~B. 1997, Publications of the
  Astronomical Society of the Pacific, 109, 907, \dodoi{10.1086/133959}

\bibitem[{Salvatier {et~al.}(2016)Salvatier, Wiecki, \&
  Fonnesbeck}]{salvatier:2016}
Salvatier, J., Wiecki, T.~V., \& Fonnesbeck, C. 2016, PeerJ Computer Science,
  2, e55, \dodoi{10.7717/peerj-cs.55}

\bibitem[{Shetrone {et~al.}(2003)Shetrone, Venn, Tolstoy, Primas, Hill, \&
  Kaufer}]{shetrone:2003}
Shetrone, M., Venn, K.~A., Tolstoy, E., {et~al.} 2003, The Astronomical
  Journal, 125, 684, \dodoi{10.1086/345966}

\bibitem[{Shetrone {et~al.}(2015)Shetrone, Bizyaev, Lawler, Allende~Prieto,
  Johnson, Smith, Cunha, Holtzman, Garc{\'\i}a~P{\'e}rez, M{\'e}sz{\'a}ros,
  Sobeck, Zamora, Garc{\'\i}a-Hern{\'a}ndez, Souto, Chojnowski, Koesterke,
  Majewski, \& Zasowski}]{shetrone:2015}
Shetrone, M., Bizyaev, D., Lawler, J.~E., {et~al.} 2015, The Astrophysical
  Journal Supplement Series, 221, 24, \dodoi{10.1088/0067-0049/221/2/24}

\bibitem[{Shetrone {et~al.}(1998)Shetrone, Bolte, \& Stetson}]{shetrone:1998}
Shetrone, M.~D., Bolte, M., \& Stetson, P.~B. 1998, The Astronomical Journal,
  115, 1888, \dodoi{10.1086/300341}

\bibitem[{Shetrone {et~al.}(2001)Shetrone, C{\^o}t{\'e}, \&
  Sargent}]{shetrone:2001}
Shetrone, M.~D., C{\^o}t{\'e}, P., \& Sargent, W. L.~W. 2001, The Astrophysical
  Journal, 548, 592, \dodoi{10.1086/319022}

\bibitem[{Shetrone {et~al.}(2009)Shetrone, Siegel, Cook, \&
  Bosler}]{shetrone:2009}
Shetrone, M.~D., Siegel, M.~H., Cook, D.~O., \& Bosler, T. 2009, The
  Astronomical Journal, 137, 62, \dodoi{10.1088/0004-6256/137/1/62}

\bibitem[{Simon \& Geha(2007)}]{simon:2007}
Simon, J.~D., \& Geha, M. 2007, The Astrophysical Journal, 670, 313,
  \dodoi{10.1086/521816}

\bibitem[{Simon {et~al.}(2015)Simon, Drlica-Wagner, Li, Nord, Geha, Bechtol,
  Balbinot, Buckley-Geer, Lin, Marshall, Santiago, Strigari, Wang, Wechsler,
  Yanny, Abbott, Bauer, Bernstein, Bertin, Brooks, Burke, Capozzi,
  Carnero~Rosell, Carrasco~Kind, D'Andrea, da~Costa, DePoy, Desai, Diehl,
  Dodelson, Cunha, Estrada, Evrard, Fausti~Neto, Fernandez, Finley, Flaugher,
  Frieman, Gaztanaga, Gerdes, Gruen, Gruendl, Honscheid, James, Kent, Kuehn,
  Kuropatkin, Lahav, Maia, March, Martini, Miller, Miquel, Ogando, Romer,
  Roodman, Rykoff, Sako, Sanchez, Schubnell, Sevilla, Smith, Soares-Santos,
  Sobreira, Suchyta, Swanson, Tarle, Thaler, Tucker, Vikram, Walker, Wester, \&
  {DES Collaboration}}]{simon:2015}
Simon, J.~D., Drlica-Wagner, A., Li, T.~S., {et~al.} 2015, The Astrophysical
  Journal, 808, 95, \dodoi{10.1088/0004-637X/808/1/95}

\bibitem[{Simon {et~al.}(2017)Simon, Li, Drlica-Wagner, Bechtol, Marshall,
  James, Wang, Strigari, Balbinot, Kuehn, Walker, Abbott, Allam, Annis,
  Benoit-L{\'e}vy, Brooks, Buckley-Geer, Burke, Carnero~Rosell, Carrasco~Kind,
  Carretero, Cunha, D'Andrea, da~Costa, DePoy, Desai, Doel, Fernandez,
  Flaugher, Frieman, Garc{\'\i}a-Bellido, Gaztanaga, Goldstein, Gruen,
  Gutierrez, Kuropatkin, Maia, Martini, Menanteau, Miller, Miquel, Neilsen,
  Nord, Ogando, Plazas, Romer, Rykoff, Sanchez, Santiago, Scarpine, Schubnell,
  Sevilla-Noarbe, Smith, Sobreira, Suchyta, Swanson, Tarle, Whiteway, Yanny, \&
  {DES Collaboration}}]{simon:2017}
Simon, J.~D., Li, T.~S., Drlica-Wagner, A., {et~al.} 2017, The Astrophysical
  Journal, 838, 11, \dodoi{10.3847/1538-4357/aa5be7}

\bibitem[{Slater {et~al.}(2015)Slater, Bell, Martin, Tollerud, \&
  Ho}]{slater:2015}
Slater, C.~T., Bell, E.~F., Martin, N.~F., Tollerud, E.~J., \& Ho, N. 2015, The
  Astrophysical Journal, 806, 230, \dodoi{10.1088/0004-637X/806/2/230}

\bibitem[{Sneden(1973)}]{sneden:1973}
Sneden, C. 1973, The Astrophysical Journal, 184, 839, \dodoi{10.1086/152374}

\bibitem[{Sneden(1974)}]{sneden:1974}
---. 1974, The Astrophysical Journal, 189, 493, \dodoi{10.1086/152828}

\bibitem[{Sneden \& Parthasarathy(1983)}]{sneden:1983}
Sneden, C., \& Parthasarathy, M. 1983, The Astrophysical Journal, 267, 757,
  \dodoi{10.1086/160913}

\bibitem[{Sneden {et~al.}(2003)Sneden, Cowan, Lawler, Ivans, Burles, Beers,
  Primas, Hill, Truran, Fuller, Pfeiffer, \& Kratz}]{sneden:2003}
Sneden, C., Cowan, J.~J., Lawler, J.~E., {et~al.} 2003, The Astrophysical
  Journal, 591, 936, \dodoi{10.1086/375491}

\bibitem[{Spencer {et~al.}(2017)Spencer, Mateo, Walker, \&
  Olszewski}]{spencer:2017}
Spencer, M.~E., Mateo, M., Walker, M.~G., \& Olszewski, E.~W. 2017, The
  Astrophysical Journal, 836, 202, \dodoi{10.3847/1538-4357/836/2/202}

\bibitem[{Spite {et~al.}(2018)Spite, Spite, Fran{\c c}ois, Bonifacio, Caffau,
  \& Salvadori}]{spite:2018}
Spite, M., Spite, F., Fran{\c c}ois, P., {et~al.} 2018, Astronomy and
  Astrophysics, 617, A56, \dodoi{10.1051/0004-6361/201833548}

\bibitem[{Starkenburg {et~al.}(2010)Starkenburg, Hill, Tolstoy,
  Gonz{\'a}lez~Hern{\'a}ndez, Irwin, Helmi, Battaglia, Jablonka, Tafelmeyer,
  Shetrone, Venn, \& de~Boer}]{starkenburg:2010}
Starkenburg, E., Hill, V., Tolstoy, E., {et~al.} 2010, Astronomy and
  Astrophysics, 513, A34, \dodoi{10.1051/0004-6361/200913759}

\bibitem[{Starkenburg {et~al.}(2013)Starkenburg, Hill, Tolstoy, Fran{\c c}ois,
  Irwin, Boschman, Venn, de~Boer, Lemasle, Jablonka, Battaglia, Groot, \&
  Kaper}]{starkenburg:2013}
---. 2013, Astronomy and Astrophysics, 549, A88,
  \dodoi{10.1051/0004-6361/201220349}

\bibitem[{Steinmetz {et~al.}(2006)Steinmetz, Zwitter, Siebert, Watson, Freeman,
  Munari, Campbell, Williams, Seabroke, Wyse, Parker, Bienaym{\'e}, Roeser,
  Gibson, Gilmore, Grebel, Helmi, Navarro, Burton, Cass, Dawe, Fiegert,
  Hartley, Russell, Saunders, Enke, Bailin, Binney, Bland-Hawthorn, Boeche,
  Dehnen, Eisenstein, Evans, Fiorucci, Fulbright, Gerhard, Jauregi, Kelz,
  Mijovi{\'c}, Minchev, Parmentier, Pe{\~n}arrubia, Quillen, Read, Ruchti,
  Scholz, Siviero, Smith, Sordo, Veltz, Vidrih, von Berlepsch, Boyle, \&
  Schilbach}]{steinmetz:2006}
Steinmetz, M., Zwitter, T., Siebert, A., {et~al.} 2006, The Astronomical
  Journal, 132, 1645, \dodoi{10.1086/506564}

\bibitem[{Str{\"o}mgren(1940)}]{stromgren:1940}
Str{\"o}mgren, B. 1940, in Festschrift f{\"u}r Elis Str{\"o}mgren, 218.
\newblock
  \url{https://ui.adsabs.harvard.edu/search/q=author%3A%22%5Estromgren%22%20year%3A1940&sort=date%20desc%2C%20bibcode%20desc&p_=0}

\bibitem[{Suda {et~al.}(2017)Suda, Hidaka, Aoki, Katsuta, Yamada, Fujimoto,
  Ohtani, Masuyama, Noda, \& Wada}]{suda:2017}
Suda, T., Hidaka, J., Aoki, W., {et~al.} 2017, Publications of the Astronomical
  Society of Japan, 69, 76, \dodoi{10.1093/pasj/psx059}

\bibitem[{Suntzeff(1981)}]{suntzeff:1981}
Suntzeff, N.~B. 1981, The Astrophysical Journal Supplement Series, 47, 1,
  \dodoi{10.1086/190750}

\bibitem[{Suntzeff {et~al.}(1993)Suntzeff, Mateo, Terndrup, Olszewski, Geisler,
  \& Weller}]{suntzeff:1993}
Suntzeff, N.~B., Mateo, M., Terndrup, D.~M., {et~al.} 1993, The Astrophysical
  Journal, 418, 208, \dodoi{10.1086/173383}

\bibitem[{Swan {et~al.}(2016)Swan, Cole, Tolstoy, \& Irwin}]{swan:2016}
Swan, J., Cole, A.~A., Tolstoy, E., \& Irwin, M.~J. 2016, Monthly Notices of
  the Royal Astronomical Society, 456, 4315, \dodoi{10.1093/mnras/stv2774}

\bibitem[{Szentgyorgyi {et~al.}(2011)Szentgyorgyi, Furesz, Cheimets, Conroy,
  Eng, Fabricant, Fata, Gauron, Geary, McLeod, Zajac, Amato, Bergner, Caldwell,
  Dupree, Goddard, Johnston, Meibom, Mink, Pieri, Roll, Tokarz, Wyatt, Epps,
  Hartmann, \& Meszaros}]{szentgyorgyi:2011}
Szentgyorgyi, A., Furesz, G., Cheimets, P., {et~al.} 2011, Publications of the
  Astronomical Society of the Pacific, 123, 1188, \dodoi{10.1086/662209}

\bibitem[{Szentgyorgyi {et~al.}(2016)Szentgyorgyi, Baldwin, Barnes, Bean,
  Ben-Ami, Brennan, Budynkiewicz, Chun, Conroy, Crane, Epps, Evans, Evans,
  Foster, Frebel, Gauron, Guzm{\'a}n, Hare, Jang, Jang, Jordan, Kim, Kim,
  Oliveira, Lopez-Morales, McCracken, McMuldroch, Miller, Mueller, Oh,
  Onyuksel, Ordway, Park, Park, Park, Paxson, Phillips, Plummer, Podgorski,
  Seifahrt, Stark, Steiner, Uomoto, Walsworth, \& Yu}]{szentgyorgyi:2016}
Szentgyorgyi, A., Baldwin, D., Barnes, S., {et~al.} 2016, in Ground-based and
  {Airborne} {Instrumentation} for {Astronomy} {VI}, Vol. 9908 (International
  Society for Optics and Photonics), 990822, \dodoi{10.1117/12.2233506}

\bibitem[{Takada {et~al.}(2014)Takada, Ellis, Chiba, Greene, Aihara, Arimoto,
  Bundy, Cohen, Dor{\'e}, Graves, Gunn, Heckman, Hirata, Ho, Kneib,
  Le~F{\`e}vre, Lin, More, Murayama, Nagao, Ouchi, Seiffert, Silverman,
  Sodr{\'e}, Spergel, Strauss, Sugai, Suto, Takami, \& Wyse}]{takada:2014}
Takada, M., Ellis, R.~S., Chiba, M., {et~al.} 2014, Publications of the
  Astronomical Society of Japan, 66, R1, \dodoi{10.1093/pasj/pst019}

\bibitem[{Tamura {et~al.}(2018)Tamura, Takato, Shimono, Moritani, Yabe,
  Ishizuka, Kamata, Ueda, Aghazarian, Arnouts, Barkhouser, Balard, Barette,
  Belhadi, Burnham, Caplar, Carr, Chabaud, Chang, Chen, Chou, Chu, Cohen,
  Almeida, Oliveira, Oliveira, Dekany, Dohlen, Santos, Santos, Ellis,
  Fabricius, Ferreira, Furusawa, Garcia-Carpio, Golebiowski, Gross, Gunn,
  Hammond, Harding, Hart, Heckman, Ho, Hope, Hover, Hsu, Hu, Huang, Jamal,
  Jaquet, Jeschke, Jing, Kado-Fong, Karr, Kimura, King, Koike, Komatsu, Brun,
  F{\`e}vre, Fur, Mignant, Ling, Loomis, Lupton, Madec, Mao, Marchesini,
  Marrara, Medvedev, Mineo, Minowa, Murayama, Murray, Ohyama, Onodera,
  Orndorff, Pascal, Peebles, Pernot, Pourcelot, Reiley, Reinecke, Roberts,
  Rosa, Rousselle, Schmitt, Schwochert, Seiffert, Siddiqui, Smee, Jr,
  Steinkraus, Strauss, Surace, Tait, Takada, Tamura, Tanaka, Tanaka, Thakar,
  Jr, Vibert, Wang, Wang, Wen, Werner, Yamada, Yan, Yasuda, Yoshida, \&
  Yoshida}]{tamura:2018}
Tamura, N., Takato, N., Shimono, A., {et~al.} 2018, in Ground-based and
  {Airborne} {Instrumentation} for {Astronomy} {VII}, Vol. 10702 (International
  Society for Optics and Photonics), 107021C, \dodoi{10.1117/12.2311871}

\bibitem[{Theler {et~al.}(2019)Theler, Jablonka, Lardo, North, Irwin,
  Battaglia, Hill, Tolstoy, Venn, Helmi, Kaufer, Lucchesi, Primas, \&
  Shetrone}]{theler:2019}
Theler, R., Jablonka, P., Lardo, C., {et~al.} 2019, arXiv e-prints,
  arXiv:1911.08627.
\newblock \url{https://ui.adsabs.harvard.edu/abs/2019arXiv191108627T/abstract}

\bibitem[{Ting {et~al.}(2016)Ting, Conroy, \& Rix}]{ting:2016}
Ting, Y.-S., Conroy, C., \& Rix, H.-W. 2016, The Astrophysical Journal, 826,
  83, \dodoi{10.3847/0004-637X/826/1/83}

\bibitem[{Ting {et~al.}(2018)Ting, Conroy, Rix, \& Asplund}]{ting:2018}
Ting, Y.-S., Conroy, C., Rix, H.-W., \& Asplund, M. 2018, The Astrophysical
  Journal, 860, 159, \dodoi{10.3847/1538-4357/aac6c9}

\bibitem[{Ting {et~al.}(2017{\natexlab{a}})Ting, Conroy, Rix, \&
  Cargile}]{ting:2017}
Ting, Y.-S., Conroy, C., Rix, H.-W., \& Cargile, P. 2017{\natexlab{a}}, The
  Astrophysical Journal, 843, 32, \dodoi{10.3847/1538-4357/aa7688}

\bibitem[{Ting {et~al.}(2019)Ting, Conroy, Rix, \& Cargile}]{ting:2019}
---. 2019, The Astrophysical Journal, 879, 69, \dodoi{10.3847/1538-4357/ab2331}

\bibitem[{Ting {et~al.}(2017{\natexlab{b}})Ting, Rix, Conroy, Ho, \&
  Lin}]{ting:2017b}
Ting, Y.-S., Rix, H.-W., Conroy, C., Ho, A. Y.~Q., \& Lin, J.
  2017{\natexlab{b}}, The Astrophysical Journal Letters, 849, L9,
  \dodoi{10.3847/2041-8213/aa921c}

\bibitem[{Tinsley(1980)}]{tinsley:1980}
Tinsley, B.~M. 1980, Fundamentals of Cosmic Physics, 5, 287.
\newblock \url{http://adsabs.harvard.edu/abs/1980FCPh....5..287T}

\bibitem[{Tolstoy {et~al.}(2009)Tolstoy, Hill, \& Tosi}]{tolstoy:2009}
Tolstoy, E., Hill, V., \& Tosi, M. 2009, Annual Review of Astronomy and
  Astrophysics, 47, 371, \dodoi{10.1146/annurev-astro-082708-101650}

\bibitem[{Tolstoy {et~al.}(2003)Tolstoy, Venn, Shetrone, Primas, Hill, Kaufer,
  \& Szeifert}]{tolstoy:2003}
Tolstoy, E., Venn, K.~A., Shetrone, M., {et~al.} 2003, The Astronomical
  Journal, 125, 707, \dodoi{10.1086/345967}

\bibitem[{Tolstoy {et~al.}(2004)Tolstoy, Irwin, Helmi, Battaglia, Jablonka,
  Hill, Venn, Shetrone, Letarte, Cole, Primas, Francois, Arimoto, Sadakane,
  Kaufer, Szeifert, \& Abel}]{tolstoy:2004}
Tolstoy, E., Irwin, M.~J., Helmi, A., {et~al.} 2004, The Astrophysical Journal
  Letters, 617, L119, \dodoi{10.1086/427388}

\bibitem[{Uns{\"o}ld(1938)}]{unsold:1938}
Uns{\"o}ld, A. 1938, Berlin, Verlag von Julius Springer, 1938.
\newblock \url{http://adsabs.harvard.edu/abs/1938psmb.book.....U}

\bibitem[{Uns{\"o}ld(1942)}]{unsold:1942}
---. 1942, Zeitschrift fur Astrophysik, 21, 22.
\newblock \url{http://adsabs.harvard.edu/abs/1942ZA.....21...22U}

\bibitem[{Vargas {et~al.}(2013)Vargas, Geha, Kirby, \& Simon}]{vargas:2013}
Vargas, L.~C., Geha, M., Kirby, E.~N., \& Simon, J.~D. 2013, The Astrophysical
  Journal, 767, 134, \dodoi{10.1088/0004-637X/767/2/134}

\bibitem[{Vargas {et~al.}(2014{\natexlab{a}})Vargas, Geha, \&
  Tollerud}]{vargas:2014a}
Vargas, L.~C., Geha, M.~C., \& Tollerud, E.~J. 2014{\natexlab{a}}, The
  Astrophysical Journal, 790, 73, \dodoi{10.1088/0004-637X/790/1/73}

\bibitem[{Vargas {et~al.}(2014{\natexlab{b}})Vargas, Gilbert, Geha, Tollerud,
  Kirby, \& Guhathakurta}]{vargas:2014b}
Vargas, L.~C., Gilbert, K.~M., Geha, M., {et~al.} 2014{\natexlab{b}}, The
  Astrophysical Journal, 797, L2, \dodoi{10.1088/2041-8205/797/1/L2}

\bibitem[{Venn {et~al.}(2004)Venn, Irwin, Shetrone, Tout, Hill, \&
  Tolstoy}]{venn:2004}
Venn, K.~A., Irwin, M., Shetrone, M.~D., {et~al.} 2004, The Astronomical
  Journal, 128, 1177, \dodoi{10.1086/422734}

\bibitem[{Venn {et~al.}(2017)Venn, Starkenburg, Malo, Martin, \&
  Laevens}]{venn:2017}
Venn, K.~A., Starkenburg, E., Malo, L., Martin, N., \& Laevens, B. P.~M. 2017,
  Monthly Notices of the Royal Astronomical Society, 466, 3741,
  \dodoi{10.1093/mnras/stw3198}

\bibitem[{Venn {et~al.}(2001)Venn, Lennon, Kaufer, McCarthy, Przybilla,
  Kudritzki, Lemke, Skillman, \& Smartt}]{venn:2001}
Venn, K.~A., Lennon, D.~J., Kaufer, A., {et~al.} 2001, The Astrophysical
  Journal, 547, 765, \dodoi{10.1086/318424}

\bibitem[{Vernet {et~al.}(2011)Vernet, Dekker, D'Odorico, Kaper, Kjaergaard,
  Hammer, Randich, Zerbi, Groot, Hjorth, Guinouard, Navarro, Adolfse, Albers,
  Amans, Andersen, Andersen, Binetruy, Bristow, Castillo, Chemla, Christensen,
  Conconi, Conzelmann, Dam, de~Caprio, de~Ugarte~Postigo, Delabre,
  di~Marcantonio, Downing, Elswijk, Finger, Fischer, Flores, Fran{\c c}ois,
  Goldoni, Guglielmi, Haigron, Hanenburg, Hendriks, Horrobin, Horville, Jessen,
  Kerber, Kern, Kiekebusch, Kleszcz, Klougart, Kragt, Larsen, Lizon, Lucuix,
  Mainieri, Manuputy, Martayan, Mason, Mazzoleni, Michaelsen, Modigliani,
  Moehler, M{\o}ller, Norup~S{\o}rensen, N{\o}rregaard, P{\'e}roux, Patat,
  Pena, Pragt, Reinero, Rigal, Riva, Roelfsema, Royer, Sacco, Santin,
  Schoenmaker, Spano, Sweers, Ter~Horst, Tintori, Tromp, van Dael, van~der
  Vliet, Venema, Vidali, Vinther, Vola, Winters, Wistisen, Wulterkens, \&
  Zacchei}]{vernet:2011}
Vernet, J., Dekker, H., D'Odorico, S., {et~al.} 2011, Astronomy and
  Astrophysics, 536, A105, \dodoi{10.1051/0004-6361/201117752}

\bibitem[{Virtanen {et~al.}(2019)Virtanen, Gommers, Oliphant, Haberland, Reddy,
  Cournapeau, Burovski, Peterson, Weckesser, Bright, van~der Walt, Brett,
  Wilson, Millman, Mayorov, Nelson, Jones, Kern, Larson, Carey, Polat, Feng,
  Moore, VanderPlas, Laxalde, Perktold, Cimrman, Henriksen, Quintero, Harris,
  Archibald, Ribeiro, Pedregosa, van Mulbregt, \& Contributors}]{virtanen:2019}
Virtanen, P., Gommers, R., Oliphant, T.~E., {et~al.} 2019, arXiv:1907.10121
  [physics].
\newblock \url{http://arxiv.org/abs/1907.10121}

\bibitem[{Voggel {et~al.}(2016)Voggel, Hilker, Baumgardt, Collins, Grebel,
  Husemann, Richtler, \& Frank}]{voggel:2016}
Voggel, K., Hilker, M., Baumgardt, H., {et~al.} 2016, Monthly Notices of the
  Royal Astronomical Society, 460, 3384, \dodoi{10.1093/mnras/stw1132}

\bibitem[{Vogt {et~al.}(1994)Vogt, Allen, Bigelow, Bresee, Brown, Cantrall,
  Conrad, Couture, Delaney, Epps, Hilyard, Hilyard, Horn, Jern, Kanto, Keane,
  Kibrick, Lewis, Osborne, Pardeilhan, Pfister, Ricketts, Robinson, Stover,
  Tucker, Ward, \& Wei}]{vogt:1994}
Vogt, S.~S., Allen, S.~L., Bigelow, B.~C., {et~al.} 1994, Instrumentation in
  Astronomy VIII, 2198, 362, \dodoi{10.1117/12.176725}

\bibitem[{Walker {et~al.}(2009{\natexlab{a}})Walker, Belokurov, Evans, Irwin,
  Mateo, Olszewski, \& Gilmore}]{walker:2009b}
Walker, M.~G., Belokurov, V., Evans, N.~W., {et~al.} 2009{\natexlab{a}}, The
  Astrophysical Journal Letters, 694, L144,
  \dodoi{10.1088/0004-637X/694/2/L144}

\bibitem[{Walker {et~al.}(2009{\natexlab{b}})Walker, Mateo, \&
  Olszewski}]{walker:2009a}
Walker, M.~G., Mateo, M., \& Olszewski, E.~W. 2009{\natexlab{b}}, The
  Astronomical Journal, 137, 3100, \dodoi{10.1088/0004-6256/137/2/3100}

\bibitem[{Walker {et~al.}(2015{\natexlab{a}})Walker, Mateo, Olszewski, Bailey,
  Koposov, Belokurov, \& Evans}]{walker:2015_m2fs}
Walker, M.~G., Mateo, M., Olszewski, E.~W., {et~al.} 2015{\natexlab{a}}, The
  Astrophysical Journal, 808, 108, \dodoi{10.1088/0004-637X/808/2/108}

\bibitem[{Walker {et~al.}(2007)Walker, Mateo, Olszewski, Gnedin, Wang, Sen, \&
  Woodroofe}]{walker:2007}
---. 2007, The Astrophysical Journal, 667, L53, \dodoi{10.1086/521998}

\bibitem[{Walker {et~al.}(2015{\natexlab{b}})Walker, Olszewski, \&
  Mateo}]{walker:2015_hectochelle}
Walker, M.~G., Olszewski, E.~W., \& Mateo, M. 2015{\natexlab{b}}, Monthly
  Notices of the Royal Astronomical Society, 448, 2717,
  \dodoi{10.1093/mnras/stv099}

\bibitem[{Walker {et~al.}(2016)Walker, Mateo, Olszewski, Koposov, Belokurov,
  Jethwa, Nidever, Bonnivard, Bailey, Bell, \& Loebman}]{walker:2016}
Walker, M.~G., Mateo, M., Olszewski, E.~W., {et~al.} 2016, The Astrophysical
  Journal, 819, 53, \dodoi{10.3847/0004-637X/819/1/53}

\bibitem[{Wallerstein {et~al.}(1987)Wallerstein, Leep, \&
  Oke}]{wallerstein:1987}
Wallerstein, G., Leep, E.~M., \& Oke, J.~B. 1987, The Astronomical Journal, 93,
  1137, \dodoi{10.1086/114396}

\bibitem[{Walt {et~al.}(2011)Walt, Colbert, \& Varoquaux}]{walt:2011}
Walt, S. v.~d., Colbert, S.~C., \& Varoquaux, G. 2011, Computing in Science \&
  Engineering, 13, 22, \dodoi{10.1109/MCSE.2011.37}

\bibitem[{Wang(2010)}]{wang:2010}
Wang, Y. 2010, Modern Physics Letters A, 25, 3093,
  \dodoi{10.1142/S0217732310034316}

\bibitem[{Wright(1948)}]{wright:1948}
Wright, K.~O. 1948, Publications of the Dominion Astrophysical Observatory
  Victoria, 8.
\newblock \url{http://adsabs.harvard.edu/abs/1948PDAO....8....1W}

\bibitem[{Xiang {et~al.}(2019)Xiang, Ting, Rix, Sandford, Buder, Lind, Liu,
  Shi, \& Zhang}]{xiang:2019}
Xiang, M., Ting, Y.-S., Rix, H.-W., {et~al.} 2019, arXiv e-prints,
  arXiv:1908.09727.
\newblock \url{https://ui.adsabs.harvard.edu/abs/2019arXiv190809727X/abstract}

\bibitem[{Yanny {et~al.}(2009)Yanny, Rockosi, Newberg, Knapp, Adelman-McCarthy,
  Alcorn, Allam, Allende~Prieto, An, Anderson, Anderson, Bailer-Jones, Bastian,
  Beers, Bell, Belokurov, Bizyaev, Blythe, Bochanski, Boroski, Brinchmann,
  Brinkmann, Brewington, Carey, Cudworth, Evans, Evans, Gates, G{\"a}nsicke,
  Gillespie, Gilmore, Nebot Gomez-Moran, Grebel, Greenwell, Gunn, Jordan,
  Jordan, Harding, Harris, Hendry, Holder, Ivans, Ivezi{\v c}, Jester, Johnson,
  Kent, Kleinman, Kniazev, Krzesinski, Kron, Kuropatkin, Lebedeva, Lee,
  French~Leger, L{\'e}pine, Levine, Lin, Long, Loomis, Lupton, Malanushenko,
  Malanushenko, Margon, Martinez-Delgado, McGehee, Monet, Morrison, Munn,
  Neilsen, Nitta, Norris, Oravetz, Owen, Padmanabhan, Pan, Peterson, Pier,
  Platson, Re~Fiorentin, Richards, Rix, Schlegel, Schneider, Schreiber,
  Schwope, Sibley, Simmons, Snedden, Allyn~Smith, Stark, Stauffer, Steinmetz,
  Stoughton, SubbaRao, Szalay, Szkody, Thakar, Sivarani, Tucker, Uomoto,
  Vanden~Berk, Vidrih, Wadadekar, Watters, Wilhelm, Wyse, Yarger, \&
  Zucker}]{yanny:2009}
Yanny, B., Rockosi, C., Newberg, H.~J., {et~al.} 2009, The Astronomical
  Journal, 137, 4377, \dodoi{10.1088/0004-6256/137/5/4377}

\bibitem[{Yoon {et~al.}(2019)Yoon, Whitten, Beers, Lee, \& Placco}]{yoon:2019}
Yoon, J., Whitten, D., Beers, T., Lee, Y., \& Placco, V. 2019, arXiv e-prints,
  arXiv:1910.10038.
\newblock \url{https://arxiv.org/abs/1910.10038}

\end{thebibliography}
\bibliographystyle{aasjournal}

\appendix


\section{Biased CRLB} \label{app:biased}
\edit1{\added{
A fundamental assumption adopted in this work is that of perfect models that accurately reproduce observed stellar spectra. However, as in most of astrophysics and as we discussed in \S\ref{sec:caveats}, this is not the case in practice. Many spectral features are poorly modeled due to 3D and non-LTE effects, miscalibrated oscillator strengths and transition wavelengths, and imperfect reductions. While these systematic errors primarily effect the accuracy of abundance measurements, they also invalidate our assumption that the MLE, $\hat{\theta}$, is an unbiased estimator of the true stellar labels and may also change the expected precision of the abundance measurements.
}}

\edit1{\added{
If the bias of a particular spectral model is known, this can be included in the prediction of stellar label precision using the ``biased" or ``misspecified" CRLB:
\begin{equation}
\sigma_{\text{biased},\alpha} = \sqrt{([I+D]F^{-1}[I+D]^{T})_{\alpha\alpha}},
\label{eq:CRLB_biased}
\end{equation}
where $F$ is the FIM as defined in Equation \ref{eq:FIM_grad}, $I$ is the identity matrix, and $D$ is the bias gradient matrix:
\begin{equation}
D=\left[\frac{\partial b}{\partial \theta_\alpha}\right]_{\hat{\theta}},
\end{equation}
where $b$ is the bias of your labels given by
\begin{equation}
b(\hat{\theta}) = E(\theta)-\hat{\theta}.
\end{equation}
Because evaluating the bias is both model and instrument dependent, it is beyond the scope of this paper. However, we note that in the simple case of a uniform bias (i.e., measuring the surface temperature of all stars to be 100 K too hot), the normal and biased CRLB are the same. In the more complicated (and realistic) case that the bias is dependent on the stellar labels (i.e., the surface temperature is measured to be 100 K too hot in giant stars but 100 K too cold in dwarf stars) the biased CRLB will differ from normal CRLB. Depending on the direction and amplitude of the bias, this may result in either better or worse precision than in the unbiased case.
}}

\edit1{\added{The main challenge in practice is not that the CRLBs cannot be used in the presence of bias, but that the bias needs to be known \textit{a priori} for the CRLB---or any forecast of precision---to be computed accurately.}}

\section{CRLB Calculation} \label{app:crlb_calc}
For instruments whose observations span noncontiguous wavelength ranges, the gradient spectra (and 1-D S/N arrays) for each of the wavelength ranges are concatenated together. This technique can also be used to combine observations from potentially complimentary instruments or observing campaigns, though we do not consider any here. All combinations of wavelength ranges examined in this work are forced to be non-overlapping to avoid more complicated treatment of the spectral covariance matrix. This is done even though it means ignoring the additional information that an overlapping region of spectrum might provide. 

From this point, the calculation of the CRLBs from the gradient spectra and spectral covariance is simply a matter of matrix multiplication and inversion. However, because the gradient spectrum for some labels is much larger than for others (i.e., Fe compares to Nb), the FIM may be near-singular and thus unstable to inversion. We take several steps to avoid matrix inversion problems and calculate robust CRLBs:

\begin{enumerate}[(i)]
    \item We divide the spectral gradient with respect to $T_\text{eff}$ by 100.
    \item If $F_{\alpha\alpha}<1$ for any label, $\alpha$, we set $F_{\alpha j} = F_{i \alpha} = 0$ and $F_{\alpha\alpha}=10^{-6}$
    \item We compute the Moore-Penrose pseudo-inverse of the FIM \citep{moore:1920, penrose:1955}.
\end{enumerate}

The purpose of (i) is to place $df / dT_\text{eff}$ on roughly the same scale as  $df / d\text{[X/H]}$. This keeps the eigenvalue of the FIM with respect to $T_\text{eff}$ from dwarfing those of the other labels. As a result, the CRLB for $T_\text{eff}$ is in units of 100 K. Step (ii) avoids zero eigenvalues for labels with very little information in the spectrum. It also removes the covariance of these labels with all other labels, which would otherwise make the matrix near-singular. This results in a CRLB of $\sim$$10^3$ for these labels, which can safely be ignored. Finally, by calculating the pseudo-inverse instead of the true inverse of the FIM in (iii), we avoid numerical instabilities when attempting to invert near-singular matrices.

When including prior information into our CRLB calculations, we add the inverse variance of these priors to the relevant diagonal entries of the FIM as outlined in Equation \ref{eq:CRLB_bayes} before inverting the FIM as before. To be rigorously Bayesian, we ought to state that we do this for all labels, including those with uninformative priors with zero inverse variance.

\section{Wavelength Sampling and Pixel Correlations} \label{app:sampling}
\edit1{\added{
To illustrate the impact of assuming the independence of all pixels on the CRLB, we consider the simple case that each resolution element is sampled by 3 pixels and all adjacent pixels are correlated by some fraction, $c$. In such a scenario, the flux covariance is no longer the diagonal matrix presented in Equation \ref{eq:covar_diag}, but now has diagonal-adjacent terms equal to $c(\sigma)^{2}$, where $\sigma=(\text{S/N})^{-1}$ at each pixel:
\begin{equation}\label{eq:covar_corr}
    \Sigma = \left[
    \begin{array}{ccccc}
    \sigma^2(\lambda_{1}) & c\sigma^2(\lambda_{1})& & & \\
    c\sigma^2(\lambda_{2}) & \sigma^2(\lambda_{2})& c\sigma^2(\lambda_{2}) & & \\
     & & \ddots &  & \\
     & & c\sigma^2(\lambda_{N-1}) & \sigma^2(\lambda_{N-1}) & c\sigma^2(\lambda_{N-1})\\
     & & & c\sigma^2(\lambda_{1}) & \sigma^2(\lambda_{N})\\
    \end{array}\right].
\end{equation}
}}

\begin{figure*}[ht!]
	\includegraphics[width=\textwidth]{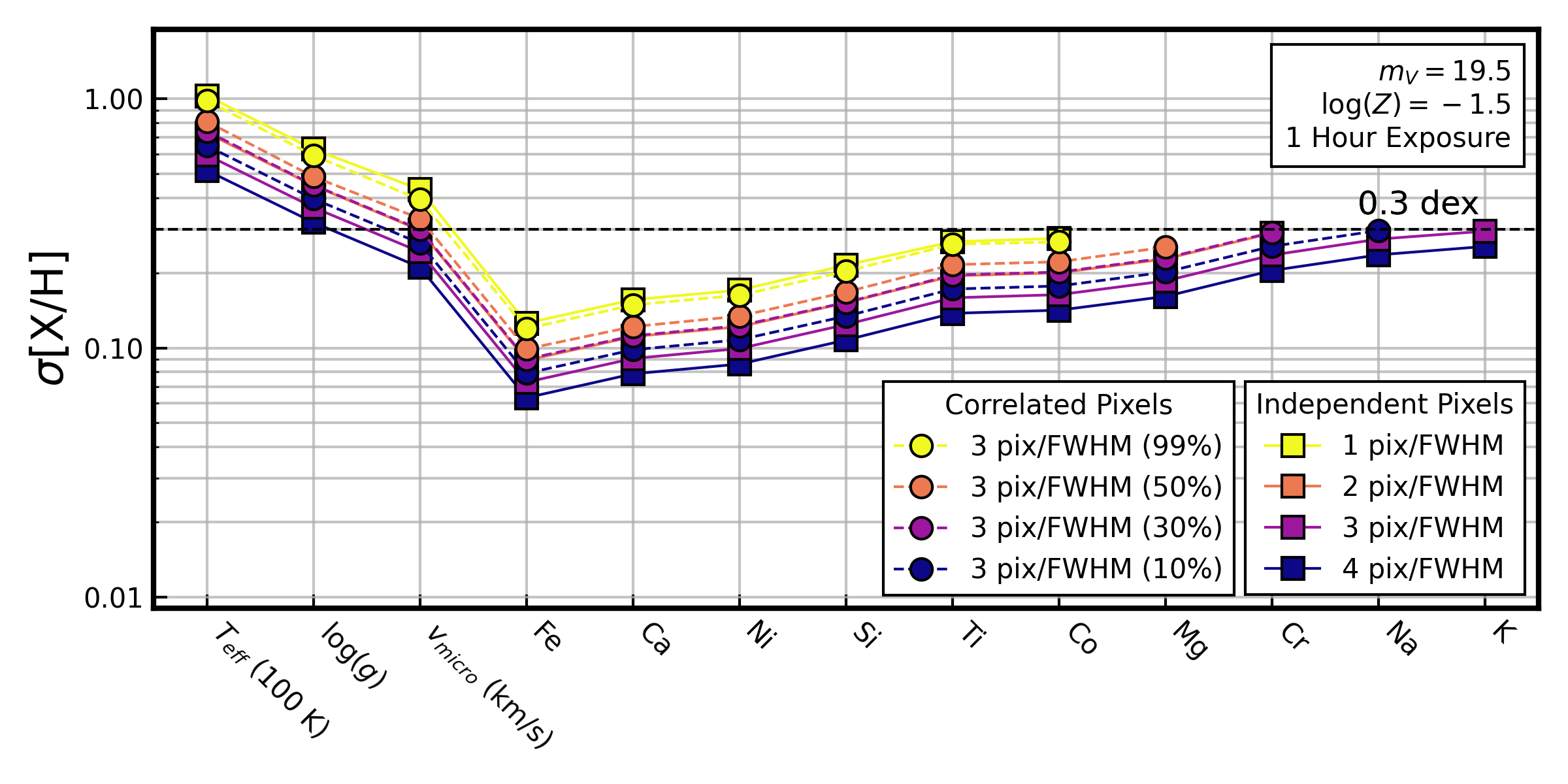}
    \caption{D1200G CRLBs for a 1 hour exposure of a $\log(Z)=-1.5$, $m_{V}=19.5$ RGB star assuming various wavelength samplings and pixel-to-pixel correlations. CRLBs assuming uncorrelated pixels but varying wavelength sampling are represented by squares and solid lines. CRLBs assuming 3 pixels/FWHM but varying degrees of correlation between adjacent pixels are represented by circles and dashed lines. For completely independent pixels, the CRLBs scale proportionally to $n^{-1/2}$, where $n$ is the number of pixels per resolution element.)} 
    \label{fig:crlb_sampling}
\end{figure*}

\edit1{\added{
Figure \ref{fig:crlb_sampling} shows the impact of assuming adjacent pixels are 10\%, 30\%, 50\%, and  99\% correlated on the CRLB as applied to our fiducial D1200G observation. For comparison, we also include the CRLBs assuming 1, 2, 3, and 4 completely uncorrelated pixels per resolution element. As expected under the assumption of independent pixels, the CRLBs scale as $n^{-1/2}$, where $n$ is the number of pixels per resolution element.
}}

\edit1{\added{
When adjacent pixels have correlations of 10\%, 30\%, and 50\%, the CRLBs are roughly 8\%, 23\%, and 35\% larger respectively than in the uncorrelated case. These CRLBs are equivalent to calculating the CRLB assuming $n=2.6$, 2.0, and 1.6 independent pixels per resolution element respectively. In the extreme case that all three pixels are nearly 100\% correlated with each other, there is effectively only one independent pixel per resolution element and the CRLB approaches the $n=1$ pixel/FWHM CRLB or $\sqrt{3}$ times what is found with uncorrelated pixels. 
}}

\edit1{\added{
A more realistic treatment of pixel correlation would require adopting a kernel describing the correlation of pixels beyond just the adjacent ones. This, however, requires a deep knowledge of each instrument, which is beyond the scope of this paper. 
}}

\section{Comparison with LAMOST DD-Payne Abundances} \label{app:lamost_compare}
\edit1{\added{
In \S\ref{sec:comparison}, we found our CRLBs for D1200G to be in good agreement with the precision reported by \citet{kirby:2018}. D1200G observations of metal-poor RGB stars, however, provide only a single point of comparison between our forecasts and what might be expected in practice.  Because so few full spectral fitting techniques are currently used in extragalactic contexts, similar comparisons are quite challenging.
}}

\edit1{\added{
Instead, we turn to an example within the Galaxy to provide an additional comparison.  Specifically, we compare our CRLBs to the internal precision reported by \citet{xiang:2019} for observations of MW stars by the LAMOST spectrograph \citep{cui:2012}. \citet{xiang:2019} employed the \texttt{DD-Payne}\footnote{The \texttt{DD-Payne} is a hybrid spectral model that is trained on high-resolution measurements from GALAH and APOGEE and regularized on \textit{ab initio} spectral gradients.} for full-spectral fitting and used repeat observations to quantify the internal precision of their measurements.
}}

\edit1{\added{
Because LAMOST observed primarily MW stars, we calculate the CRLBs for a typical solar-metallicity K-Giant star ($T_{\text{eff}}=4800$ K, $\log(g)=2.5$, $v_{\text{micro}}=1.7$ km/s, $\log(Z)=0$, and solar abundance patterns). To estimate the S/N of the LAMOST spectra, we use the mean flux variance from several LAMOST spectra of giant stars with a g-band S/N of 50 pixel$^{-1}$. As in our comparison to \citet{kirby:2018}, we make several cuts on the sample in order to fairly compare the reported precision with our CRLBs, which we list in Table \ref{tab:lamost_cuts}. These cuts leave the reported precision for approximately 6000 stars.
}}

\begin{table}[ht!] \label{tab:lamost_cuts}
\begin{center}
	\caption{Cuts on LAMOST DR5} 
    \begin{tabular}{c}
    \hline \hline
	$4600 < T_{\text{eff}}~(\text{K}) < 5000$\\
    $2.3 < \log(g) < 2.7$ \\
    $-0.1 < \text{[Fe/H]} < 0.1$ \\
    $-0.1 < [\alpha/\text{Fe}] < 0.1$ \\
    $40 < \text{g-band S/N (pixel}^{-1})< 60$ \\
    $\chi^2~\text{Flag} = \text{good}$ \\
    $\text{[X/Fe] Flag} = 1$ \\
    \tableline
	\end{tabular}
\end{center}
\end{table}

\edit1{\added{
Because \citet{xiang:2019} reports their abundance precision in terms of [X/Fe], we add $\sigma$[Fe/H] in quadrature to $\sigma$[X/Fe] so that the CRLBs are on the same scale. \citet{xiang:2019} does provide estimated systematic uncertainties for their measurements, but since CRLBs are a measure of precision and not accuracy, we do not include them in this comparison. 
}}

\begin{figure*}[ht!]
	\includegraphics[width=\textwidth]{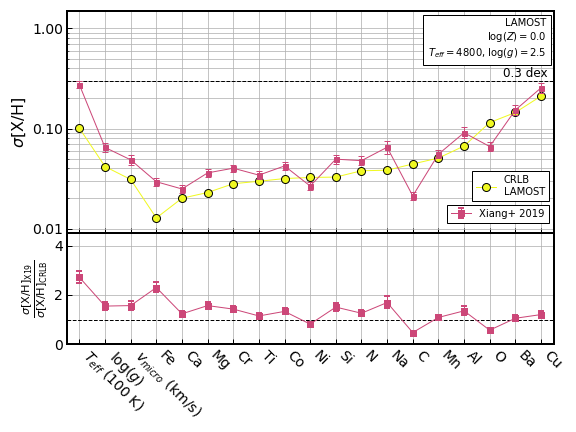}
    \caption{(Top) LAMOST CRLBs for a typical solar-metallicity K-Giant with a g-band S/N of 50 pixel$^{-1}$ over-plotted with the internal precision of $\sim$6000 comparable stars report by \citet{xiang:2019}. Error bars denote the upper and lower quartiles of the sample's precision. (Bottom) The ratio of the forecasted LAMOST CRLBs to the reported precision for each stellar label. As found with the comparison to \citet{kirby:2018} in Figure \ref{fig:DEIMOS_lit}, the measurement uncertainties for most elements are generally a factor of $\lesssim$2 larger than the CRLBs. The reported precision for Ni, C, and O slightly out-perform the CRLBs, which may be the result of additional spectral information included by the data-driven model of \citet{xiang:2019} that is not incorporated in our purely \textit{ab initio} model.} 
    \label{fig:LAMOST_lit}
\end{figure*}

\edit1{\added{
Figure \ref{fig:LAMOST_lit} shows the reported measurement precision of these stars compared to our LAMOST CRLBs. Similar to our comparison with \citet{kirby:2018}, we find that most abundances reported by \citet{xiang:2019} are within a factor of $\sim$2 of our CRLBs. The largest difference is in the precision of $T_\text{eff}$, which is reported to be 27 K, nearly 3 times larger than our predicted precision (10 K). This is not wholly unreasonable given the subtle and highly model-dependent effects that $T_\text{eff}$ has on spectral features. The reported precision for Fe (0.029 dex) is also more than a factor of two larger than our forecast (0.013 dex)---though the absolute difference is quite small. We suspect this is driven by the larger uncertainties found for $T_\text{eff}$ and $\log(g)$ by \citet{xiang:2019} and the substantial correlation these labels have with Fe in giant stars. 
}}

\edit1{\added{
Interestingly, we find that the precision reported for Ni, O, and C outperforms the CRLB by a factor of 1.2, 1.7, and 2.1. We suspect that this might be the result of ``gradient aliasing" in the \texttt{DD-Payne}, whereby the model picks up spectral gradient features from elements other than the one it attributes them to. This is a common challenge in data-driven methods, and while \citet{xiang:2019} attempted to mitigate it by regularizing the model with \textit{ab initio} spectral gradients, some gradient aliasing may remain.
For the remaining abundances, there are several reasons why slightly poorer precision might be expected in practice, including model fidelity and imperfect calibrations (see \S\ref{sec:caveats} for further discussion).
}}

\edit1{\added{
Together, the comparisons conducted here and in \S\ref{sec:comparison} illustrate that the CRLBs are quite reasonable representations of contemporary abundance measurements.
}}

\section{Validation of CRLBs} \label{app:validation}
To validate the robustness of the CRLBs, we infer the stellar labels of a mock spectrum at various S/N using an \textit{ab initio} trained spectral model and Hamiltonian-Monte Carlo sampling method and compare the precision of this inference with the precision forecasted by the CRLBs. We outline the process of training the spectral model in \S \ref{app:train} and fitting the mock spectrum in \S \ref{app:hmc}. The results of this comparison is presented in \S \ref{app:crlb_hmc_compare}.

\subsection{Training a Spectral Model}\label{app:train}
Training a spectral model requires a large set of stellar spectra with known labels that span the relevant parameter space. To generate this training set, we randomly drew $10^4$ stellar labels from the following uniform distribution\footnote{$\mathcal{O}(10^3)$ stellar spectra would likely have been sufficient, but opted to generate $10^{4}$ to further reduce emulation errors.}:
\begin{align*}
    T_\text{eff} &\sim \mathcal{U}(4500 \text{~K}, 5000\text{~K}), \\
    \log(g) &\sim \mathcal{U}(1.5, 2.1), \\
    v_{micro} &\sim \mathcal{U}(1.4\text{~km/s}, 2.4\text{~km/s}), \text{~and} \\
    [\text{X/H}] &\sim \mathcal{U}(-0.5, 0.5),\\
\end{align*}
where in this case X refers to a smaller subset of elements: Fe, Ca, Ni, Si, Ti, Mg, and Co. We only considered 7 elements, limiting the model to 10 stellar labels, to simplify the training process. These specific elements were chosen as they are the most precisely recovered elements by the D1200G setup (see \S \ref{sec:d1200g} and Table \ref{tab:instruments}). The bounds of the uniform distributions are chosen to center on the parameters of our fiducial RGB star (Table \ref{tab:ref_stars}) and span roughly 2 times the D1200G ($\text{S/N}=50$) CRLB for each stellar label, assuming the Gaussian priors of $\sigma_{T_\text{eff}}=100$ K, $\sigma_{\log(g)}=0.15$, and $\sigma_{\text{micro}}=0.25$ km/s used previously in \S \ref{sec:d1200g_priors}. Spectra were generated and convolved to instrumental resolution as previously described in \S \ref{sec:gradients}. 

Withholding 2500 spectra for validation, we train an updated version of \texttt{The Payne}\footnote{\url{https://github.com/tingyuansen/The_Payne}} (details in Table \ref{tab:model_par}). To aid the training process, the labels are normalized according to
\begin{equation}
    \theta_{i,\text{scaled}} = \frac{\theta_{i} - \theta_{i, \text{min}}}{\theta_{i, \text{max}} - \theta_{i, \text{min}}} - 0.5,
\end{equation}
where $\theta_{i, \text{min}}$ and $\theta_{i, \text{max}}$ are the minimum and maximum values represented in the training and validation datasets.
After $10^5$ training steps, which takes roughly 4 hours on an NVIDIA K80 GPU, the model which minimized the L1 mean loss on the validation spectra is chosen as the final model.

\begin{table}[ht!] \label{tab:model_par}
\begin{center}
	\caption{Details of \texttt{The Payne}.} 
    \begin{tabular}{lc}
    \hline \hline
		 \# Training Spectra & 7500 \\
		 \# Validation Spectra & 2500 \\
		 \# Spectra / Batch & 512 \\
		 \# Hidden Dense Layers & 2 \\
		 \# Neurons / Layer & 300 \\
		 Activation Function & Leaky ReLU \\
		 \# Training Steps & $10^{5}$ \\ 
		 Loss Function & L1 Mean\\
		 Optimizer & Rectified Adam\\
		 Learning Rate & $10^{-3}$ \\
		 Interpolation Errors & $<0.1\%$ \\
    \tableline
	\end{tabular}
\end{center}
\end{table}

We compare \textit{ab initio} spectra from our validation set to spectra generated with the same labels using the Payne and find mean interpolation errors of individual pixels to be less than 0.1\%. These errors are much smaller than typical observational uncertainties in the normalized spectra.

\subsection{Fitting Mock Spectra with HMC Sampling} \label{app:hmc}
The mock spectrum is generated using \texttt{The Payne} at the labels of the fiducial $\log(Z)=-1.5$ RGB star to avoid introducing any bias that may have been introduced in the training of the spectral model---recall that we are interested in precision, not accuracy, here. We assume a constant S/N across the entire spectrum, which manifests as an uncertainty in each pixel of $\sigma=f(\lambda)/(\text{S/N})$, where $f(\lambda)$ is the normalized flux of the model. With the same mock spectrum, we perform the fitting assuming a range in S/N from 5 to 200 pixel$^{-1}$ that is constant across the entire wavelength coverage.

With only 10 stellar labels and likelihoods that we believe to be close to Gaussian, using a Markov-Chain Monte Carlo (MCMC) sampling technique would likely be adequate for this scenario. However, because our neural network spectral emulator is differentiable, we opt to use a Hamiltonian Monte Carlo (HMC) sampler, making it readily adapted for inference with many more labels where an MCMC sampler might face convergence problems.

We adopt the Gaussian likelihood function in Equation \ref{eq:loglike} and the following priors:
\begin{align*}
    T_\text{eff} &\sim \mathcal{N}^*(4750 \text{~K}, 100\text{~K}), \\
    \log(g) &\sim \mathcal{N}^*(1.8, 0.15), \\
    v_{micro} &\sim \mathcal{N}^*(1.9\text{~km/s}, 0.25\text{~km/s}), \text{~and} \\
    [\text{X/H}] &\sim \mathcal{U}(-0.5, 0.5), \text{~and} \\
    [\text{X}^*\text{/H}] &\sim \delta(0.0), \\
\end{align*}
where $\mathcal{N}^*(\mu,\sigma)$ represents a normal distribution truncated at the limits of the training set so that the model does not extrapolate. Here, X$^*$ refers to elements that the CRLB predict cannot be recovered to better than 0.3 dex at the given S/N. These elements are held fixed at Solar value, which is equivalent to applying a delta function prior at [X/H$]=0.0$. The fixed labels at each S/N are displayed in Table \ref{tab:free_par}.

\begin{table}[ht!] \label{tab:free_par}
\begin{center}
	\caption{Fixed stellar labels at each S/N.} 
    \begin{tabular}{ll}
        \hline \hline
        S/N (pix$^{-1}$)& Fixed Labels\\
        \hline
	    5, 10 & [Ni/H], [Si/H], [Ti/H], [Co/H], [Mg/H] \\
	    15 & [Si/H], [Ti/H], [Co/H], [Mg/H] \\
	    20 & [Co/H], [Mg/H] \\
	    30, 50, 100, 200 & None \\
    \tableline
	\end{tabular}
\end{center}
\end{table}

For each S/N we perform the HMC sampling using 24 parallel chains. Each chain begins with 3000 burn-in samples, which are discarded, followed by another 3000 samples, which constitute our posterior sample. 


\subsection{Comparison to CRLB} \label{app:crlb_hmc_compare}
\begin{figure*}[ht!]
	\includegraphics[width=\textwidth]{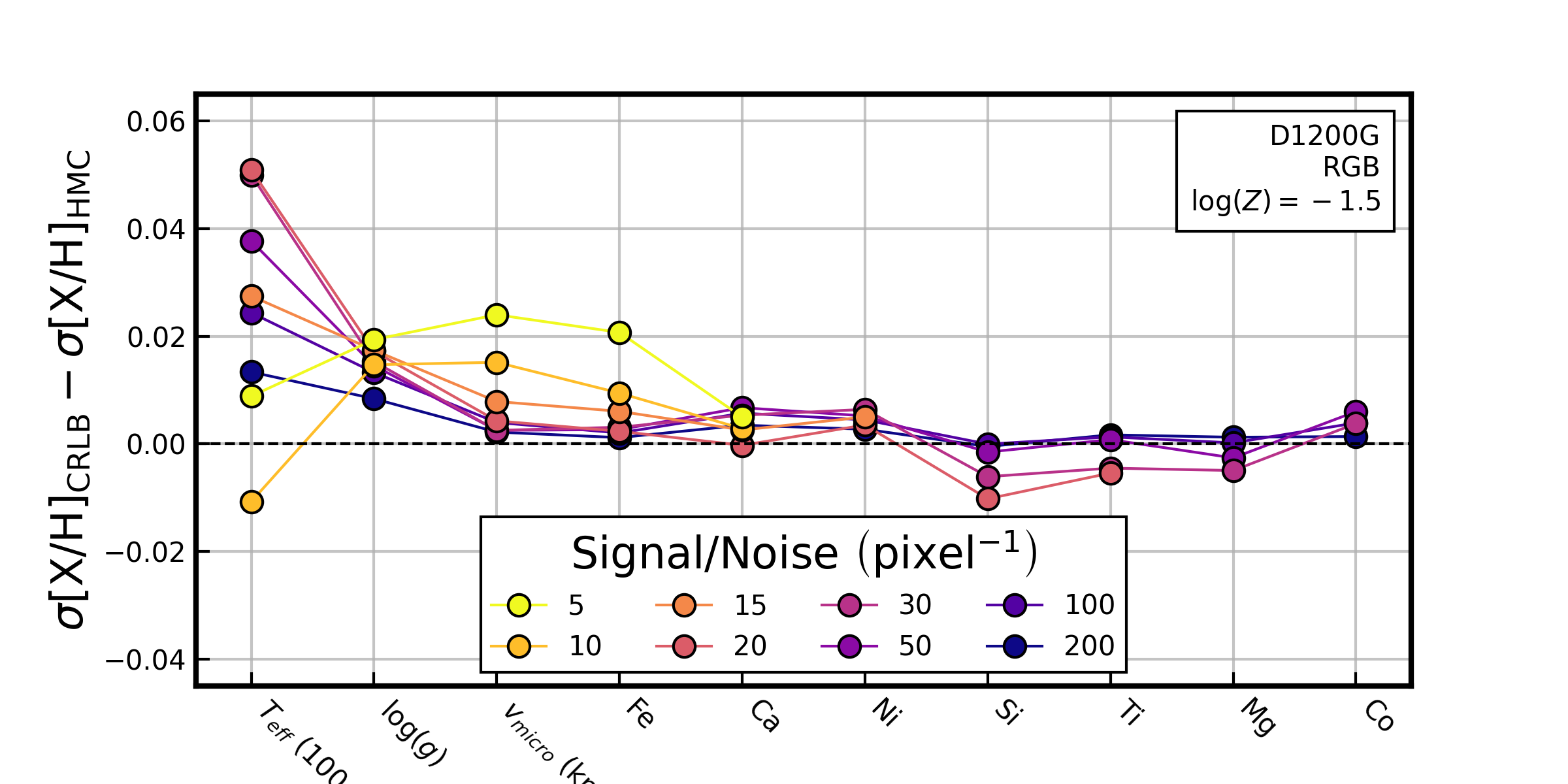}
    \caption{The difference between the CRLB and the stellar label precision found through HMC sampling for a $\log(Z)=-1.5$ RGB star observed with the D1200G setup. A constant S/N across the wavelength coverage was assumed. Differences are small ($\lesssim 5$ K for $\sigma_{T_\text{eff}}$; $\lesssim 0.02$ dex for $\sigma_{\log(g)}$; $\lesssim 0.02$ km/s for $\sigma_{v_\text{micro}}$; and $\lesssim 0.02$ dex for $\sigma_{\text{[X/H]}}$), indicating that the CRLB is a robust predictor of stellar label precision down to at least \edit1{\replaced{$\text{S/N}\sim5$}{$\text{S/N}\sim15$}} pixel$^{-1}$.} 
    \label{fig:crlb_valid}
\end{figure*}

In Figure \ref{fig:crlb_valid}, we plot the difference between the precision predicted by the CRLBs and the standard deviation of the mock fit posteriors for each S/N. In the calculation of the CRLBs, we include the same priors on $T_\text{eff}$, $\log(g)$, and $v_\text{micro}$ used in the HMC sampling. In addition, for each S/N, we only consider the gradients for the stellar labels that are left free in the sampling (see Table \ref{tab:free_par}), thus holding all other labels fixed at Solar values. Instead of calculating spectral gradients from \textit{ab initio} spectra, we calculate the gradients from our trained spectral model to exclude any systematics introduced by interpolation errors of the model.

In general, we find the CRLBs and the standard deviations of the mock fits to be in agreement at the 0.01 dex level down to a S/N of 10 and at the 0.02 dex level down to a S/N of 5. At very high S/N (200 pixel$^{-1}$), the CRLBs accurately predict the precision of the $v_\text{micro}$ and all chemical abundances, only very slightly under-predicting the precision of $T_\text{eff}$ by 1 K and $\log(g)$ by 0.01 dex. As the S/N decreases to 20 pixel$^{-1}$ the difference grows to 5 K and 0.02 dex in $T_\text{eff}$ and $\log(g)$ respectively, and the CRLBs slightly over-predict the precision for Si, Ti, and Mg by no more than 0.01 dex. All of these differences remain relatively small compared to the typical precision found for these labels and are the result of the posteriors of these labels being slightly non-Gaussian (negatively skewed).

As the S/N decreases further, the precision of both the mock fit and the CRLB become prior-dominated for $T_\text{eff}$, $\log(g)$, and $v_\text{micro}$, resulting in a smaller difference in the precision of $T_\text{eff}$. This is not the case for the precision of $\log(g)$ and $v_\text{micro}$ due to the difference between the Gaussian prior included in the CRLB calculation and the truncated Gaussian included in the HMC sampling. Still, the differences are only $\sim$ 0.02 dex, which is quite minor in relation to the expected precision at $\text{S/N}<15$. Thus we find that the CRLB is a robust predictor of stellar label precision down to at least a S/N of 15 pixel$^{-1}$.

\section{DESI CRLBs} \label{app:desi}
The Dark Energy Spectroscopic Instrument (DESI) is a fiber-fed MOS that covers a wavelength range from 3600 to 9800 $\text{\AA}$ with a resolving power of 2000-5000. The primary science goal of the DESI survey is not galactic archaeology, nor is the 4-m Mayall telescope it's mounted on large enough to efficiently observe resolved stars in dwarf galaxies. Nevertheless, it is a particularly interesting spectrograph for stellar chemical abundance measurements. When observing conditions are too poor for faint galaxy work, DESI will target bright galaxies, filling unused fibers with MW stars. This will yield spectra for roughly 10 million MW stars. In addition to many thin and thick disk stars, these deep observations are expected to reach MSTO stars in the MW's halo out to 30 kpc, allowing for a dramatically improved understanding of the stellar halo's chemical composition. In addition, DESI's instrumental design has been a major inspiration for current and next-generation survey instruments that will be targeting stars in dwarf galaxies.

Thus, while DESI will not be observing dwarf galaxy stars, we still think it valuable to present the theoretical abundance precision achievable by DESI in the MW halo. For these calculations we assume a uniform S/N of 30 pixel$^{-1}$, which should be achievable for stars of $m_r=16.5$-18 in a short 5-10 minute exposure \cite{desi:2016_sci}. The spectroscopic configuration used is given in Table \ref{tab:instruments}. Because DESI will be able to observe down to the MSTO in the halo, we calculate the CRLBs for MSTO, RGB, and TRGB stars as done for D1200G in \S \ref{sec:d1200g_phase}.

\begin{figure*}[ht!]
	\includegraphics[width=\textwidth]{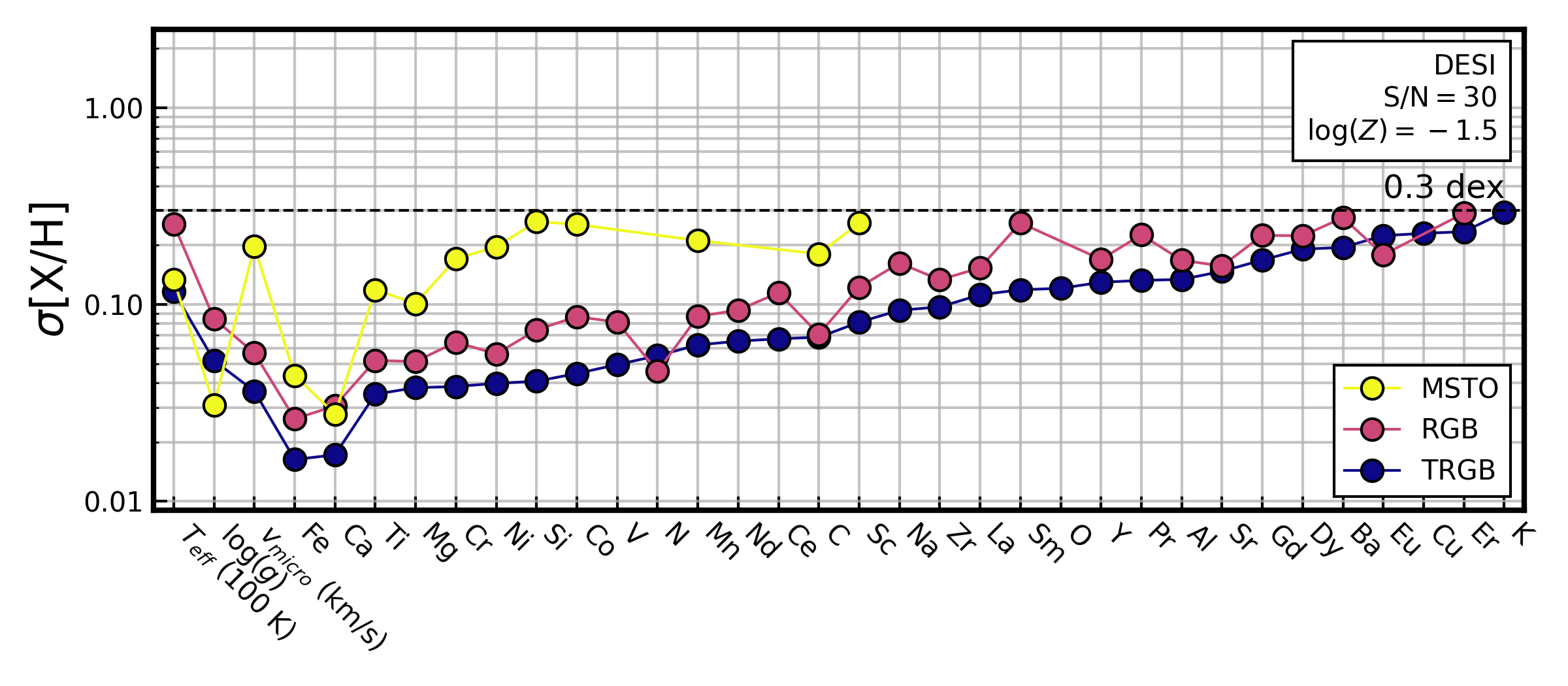}
    \caption{DESI CRLBs of $\log(Z)=-1.5$ MSTO, RGB, and TRGB stars with a constant S/N of 30 pixel$^{-1}$. The atmosphere parameters for each star can be found in Table \ref{tab:ref_stars}. Just as for D1200G, abundance recovery is more precise for cool giants and less precise for hot sub-giants.} 
    \label{fig:crlb_desi}
\end{figure*}

In Figure \ref{fig:crlb_desi}, we plot the CRLBs for DESI, illustrating its capability to extend the precise chemical abundance measurements of MW-disk surveys out to the MW's halo. As seen for D1200G in Figure \ref{fig:crlb_phase}, abundance recovery is more precise for cool giants due to stronger absorption features and less precise for hot sub-giants, which have weaker absorption features.

%
%
%
%


\end{document}